\documentclass[longauth]{aa}   

\usepackage{graphicx}
\usepackage{txfonts}
\usepackage{subcaption}         
\usepackage{lscape}             
\usepackage{placeins}                                            
\usepackage{natbib}
\usepackage[utf8]{inputenc}
\usepackage{amsmath}
\usepackage{enumitem} 
\usepackage[colorlinks=True]{hyperref}
\hypersetup{
    linkcolor=blue,  
    citecolor=blue,
    }

\newcommand{\HalphaNII}{H$\alpha$+[N II]}
\newcommand{\NII}{[N II]}
\newcommand{\OII}{[O II]}
\newcommand{\Rvir}{$R_\mathrm{vir}$}

\begin{document}

   \title{The S-PLUS Fornax Project (S+FP): Mapping \HalphaNII \,emission in 77 Fornax galaxy members reaching $\sim$4 Rvir}
   
   \author{A. R. Lopes\inst{\ref{inst.IALP}}\thanks{\email{amandalopes1920@gmail.com}}
   \and A. V. Smith Castelli\inst{\ref{inst.IALP},\ref{inst.FCAL}}
   \and A. C. Krabbe\inst{\ref{inst.IAG}}
   \and J. A. Hernandez-Jimenez\inst{\ref{inst.UID}}
   \and D. Pallero\inst{\ref{inst.UTFSM}}
   \and S. Torres-Flores\inst{\ref{inst.ULS}}
   \and E. Telles\inst{\ref{inst.ON}}
   \and M. Sarzi\inst{\ref{inst.Armagh}}
   \and A. Cortesi\inst{\ref{inst.IF-UFRJ},\ref{inst.OV}}
   \and J. Thain\'{a}-Batista\inst{\ref{inst.UFSC}}
   \and R. Cid Fernandes\inst{\ref{inst.UFSC}}
   \and E. A. D. Lacerda\inst{\ref{inst.IAG}}
   \and M. Sampaio\inst{\ref{inst.IAG}}
   \and V. H. Sasse\inst{\ref{inst.UFSC}}
   \and F. R. Herpich\inst{\ref{inst.CASU}}
   \and I. Andruchow\inst{\ref{inst.FCAL},\ref{inst.IAR}}
   \and R. Demarco\inst{\ref{inst.UAB}}
   \and L. A. Gutiérrez Soto\inst{\ref{inst.IALP}}   
   \and M. Grossi\inst{\ref{inst.OV}}
   \and R. F. Haack\inst{\ref{inst.IALP},\ref{inst.FCAL}}
   \and P. K. Humire\inst{\ref{inst.IAG}}
   \and C. Lima-Dias\inst{\ref{inst.IMIP}}
   \and G. Limberg\inst{\ref{inst.Kavli}}
   \and C. Lobo\inst{\ref{inst.IACE},\ref{inst.UP}}
   \and L. Lomel\'i-N\'u\~nez\inst{\ref{inst.OV}}
   \and P. A. A. Lopes\inst{\ref{inst.OV}}
   \and D. E. Olave-Rojas\inst{\ref{inst.UTalca}}
   \and S. V. Werner\inst{\ref{inst.Durham}}
   \and F. Almeida-Fernandes\inst{\ref{inst.OV},\ref{inst.INPE}}
   \and G.B. Oliveira Schwarz\inst{\ref{inst.IAG},\ref{inst.Mack}}
   \and W. Schoenell\inst{\ref{inst.GMT}}
   \and T. Ribeiro\inst{\ref{inst.Rubin}}
   \and A. Kanaan\inst{\ref{inst.UFSC}}
   \and C. Mendes de Oliveira\inst{\ref{inst.IAG}}
    }

   \institute{\label{inst.IALP}Instituto de Astrof\'{i}sica de La Plata, CONICET-UNLP, Paseo del Bosque s/n, B1900FWA, Argentina
   \and \label{inst.FCAL}Facultad de Ciencias Astron\'{o}micas y Geof'{i}sicas, Universidad Nacional de La Plata, Paseo del Bosque s/n, B1900FWA, Argentina
   \and \label{inst.IAG}Universidade de S\~ao Paulo, IAG, Rua do Mat\~ao 1226, S\~ao Paulo, SP, Brazil
   \and \label{inst.UID}Universidad de Investigaci\'{o}n y Desarrollo, Departamento de Ciencias Básicas y Humanas, Grupo FIELDS, Calle 9 No. 23-55, Bucaramanga, Colombia
   \and \label{inst.UTFSM}Departamento de F\'{i}sica, Universidad T\'{e}cnica Federico Santa María, Avenida España 1600, 2390123, Valpara\'{i}so, Chile
   \and \label{inst.ULS}Departamento de Astronom\'{i}a, Universidad de La Serena, Av. J. Cisternas 1200 N, 1720236 La Serena, Chile
   \and \label{inst.ON}Observatório Nacional, Rua General Jos\'{e} Cristino, 77, S\~ao Crist\'{o}v\~ao, 20921-400 Rio de Janeiro, RJ, Brazil
   \and \label{inst.Armagh}Armagh Observatory and Planetarium, College Hill, Armagh BT61 9DG, UK
   \and \label{inst.IF-UFRJ}Instituto de F\'{i}sica, Universidade Federal do Rio de Janeiro, 21941-972, Rio de Janeiro, Brazil
   \and \label{inst.OV}Observat\'{o}rio do Valongo, Ladeira Pedro Antonio, 43, 20080-090, Rio de Janeiro, Brazil 
   \and \label{inst.UFSC}Departamento de F\'{i}sica - CFM - Universidade Federal de Santa Catarina, PO BOx 476, 88040-900, Florian\'{o}polis, SC, Brazil
   \and \label{inst.CASU}Cambridge Survey Astronomical Unit (CASU), Insitute of Astronomy, University of Cambridge, Madingley Road, Cambridge, CB3 0HA, GB, UK
   \and \label{inst.IAR}Instituto Argentino de Radioastronom\'{i}a, CONICET-CICPBA-UNLP, CC5 (1894) Villa Elisa, Provincia de Buenos Aires, Argentina
   \and \label{inst.UAB}Institute of Astrophysics, Facultad de Ciencias Exactas, Universidad Andr\'es Bello, Sede Concepci\'on, Talcahuano, Chile
   \and \label{inst.IMIP}Instituto Multidisciplinario de Investigaci\'{o}n y Postgrado, Universidad de La Serena, Raúl Bitrán 1305, 1700000 La Serena, Chile
   \and \label{inst.Kavli}Kavli Institute for Cosmological Physics, University of Chicago, Chicago, IL 60637, USA
   \and \label{inst.IACE} Instituto de Astrof\'{i}sica e Ci\^{e}ncias do Espaço, Universidade do Porto, CAUP, Rua das Estrelas, 4150-762, Porto, Portugal
   \and \label{inst.UP} Departamento de F\'{i}sica e Astronomia, Faculdade de Ci\^{e}ncias, Universidade do Porto, Rua do Campo Alegre 687, 4169-007, Porto, Portugal
   \and \label{inst.UTalca}Departamento de Tecnolog\'{i}as Industriales, Facultad de Ingenier\'{i}a, Universidad de Talca, Los Niches km 1, Curic\'{o}, Chile
   \and \label{inst.Durham}Centre for Extragalactic Astronomy, Department of Physics, Durham University, South Road, Durham DH1 3LE, UK
   \and \label{inst.INPE}Instituto Nacional de Pesquisas Espaciais, Av. dos Astronautas 1758, Jardim da Granja,12227-010 S\~ao Jos\'e dos Campos, SP, Brazil
   \and \label{inst.Mack}Universidade Presbiteriana Mackenzie, Rua da Consola\c{c}\~ao, 930 - Consola\c{c}\~ao, São Paulo, Brazil
   \and \label{inst.GMT}GMTO Corporation 465 N. Halstead Street, Suite 250 Pasadena, CA 91107, USA
   \and \label{inst.Rubin}Rubin Observatory Project Office, 950 N. Cherry Ave., Tucson, AZ 85719, USA   
   }

   \date{Received September 30, 20XX}

  \abstract
   {The Fornax cluster, the second-largest galaxy cluster within 20 Mpc, presents an ideal environment for studying environmental effects on galaxy evolution. Utilizing data from the Southern Photometric Local Universe Survey (S-PLUS), this study explores the \HalphaNII{} emission maps in Fornax and its outskirts.}
   {By mapping emission features across an area of approximately 208 square degrees around NGC 1399, this work aims to identify and characterize emission-line galaxies (ELGs) and analyze their spatial distribution, morphology, and their projected phase space (PPS) diagram. } 
   {To achieve these objectives, a dedicated semi-automated pipeline, Pixel-to-Pixel Emission Line Estimate (PELE), was developed to generate emission line maps by processing S-PLUS images using the Three Filter Method. A morphological analysis was conducted using the ASTROMORPHLIB package to determine whether \HalphaNII{} emitters exhibit perturbed features.}
   {The study successfully detected 77 \HalphaNII{} emitters with $r_{AB}<18$ mag, extending to four times the virial radius of the Fornax cluster. PELE demonstrated its ability to recover flux down to $\sim2\times10^{-17}$ erg s$^{-1}$ cm$^{-2}$ when compared to H$\alpha$ maps from MUSE/VLT. Among the emitters, 25\% are early-type galaxies (ETG) and 75\% late-type galaxies (LTG). Signs of morphological perturbation or merger activity are observed in 44\% of the LTG and in three ETG located beyond the cluster’s virial radius. A significant fraction (91\%) of the emitters are identified as recent infallers, primarily located in the northwestern region of the cluster, while others are associated with the infalling group Fornax\,A in the southwest. Disturbed, low-mass galaxies at larger cluster-centric distances provide evidence of galaxies begin transforming before entering the main cluster.}
   {This study demonstrates S-PLUS’s effectiveness in detecting ELGs, whose distribution reflects the Fornax cluster’s assembly history, with LTG linked to recent infall from the field, possibly along a Fornax-Eridanus filament, and ETG may have evolved prior to entry.}

\titlerunning{\HalphaNII \,emission reaching $\sim$ 4 Rvir in the Fornax cluster}
   \maketitle

\section{Introduction}
The gas content of galaxies located in the cores of rich clusters is expected to be depleted due to the interaction with the hot intracluster medium \citep[e.g.][]{1972/Gunn,1983/Giovanelli,2017/Jachym,2022/Pedrini}. Additional processes, such as tidal interactions between galaxies and galaxy-galaxy collisions, also contribute to this depletion \citep[e.g.][]{1988/Combes,1995/Kenney,2009/Vollmer,2022/Spilker}. As a consequence, galaxies placed in the central regions of galaxy clusters are unable to sustain intense star formation activity and, in comparison with galaxies in low-density environments, tend to display poor amounts of cold gas. However, the quenching found in the outskirts of galaxy clusters cannot be explained by infalling processes. In contrast, such a suppression of star formation is more likely linked to a pre-processing effect, i.e. galaxies are organized in groups or subclusters that later fell into the main cluster \citep{1996/Zabludoff,2003/Fujita,2013/Haines,2015/Haines,2019/Fossati,2024/Lopes}. 

In order to disentangle the time scales involved in the star formation activity of a galaxy, we can take advantage of several spectroscopic features. As an example, the star formation occurred in the last 10 Myr is traced by the H$\alpha$ emission line detected in H II regions, while the H$\delta$ (D4000) absorption feature together with the $(g-r)$ color index can trace the mean star formation rate (SFR) of the last 800 Myr \citep{1999/Balogh,2003/Kauffmann}. These features allow us to distinguish between different quenching processes, as ram-pressure stripping is expected to stop star formation in the order of a few hundreds of Myr in contrast to, for example, starvation which takes of the order of a few Gyr to produce a similar effect (e.g. \citealt{2013/Wetzel,2014/Boselli}).

The Fornax galaxy cluster is a remarkable place to analyze the influence of a high density environment in the star formation history of the galaxies. With a low virial mass ($M_\mathrm{vir}$) of $(7 \pm 2)\times 10^{13} M_\odot$ within a projected radius of 1.4 Mpc \citep{2001/Drinkwater}, it is the second well populated galaxy cluster within 20 Mpc \citep{2009/Blakeslee} after Virgo. Due to its also low total X-ray luminosity \citep[$5 \times 10^{41}$ erg s$^{-1}$, ][]{1997/Jones}, the presence of a large amount of hot intracluster gas is not expected. As a consequence, ram-pressure stripping would not be the dominant mechanism of gas loss in Fornax' galaxies \citep[][and references therein]{1995/Horellou}. In contrast, Fornax displays a higher galaxy number density in its center and a lower velocity dispersion than Virgo pointing to an environment in which tidal interactions are more effective than hydrodynamical effects \citep{2019/Maddox}. However, it has been found that H I-rich galaxies in Fornax display weak H$\alpha$ emission which is consistent with a general low star formation activity as revealed by low far-infrared and nonthermal radiocontinuum emissions \citep[][and references therein]{1995/Horellou}. 

Fornax is still growing by accreting new galaxies and nearby galaxy groups. As an example, Fornax\,A group, centered around NGC\,1316, is a galaxy group at the edge of the cluster and moving towards its center, possibly along a large-scale filament \citep{2005/Scharf,2019/Venhola}. According to \citet{2020/Raj}, Fornax\,A is in an early stage of assembly, as shown by the fact that there is no clear trend between the photometric properties of its galaxies and the group-centric distances. In this context, there is observational evidence indicating that NGC\,1316 arose  from  a merger between a gas-poor early-type galaxy and a smaller, gas-rich spiral, just 1 to 3 Gyr ago \citep{2019/Serra}. Additionally, \citet{2021/Kleiner} analyzed 12 galaxies in the Fornax A group using MeerKAT H I observations and found evidence of pre-processing in 9 of them. These galaxies were classified according to their pre-processing stage, categorized as early, ongoing, or advanced. Their study highlighted that pre-processing stages could vary significantly even within the same group, revealing complex interactions that influence galaxy evolution before they merge with the main cluster.

The Fornax system (Fornax and Fornax\,A) has been extensively studied through complementary surveys. The Fornax Deep Survey (FDS) has performed deep imaging in four broad bands ($u$, $g$, $r$, $i$) using the Very Large Telescope (VLT) located at Paranal Observatory, Chile. This survey covers an area of $\sim$26 square degrees around NGC\,1399, including the Fornax\,A group. This effort led to a series of publications exploring various components of the Fornax system, such as the extended and diffuse stellar halo of NGC\,1399 \citep{2016/Iodice}, low surface brightness galaxies \citep{2017/Venhola, 2022/Venhola}, dwarf galaxies \citep{2018/Venhola, 2019/Venhola}, bright early type galaxies \citep{2019a/Iodice, 2020/Spavone}, late type galaxies \citep{2019/Raj}, globular clusters \citep{2020/Cantiello}, Fornax\, A galaxy assemble and its intra-group light \citep{2017/Iodice, 2020/Raj}, and the signs of pre-processing in the Fornax system \citep{2021/Su}. In parallel, the Fornax3D project (F3D; \citealt{2018/Sarzi}) observed 31 galaxies in the Fornax cluster using the Multi Unit Spectroscopic Explorer \citep[MUSE,][]{2010/Bacon} to create a rich dataset that has allowed numerous studies. These includes analyses of kinematic and line-strength maps \citep{2019/Iodice}, planetary nebulae \citep{2020/Spriggs,2021/Spriggs}, globular clusters \citep{2020/Fahrion,2020B/Fahrion}, stellar kinematics, dynamical and population analysis for lenticular galaxies \citep{2019/Pinna,2019B/Pinna,2021/Poci}, gas metallicity gradients \citep{2022/Lara}, and the assembly history of massive early-type galaxies \citep{2022/Spavone}. However, the F3D sample is limited to galaxies brighter than $m_{B}=15$ mag, introducing a bias against fainter galaxies. In addition, by design, it does not include objects located outside the \Rvir{} of the cluster and its group, Fornax\,A. Together, FDS and F3D have significantly advanced our understanding of galaxy evolution and the assembly history of the Fornax cluster, particularly within its virial radius \citep[\Rvir $=2$ deg, ][]{2001/Drinkwater}, shedding light on the formation of substructures in galaxy clusters.

In order to assess the recent star-forming activity in the Fornax cluster, its outskirts, and the Fornax\,A group following a comprehensive standard method, we search for galaxies whose \HalphaNII{} emission can be spatially resolved. The prime data for such analysis is obtained from integral field spectroscopic (IFS) surveys, such as MUSE. However, IFS surveys do not present extensive contiguous observed area, which renders difficult the analysis of the surrounding environment of nearby galaxies. In addition, the limited field of view (FOV) of instruments in these surveys creates a limitation for larger galaxies, such as NGC\,1365, that extends to a diameter higher than 10 arcmin and requires multiple pointings to cover the whole galaxy. 

To circumvent the aforementioned limitations of IFS, one effective approach is to use multi-band surveys. such as the Classifying Objects by Medium-Band Observations in 17 Filters \citep[COMBO-17,][]{2003/Wolf}, the Large Area Multimedium-Band Optical and Near-Infrared Photometric Survey \citep[ALHAMBRA,][]{2008/Moles}, the Multiwavelength Survey by Yale-Chile \citep[MUSYC,][]{2010/Cardamone}, the Survey for High-z Absorption Red and Dead Sources \citep[SHARDS,][]{2013/PerezGonzalez},  Cosmic Evolution Survey \citep[COSMOS]{2015/Taniguchi}, Javalambre Photometric Local Universe Survey \citep[J-PLUS]{2019/Cenarro} and Javalambre Physics of the Accelerating Universe Astrophysical Survey \citep[J-PAS]{2014/Benitez}. 
These surveys are designed with a set of intermediate and/or narrow-band filters that provide the intended spectral information while simultaneously encompassing large contiguous regions. Therefore, studies focusing in the detection of emission line galaxies can be performed using the released photometric catalogs through the analysis of, for example, color-color diagrams, or by building emission-line maps from narrow-band (NB) images \citep[e.g.,][]{2003/Fujita, 2004/Glazebrook, 2008/Geach, 2009/Sobral, 2011/Ly, Kellar2012, 2019/Cook, 2020/Khostovan, 2023/Salzer}. 

The Southern Photometric Local Universe Survey (S-PLUS; \citealp{2019/Mendes}) is an ongoing imaging project with 12-filters, aiming to cover $\sim 9300$ square degrees of the Southern sky. It uses a 0.8-meter robotic telescope, the T80-South, located at the Cerro Tololo Inter-American Observatory (CTIO) in Chile. It is equipped with the Javalambre filter system \citep{2019/Cenarro} which includes 5 SDSS-like broad bands ($u$, $g$, $r$, $i$, $z$), and 7 narrow bands ($J0378$, $J0395$, $J0410$, $J0430$, $J0515$, $J0660$, $J0861$), covering different spectral features. In particular, the $J0660$ filter catches \HalphaNII{} in galaxies at z $\lesssim$ 0.019, making it an invaluable tool to study the star formation history in the local Universe on a large scale. 

In that context, and as part of the S-PLUS Fornax Project (S+FP; \citealt{PaperI}, hereafter, Paper\,I), we present the analysis of the \HalphaNII{} emission in the Fornax galaxy cluster using images of S-PLUS. In its first stage, the S+FP is exploring Fornax up to $\sim$ 5 \Rvir{} in right ascension (RA) and up to $\sim$ 2.5 \Rvir{} in declination (Dec). That represents a sky projected area of $\sim$ 208 deg$^2$ around NGC\,1399, the dominant galaxy of the cluster. Given the unprecedented sky coverage of Fornax provided by S-PLUS, we are reaching the outskirts of the cluster where pre-processing is expected to play a key role. In this paper we focus on the identification of \HalphaNII{} spatially resolved emitters among the reported galaxy members of Fornax. This detection is performed using \HalphaNII{} maps generated with a code specially developed to that purpose. In addition, we analyze the morphologies displayed by the identified emitters in relation to their spatial distribution within the cluster and in the Projected Phase Space (PPS) diagram \citep[e.g. ][]{2017/Rhee}. The combination of such analysis allows us to identify recent and ancient infallers as well as perturbed and merged galaxies, which represents valuable information to give clues about the influence of the environment in the star formation history and morphological transformation of \HalphaNII{} emitters in Fornax. 

The paper is organized as follows. In Section \ref{sec:data} we describe the input galaxy catalog and the data images used here. Section \ref{sec:PELE} introduces the code applied to create emission line maps and its validation using IFS results. In Section \ref{sec:emitters}, we analyze the general characteristics of galaxies where the \HalphaNII{} maps were successfully obtained, such as optical and emission morphologies, spatial distribution, location in the phase-space diagram, and compare with literature results and preliminary analysis from simulations. In Section \ref{sec:conclusions} we present our conclusions.

Throughout this work we use AB magnitudes and a distance modulus of $(m-M) = 31.51$ mag is assumed for Fornax \citep{2009/Blakeslee}, and a mean redshift of $z = 0.0048$ \citep[considering a mean velocity $V_\mathrm{cluster}=1442$ km s$^{-1}$ around NGC\,1399; ][]{2019/Maddox}. We will consider, for the main structure of Fornax, \Rvir{} = 0.7 Mpc centered on NGC\,1399, and for the Fornax\,A group, \Rvir{} = 0.35 Mpc centered on NGC\,1316 \citep{2001/Drinkwater}. At the assumed Fornax distance, 1 arcsec subtends 0.1 kpc.

\section{Data description}
\label{sec:data}
\subsection{Fornax galaxy members}
\label{subsec:fornax_members}

As outlined in the previous section, the goal of the S+FP is to perform a comprehensive analysis of the Fornax galaxy cluster over more than 200 deg$^2$, using homogeneous data taken through the 12 optical filters of S-PLUS. In its first stage, the S+FP consists of 106 S-PLUS contiguous fields around NGC\,1399, all of them observed in the S-PLUS Main Survey (for more details in the different sub-surveys of S-PLUS, see \citealt{2019/Mendes}). This set of fields extends out till $\sim$ 5\Rvir{} in RA and $\sim$ 2.5\Rvir{} in Dec, covering the region 40$^{\circ}\lesssim$ RA $\lesssim70^{\circ}$, $-38^{\circ}\lesssim$ Dec $\lesssim-28^{\circ}$ (see Figure\,1 in Paper\,I), and reaching the outskirts of the cluster where pre-processing is expected to occur. 

From the literature, we have compiled a list of 1\,005 galaxies reported as members of Fornax (see Paper\,I). Among them, and within the region covered by the S+FP, only 233 objects have radial velocities ranging between 600 and 3\,000 km s$^{-1}$, consistent with being spectroscopically confirmed cluster members \citep{2019/Maddox}. 

\begin{figure}
    \centering
    \includegraphics[width=\hsize]{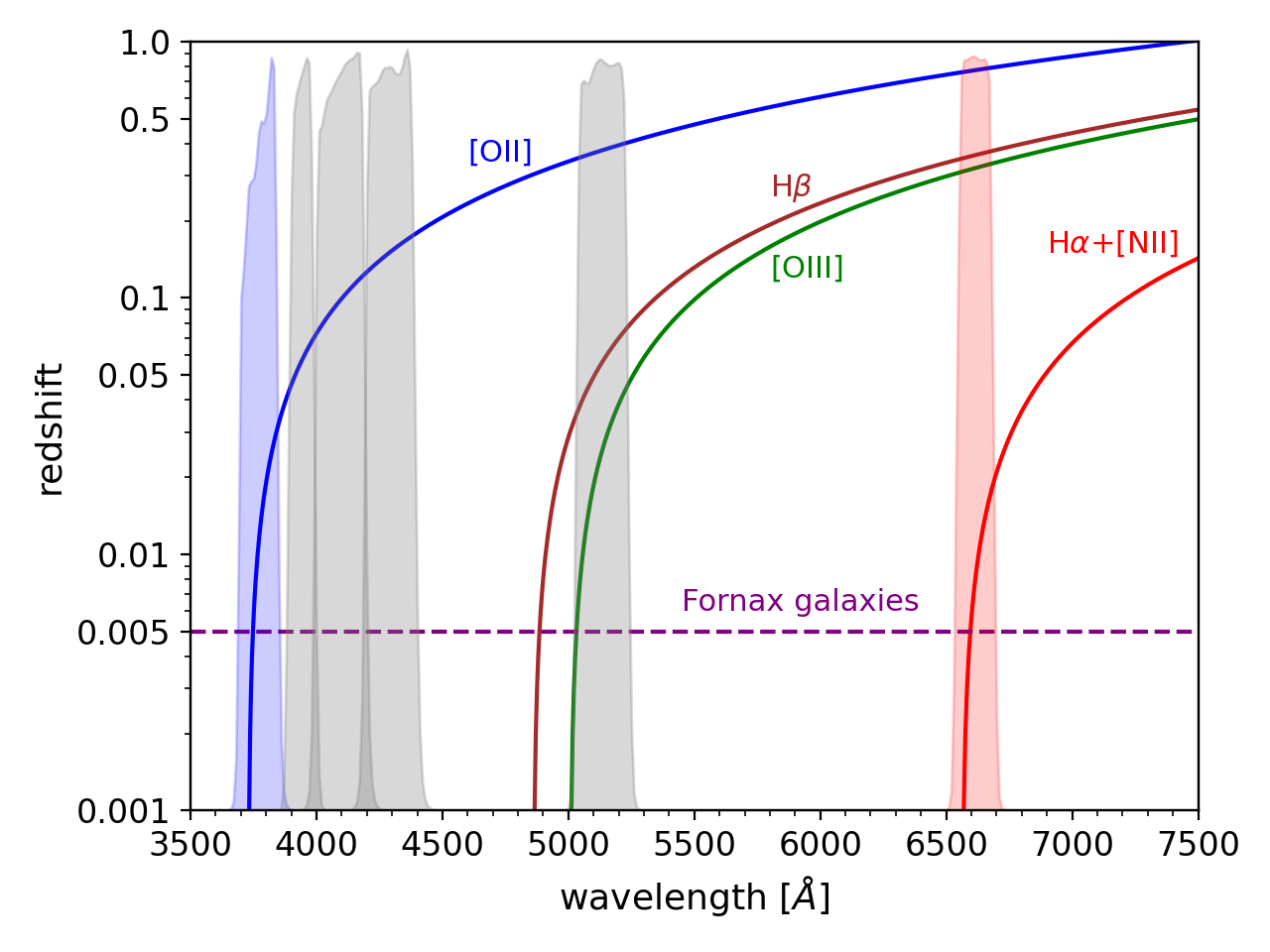}
    \caption{The redshift evolution of the observed wavelength of the \OII, [O III]$\lambda$5007, \HalphaNII{} and H$\beta$ emission lines. At the distance of the Fornax cluster (indicated by the purple dashed line), we can potentially detect two of these emission lines, \OII{} and \HalphaNII, in the narrow-band filters $J0378$ (blue region) and $J0660$ (red region), respectively. The grey shaded regions are the remaining narrow-band filters of SPLUS.}
    \label{fig:el_splus_filters}
\end{figure}

\subsection{S-PLUS datacubes}
\label{subs.:data}
We create 12-band datacubes for all galaxies in the literature described in Subsec. \ref{subsec:fornax_members} using the python package \texttt{S-Cubes}\footnote{Available at \url{https://splus-collab.github.io/s-cubes/}}. Given each object´s RA and Dec, the code automatically downloads cutout images and their corresponding weight images in all filters from the S-PLUS database\footnote{\url{https://splus.cloud}}. Then each image is converted to flux and calibrated using the zero-points from S-PLUS Data Release 4 \citep{2024/Herpich}, with their corresponding flux errors estimated. The resulting 12 flux and 12 error images are compiled into an output datacube file.

At the distance of the Fornax cluster, the filter configuration of S-PLUS allows the detection of the \OII{} and \HalphaNII{} emission lines by J0378 and J0660 bands, respectively, as shown in Fig.\,\ref{fig:el_splus_filters}. For this paper, we will focus on the spatial detection of \HalphaNII{}, in order to be able to validate our approach using MUSE data, which does not reach to \OII{} wavelength.

\subsection{Fornax3D/MUSE data}
\label{subsec:MUSE}
The Fornax3D project \citep[F3D,][]{2018/Sarzi} is an IFS survey observed with MUSE \citep[][]{2010/Bacon} attached at the Very Large Telescope of the European Southern Observatory. The sample contains 31 galaxies brighter than $m_{B}=15$ within the $R_\mathrm{vir}$ ($\sim2$ deg) of the Fornax cluster. The observation dates are between July 2016 and December 2017. The FOV is $1\times1$ arcmin with a pixscale of 0.2 arcsec. The spectral wavelength range goes from 4650 to 9300 \AA{} with a sampling of 1.25 \AA{} pixel$^{-1}$. Extended emission-line gas was detected in 13 of those objects \citep{2019/Iodice}, with most of them being late-type galaxies. The H$\alpha$ map obtained using these data will be used to validate our technique and to detect the limit associated to our photometric emission line estimates.

\begin{figure}
    \centering
    \includegraphics[width=\hsize]{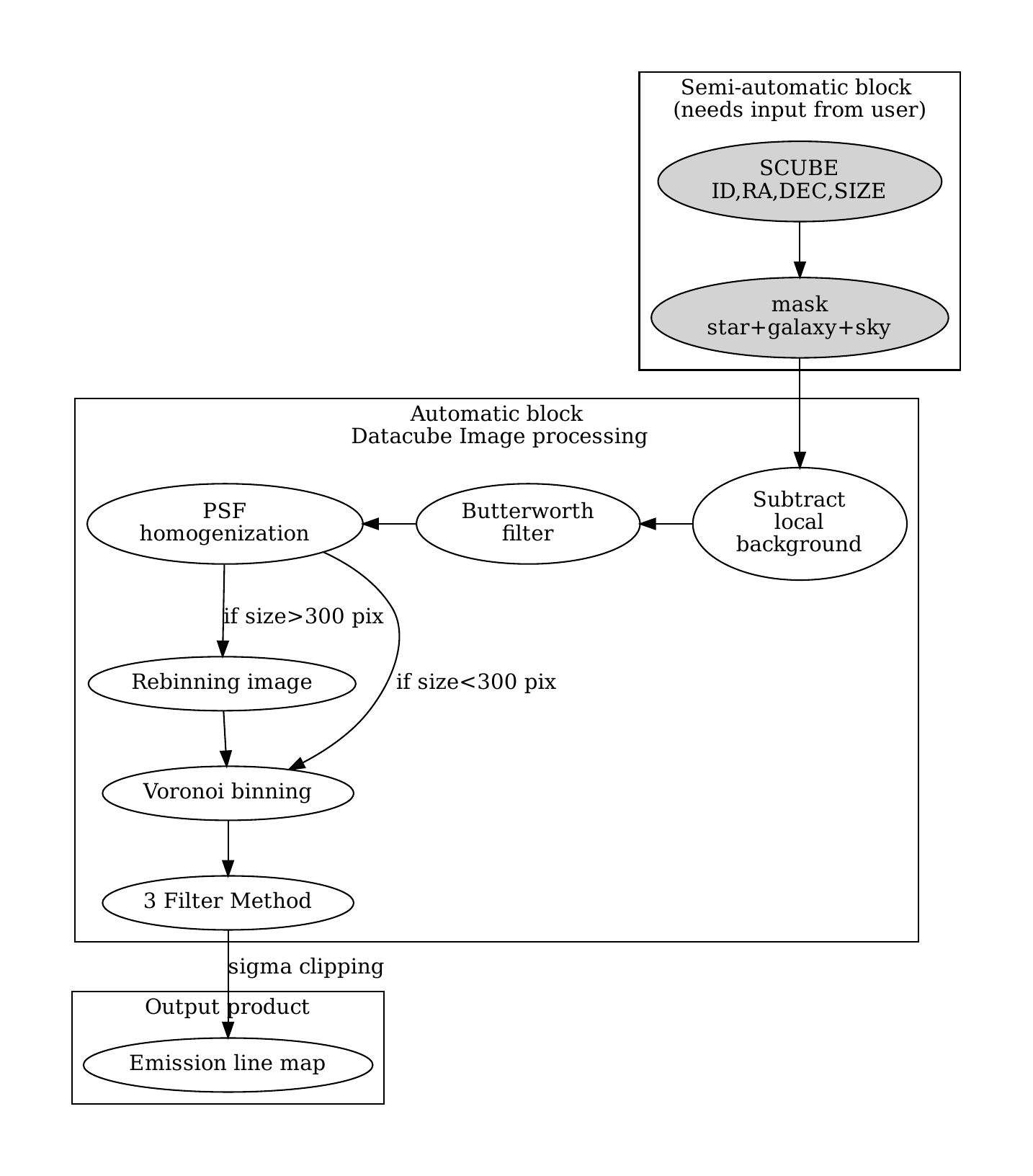}
    \caption{The workflow associated with the \texttt{PELE} code. The input information is a list of galaxies with ID, RA, Dec and size of the images, followed by the creation of masks, in which the user can check the objects to be masked and labeled as stars and main galaxy. Then the code automatically performs the image processing, including PSF homogenization, Butterworth filtering, binning functions and the Three Filter Method. Prior to generating the final emission line map, sigma clipping is applied to remove outliers that are overestimated, as discussed in \ref{subsec.:validation}.}
    \label{fig:workflow}
\end{figure}

\begin{figure*}
    \centering
    \includegraphics[width=0.99\textwidth]{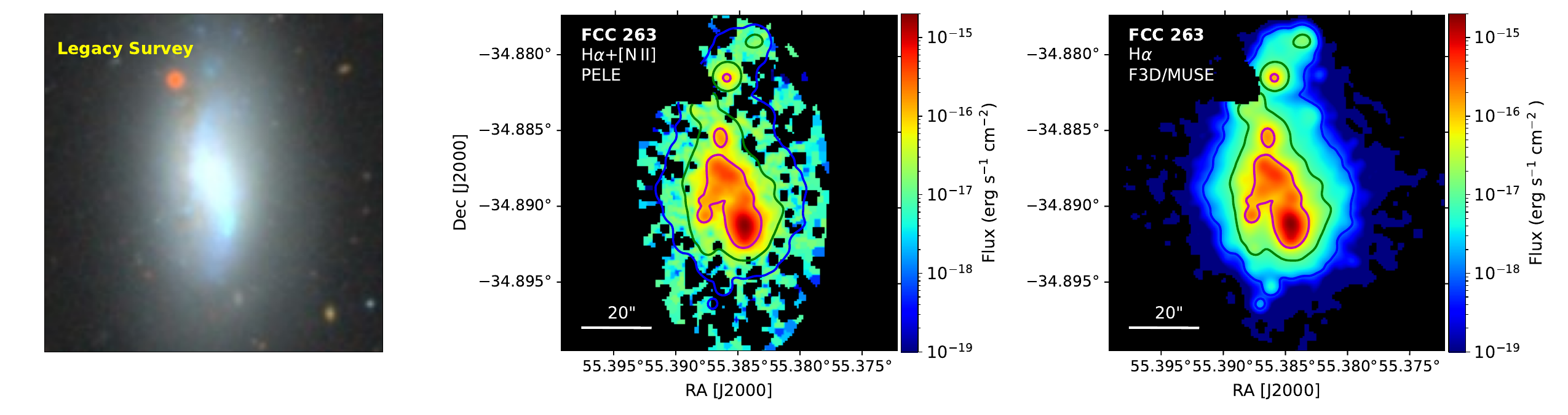}
    \includegraphics[width=0.99\textwidth]{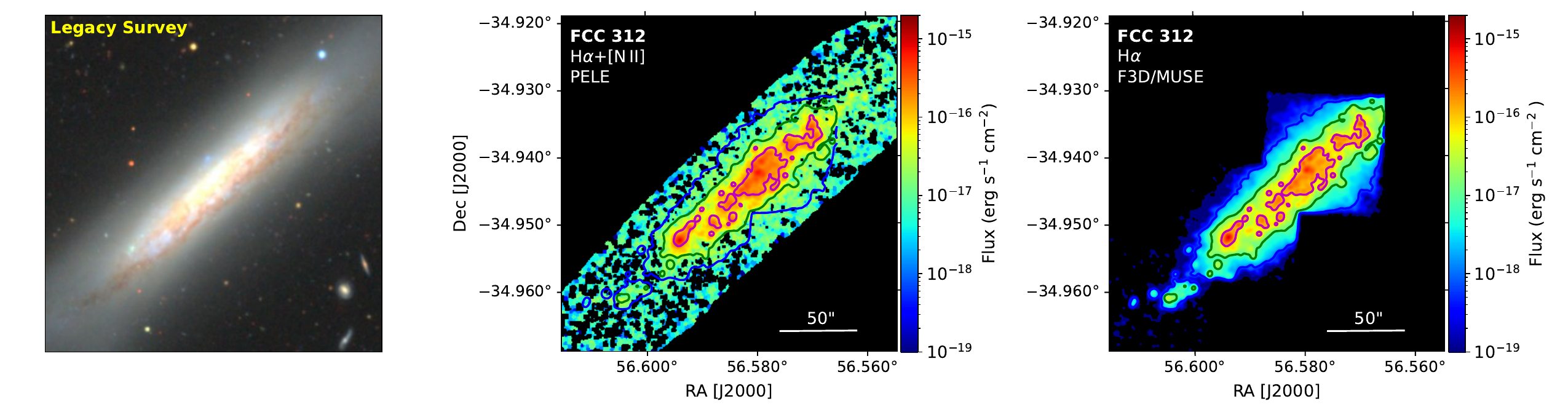}
    \caption{Comparison of the \HalphaNII{} maps derived from S-PLUS data with the H$\alpha$ maps obtained from F3D/MUSE for FCC 263 (top) and FCC\,312 (bottom). The left panels display the Legacy Survey images (a combination of $g$, $r$, and $z$ bands) at the same scale as the \HalphaNII{} maps from S-PLUS (middle panels) and MUSE (right panels).The contours are based on the MUSE H$\alpha$ map, corresponding to isophotes of $1\times10^{-18}$ (blue), $1\times10^{-17}$ (green) and $1\times10^{-16}$ (magenta) erg s$^{-1}$ cm$^{-2}$. No internal dust correction was applied to any map.}
    \label{fig:musexsplus}
\end{figure*}

\begin{figure}
    \centering
    \includegraphics[width=0.47\textwidth]{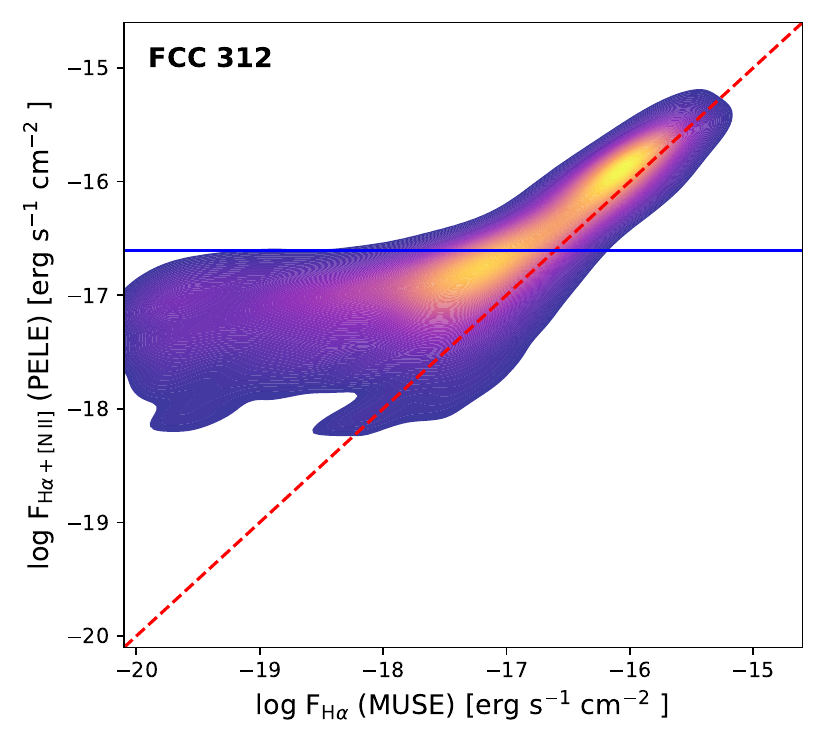}
    \caption{An example of a pixel-by-pixel analysis of the \HalphaNII{} flux derived from S-PLUS images using \texttt{PELE} and the H$\alpha$ flux observed with MUSE for FCC\,312. The lighter color (yellow) indicates a higher number of pixels, while the darker shade (purple) represents lower data density. The red dashed line is the 1-to-1 relation. The linear correlation diverges at $\sim 2 \times 10^{-17}$ erg s$^{-1}$ cm$^{-2}$, marked by the blue solid line.}
    \label{fig:p2p-splus-muse}
\end{figure}

\section{Pixel-to-pixel emission line estimate}
\label{sec:PELE}

Analyzing emission lines using images involves several intermediate procedures in image processing. For such, we developed a code named Pixel-to-pixel Emission Line Estimate (\texttt{PELE}). This script integrates the S-Cubes program, which creates S-PLUS datacubes, given the RA, Dec and the size for the images in pixels. The process then generates star and galaxy masks, requiring the user to review intermediate mask images to ensure that critical features, like H II regions within the primary galaxy, are not incorrectly masked. The masking process employs photutils.detection.StarFinder on an  image from the $r$-band filter to identify foreground stars and background objects. To apply this function, we first obtain a rough estimate of the image's background mean using a sigma clipping function with a 3 sigma threshold to exclude extreme pixel values. This initial background estimation is followed by an interactive validation step, in which saturated stars and other spurious sources are manually added. A segmentation map is also used to enhance the masking of diffraction spikes from saturated stars. For most galaxies in our sample, the main object mask is based on the 24 mag arcsec$^{-2}$ isophote in the $r$-band (Thainá-Batista et al., in prep.). Note that this initial background estimate is used solely for the purpose of creating the mask and does not modify the original image data.

Once all sources are masked, the remaining pixels are considered background. To refine the background estimation, we apply a sigma-clipping routine that removes pixel values lying beyond 3 sigma from the mean of the distribution. After five iterations, we compute the mean of the remaining pixels, which is then subtracted from the original image to obtain the background-corrected version. The following tasks are the application of the Butterworth spatial filter \citep[e.g.][]{2014/Menezes} to remove high spatial-frequency noise, point spread function (PSF) homogenization, and Voronoi binning \citep{2003/Cappellari}. These last two steps ensure that the emission signal detection is not affected by any PSF differences between the different filters and that each bin meets a minimum signal-to-noise ratio (S/N). As the process of Voronoi binning can be very time-consuming, for images larger than 300 pixels (165") a rebinning is performed, changing the pixel scale from the original 0.55" to 1.1". For our purpose, the selected Voronoi target S/N is 10 in the filter $J0660$. The binning threshold was optimized to balance processing efficiency and spatial resolution. Specifically, it was selected to minimize computation time for large galaxies while preserving sufficient spatial sampling in the binned regions of smaller galaxies. Subsequently, we estimate the emission line flux using a combination of images following the Three Filter Method. A schematic summary of the analysis is presented in Fig.\,\ref{fig:workflow}.
 
\subsection{Three Filter Method}
This approach estimates the contribution of the emission line (EL) fluxes within a given narrow band filter (NB) by analyzing its contrast to the continuum. This methodology assumes that the EL can be approximated by a Dirac delta function centered on the emission line wavelength $\lambda_\mathrm{EL}$, and the object continuum is well represented by a linear function over the three filters of interest, described below. A complete discussion can be found in \cite{2007/Pascual}. 

Considering the S-PLUS filter system and our target EL, \HalphaNII{}, we included in our analysis the NB that contains the EL ($J0660$), the broad-band that overlaps with the NB ($r$), and a second broad-band that is the closest possible to the NB wavelength range, for further assessment of the continuum ($i$). Based on these observational fluxes $F^\mathrm{obs}$, the photometric flux of the EL can be derived by the following equation presented in \cite{2015/VilellaRojo},
\begin{equation}
    F^\mathrm{photo}_{\mathrm{H}\alpha+[\mathrm{N}\,\mathrm{II}]} = \frac{\left(F^\mathrm{obs}_{r} - F^\mathrm{obs}_{i}\right) - \left(\displaystyle\frac{\alpha_\mathrm{r} - \alpha_\mathrm{i}}{\alpha_\mathrm{J0660}-\alpha_\mathrm{i}}\right) \left(F^\mathrm{obs}_\mathrm{J0660}-F^\mathrm{obs}_\mathrm{i}\right)}{\beta_{J0660} \left(\displaystyle\frac{\alpha_{i}-\alpha_{r}}{\alpha_{J0660}-\alpha_{i}}\right)+ \beta_{r}}
    \label{eq.flux}
\end{equation}
with 
\begin{equation}
    \alpha_\mathrm{band} = \frac{\int \lambda^{2} T_\mathrm{band}(\lambda) \mathrm{d}\lambda}{\int T_\mathrm{band}(\lambda) \lambda \mathrm{d}\lambda},
    \label{eq.alpha}
\end{equation}
\begin{equation}
    \beta_\mathrm{band} = \frac{\lambda_\mathrm{EL} T_\mathrm{band}(\lambda=\lambda_\mathrm{EL})}{\int T_\mathrm{band}(\lambda) \lambda \mathrm{d}\lambda}
    \label{eq.beta}
\end{equation}
where $T_\mathrm{band}$ is the transmission curve for a given band ($J0660$, $r$ or $i$), and $\lambda_\mathrm{EL}$ is the wavelength of the emission line we want to measure. We assume that \HalphaNII{} flux contribution primarily stems from the H$\alpha$ line, hence our estimates are obtained at $\lambda_\mathrm{EL} = \lambda_{\mathrm{H}\alpha,0} \, (1+z_\mathrm{EL})$, where $\lambda_{\mathrm{H}\alpha,0}=6562.8$\AA{} is the rest-frame wavelength of H$\alpha$, and $z_\mathrm{EL}=0.0048$, which is the mean redshift of the cluster \citep{2019/Maddox}. 

As the internal dust and \NII{} corrections presented in \cite{2015/VilellaRojo} are only valid within star formation regions, we chose not to apply them in the present paper. The main reason is that we cannot guarantee the origin of the H$\alpha$ emission is only due to star formation. Indeed, some of our emitters are known AGN sources with X-ray measurements \citep[e.g.,][]{2024/Hou}

It is important to note that the origin of H$\alpha$ emission can vary, including star formation activity, the presence of an AGN, and from old stars. Notably, the diffuse ionized gas (DIG) is known to contribute up to 60\% of the total H$\alpha$ flux in local spiral galaxies \citep[e.g.][]{2016/Kreckel,2020/Chevance}. Since we are not providing pure H$\alpha$ fluxes or their equivalent widths, this type of discrimination is not feasible. In forthcoming papers, we will incorporate corrections for internal dust attenuation and \NII{} contamination, allowing for a better understanding of the emission sources.

After setting up \texttt{PELE}, the next step is to assess how efficient our detection of \HalphaNII{} flux is. Indeed, we expect our detection to be limited when compared to spectroscopic measurements, due to both the technique and the depth of S-PLUS images. As presented in the Figure 6 of \cite{PaperI}, the depth of S-PLUS images varies significantly across different fields (up to 1 mag). However, Table 2 of \cite{2024/Herpich} reports that the depth of $r$, $J0660$ and $i$ filters (at S/N > 3) are 21.18, 20.98 and 20.79, respectively. These variations in depth introduce additional complexity to the analysis.

\subsection{Validation of the code}
\label{subsec.:validation}

\begin{figure*}
    \centering
    \includegraphics[width=0.23\textwidth]{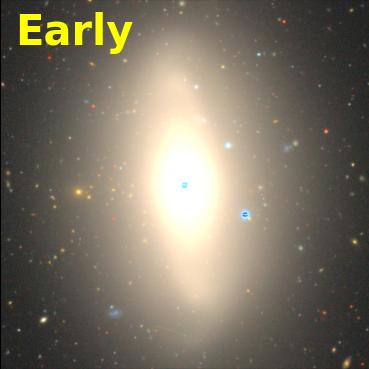}
    \includegraphics[width=0.23\textwidth]{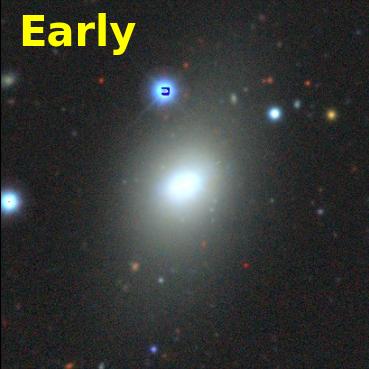}
    \includegraphics[width=0.23\textwidth]{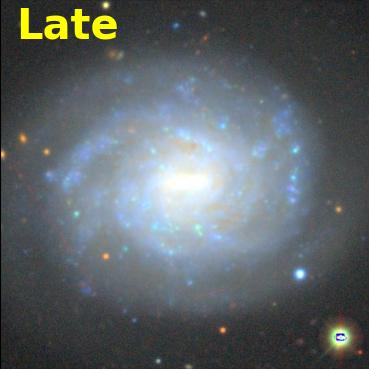}
    \includegraphics[width=0.23\textwidth]{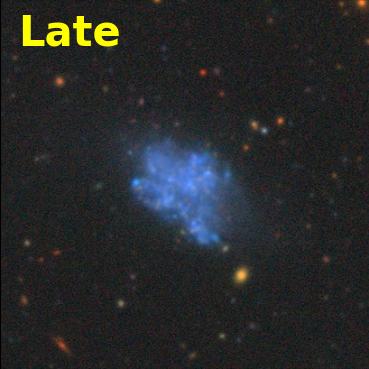}
    \caption{Examples of galaxies classified as early- and late-type using images from the DESI Legacy Imaging Surveys. From left to right, the galaxies shown are NGC\,1380, FCC\,90, NGC\,1310 and FCC\,9.}
    \label{fig:examples-early-late}
\end{figure*}

A preliminary evaluation of \texttt{PELE} was presented in \cite{2023/ThainaBatista}, comparing its results with those obtained from spectral energy distribution fitting using the \texttt{AlSTAR} code. The study found a strong agreement between both methods. Here, we further assess the effectiveness of our approach using 13 galaxies from MUSE data obtained in the context of the F3D project (see Subsection \ref{subsec:MUSE}). The input H$\alpha$ maps are the same as those presented in \cite{2019/Iodice}\footnote{Provided via private communication by Dr. Enrichetta Iodice.}. For a proper comparison, we aligned and degraded the MUSE images to match the observations of S-PLUS (i.e. pixscale and PSF matching). Fig. \ref{fig:musexsplus} shows two examples of the maps derived with \texttt{PELE} and their corresponding MUSE maps for the galaxies FCC\,263 and FCC\,312, along with a composite image $g$, $r$, and $z$ bands from the Legacy Survey. Since we are comparing H$\alpha$ maps with \HalphaNII{} maps, we expect a general overestimation of our fluxes relative to MUSE due to the contribution of the \NII{} line. Indeed, by visual inspection, we observe a systematic overestimation of flux in the outskirts of the galaxies, where the values based on MUSE data are lower. 

With the purpose of obtaining a deep understanding of our results, we perform a pixel-by-pixel analysis of the maps generated by \texttt{PELE} and MUSE. As we have the same number of pixels in both \texttt{PELE} and MUSE, we can directly compare the maps, as shown in Fig. \ref{fig:p2p-splus-muse}. We find that our method tends to overestimate fluxes below a specific threshold. Therefore, such detections are unreliable. After performing a $3\sigma$ clipping, we are able to remove the overestimation. We extended this analysis to all 13 galaxies and consistently found comparable detection thresholds. Consequently, we confirm that our procedure can reliably detect fluxes down to approximately $2 \times 10^{-17}$ erg s$^{-1}$ cm$^{-2}$. As lower values are usually found on the outskirts of galaxies, as shown in Fig.\,\ref{fig:musexsplus}. This indicates that accessing and studying the peripheral areas of galaxies presents significant challenges.

\section{\HalphaNII{} emitters in Fornax}
\label{sec:emitters}

From the literature sample outlined in Subsection\,\ref{subsec:fornax_members}, we have identified 77 \HalphaNII{} emitters with sufficient spatial resolution to generate emission line maps. Of these, only two (ESO 358-11 and ESO 359-25) lack spectroscopic confirmation of membership. Figs. \ref{fig:0Rvir_0.7Rvir}-\ref{fig:1.6Rvir_1.99Rvir} show the maps derived with \texttt{PELE} and composite image $g$, $r$, and $z$ bands from the Legacy Survey for the entire sample.

Additionally, compare to the H$\alpha$ emitter catalog provided by \cite{2001/Drinkwater}, our method successfully recovers 78\% of galaxies. Several of the missed objects are dwarf and/or elliptical galaxies with reported EW(H$\alpha)\lesssim5$\AA. We then cross-matched our sample with the spectroscopic study conducted by \cite{2024/Lousber}, which analyzed 10 galaxies within the Fornax\,A group. We successfully obtained \HalphaNII{} maps for all except for one, FCC\,46, which has an EW(H$\alpha$)<1\AA. In conclusion, \texttt{PELE} is particularly effective at identifying concentrated emissions, whether located centrally or in H II regions. However, capturing more diffuse, extended features is less consistent and depends on both flux levels and equivalent width.

\subsection{Optical morphology}
\label{subs:morphology}
Analyzing the morphologies of these emitters is insightful for understanding their role in the evolutionary path of the cluster and also in the context of galaxy evolution. For such, we use images from the DESI Legacy Imaging Surveys (hereafter, Legacy)\footnote{https://www.legacysurvey.org/}, which are deeper than those from S-PLUS, reaching a median $5\sigma$ detection limit for galaxies at $r \sim 23.4$ AB mag \citep{2019/Dey}. This survey is well-suited for morphological studies, as it allows for the detection of fainter and more extended features. Although FDS provides much deeper imaging than Legacy in the Fornax region, it does not encompass all 77 galaxies in our sample. Indeed, $\sim32\%$ of our selected objects are presented in previous papers from FDS. Therefore, to ensure a homogeneous morphological analysis across the entire dataset, we opted to use the Legacy images.

For the initial analysis, we begin with a basic visual inspection to distinguish between early- and late-type galaxies. Early-type galaxies (hereafter ETG), including ellipticals and lenticulars, are characterized by a smooth light distribution with no distinct features. In contrast, late-type galaxies (hereafter, LTG), such as spirals and irregulars, display visible spiral arms or multiple knots. Fig. \ref{fig:examples-early-late} presents examples of both classifications. Among our detected emitters, 58 (75\%) are LTG, while 19 (25\%) are ETG. 

The next step is to assess whether these galaxies display any unique features, such as being part of a merger system or showing signs of disturbance. To do so, we employ the \texttt{ASTROMORPHLIB} code \citep{2022/Hernandez}, which conducts a non-parametric morphological analysis on the $g$-band images of the DESI Legacy Imaging Surveys. This code follows several steps: it calculates a 2D sky background model, masks foreground stars, creates a segmentation map, and executes the \texttt{STATMORPH} Python package \citep{2019/RodriguezGomez}. In the final phase, it fits a 2D Sérsic model to the galaxy’s disk component and computes non-parametric metrics, including concentration ($C$), asymmetry ($A$), the Gini index, and the second moment of the galaxy's brightest regions ($M_{20}$).  In a forthcoming paper, the complete list of confirmed members of the Fornax cluster will have their morphology analyzed (Hernandez-Jimenez et al. in prep.). 

In the literature, these parameters have been widely employed as key indicators of galaxy morphology. For instance, \cite{2003/Conselice} demonstrated that the concentration–asymmetry–smoothness (CAS) system correlates with critical phases of galaxy evolution. The study revealed a strong correlation between the H$\alpha$ equivalent width and the clumpiness parameter, with star-forming galaxies displaying star-forming regions in distinct clumps. This underscores the parameter's effectiveness in tracing ongoing or recent star formation activity. Additionally, \cite{2004/Lotz, 2008/Lotz} combined the Gini index and $M_{20}$ to classify galaxies into three groups: E/S0/Sa, Sb-Ir, and mergers. More recently, \cite{2024/Krabbe} introduced the $C-A$ diagram as a diagnostic tool to distinguish galaxies with disturbed features, possibly undergoing ram pressure stripping or tidal interactions. 

\begin{figure}
    \centering
    \includegraphics[width=0.5\textwidth]{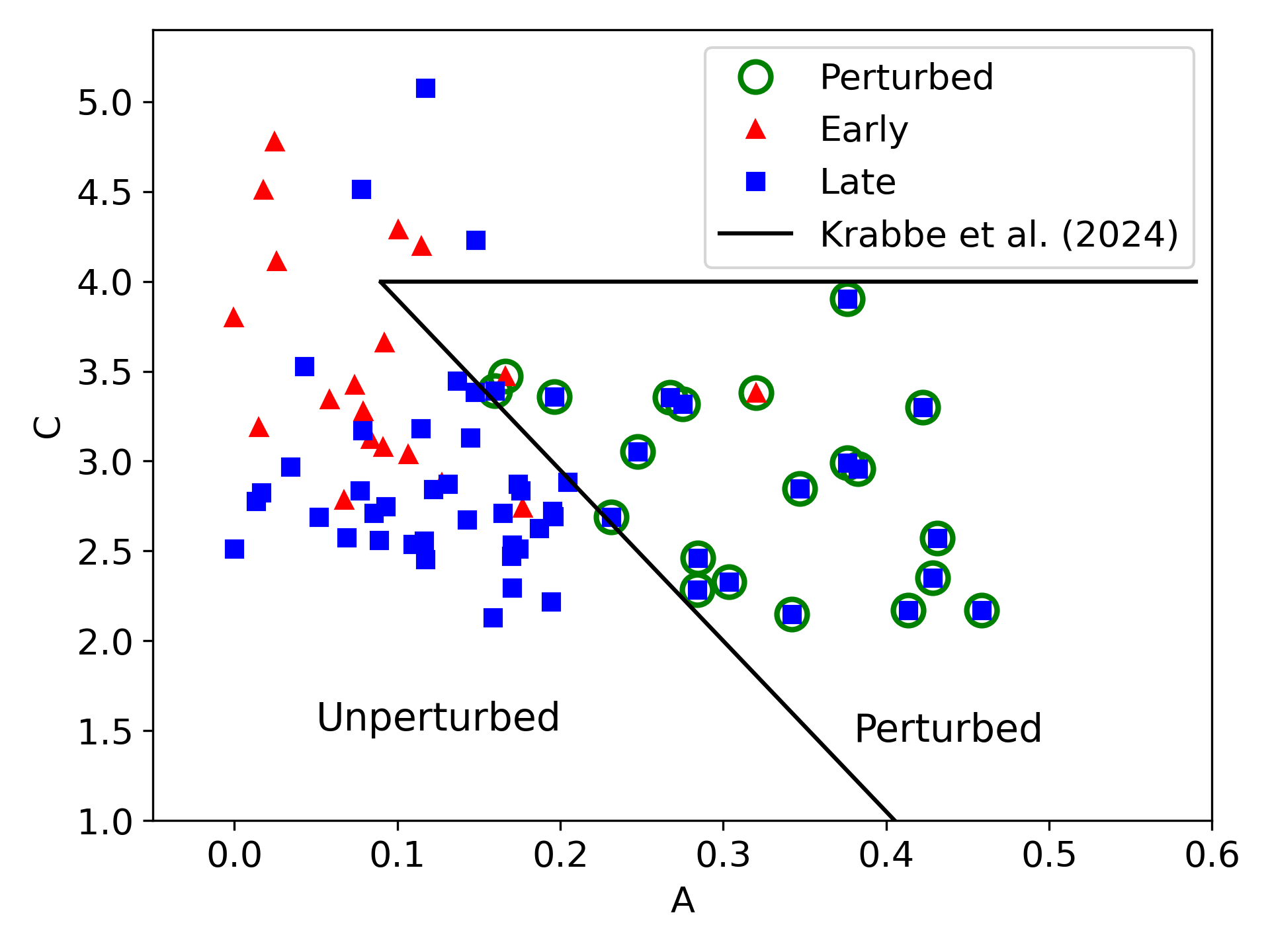}
    \includegraphics[width=0.49\textwidth]{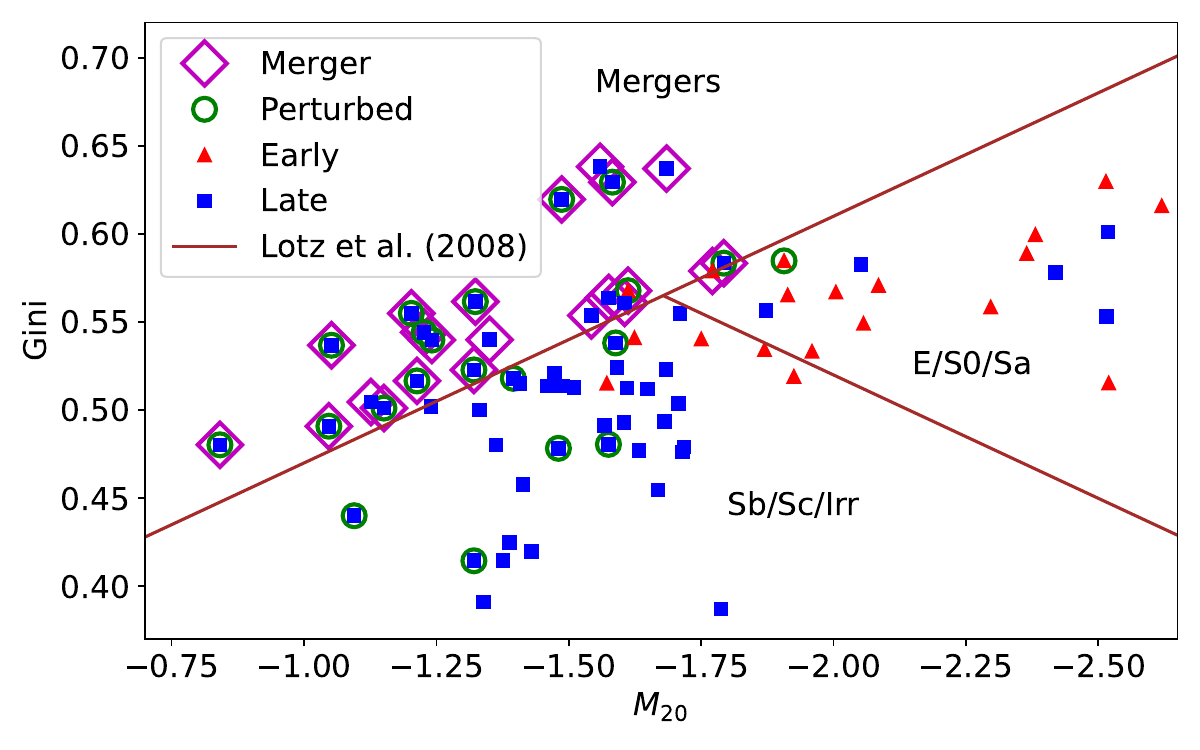}
    \caption{Concentration versus Asymmetry diagram {\it (top panel)} and  Gini versus $M_{20}$ diagram {\it (bottom panel)} for 77 \HalphaNII{} emitters in the Fornax cluster. The black solid line in the top panel delineates the morphological transition zone boundaries, where the disturbed galaxies lie, given by $C = -9.5 A+ 4.85$ and $C<4$ \citep{2024/Krabbe}. The brown solid lines in the bottom panel are defined by \citet{2008/Lotz}. The galaxies belonging to merger systems are shown as magenta open diamonds and the perturbed ones with green open circles.}
    \label{fig:CAGM20}
\end{figure}

Here we employ the classifications from \cite{2024/Krabbe} and \cite{2008/Lotz} to identify possibly perturbed and merger galaxies within our sample, as shown in Fig. \ref{fig:CAGM20}. Out of the 77 \HalphaNII{} emitters, we found that 29 (38\%)  display one or both designations, with 62\% categorized as unperturbed objects. Among those classified as presenting any type of disturbance, 7 (24\%) are seen as perturbed, 8 (28\%) as mergers, and 14 (48\%) exhibit both characteristics. From the 19 ETG analyzed, 1 (5\%) is perturbed, 1 (5\%) is merger, 1 (5\%) displays both traits, and 16 (85\%) are undisturbed. For the 58 LTG, 6 (10\%) are perturbed, 7 (12\%) are mergers, 13 (22\%) exhibit both features, and 32 (56\%) are labeled as unperturbed. This means that 15\% of the early-type emitters display evidence of perturbation and/or merging, whereas the same occurs for 44\% of the late-type emitters.

\subsection{Spatial distribution}

\begin{figure*}
    \centering
    \includegraphics[width=0.49\textwidth]{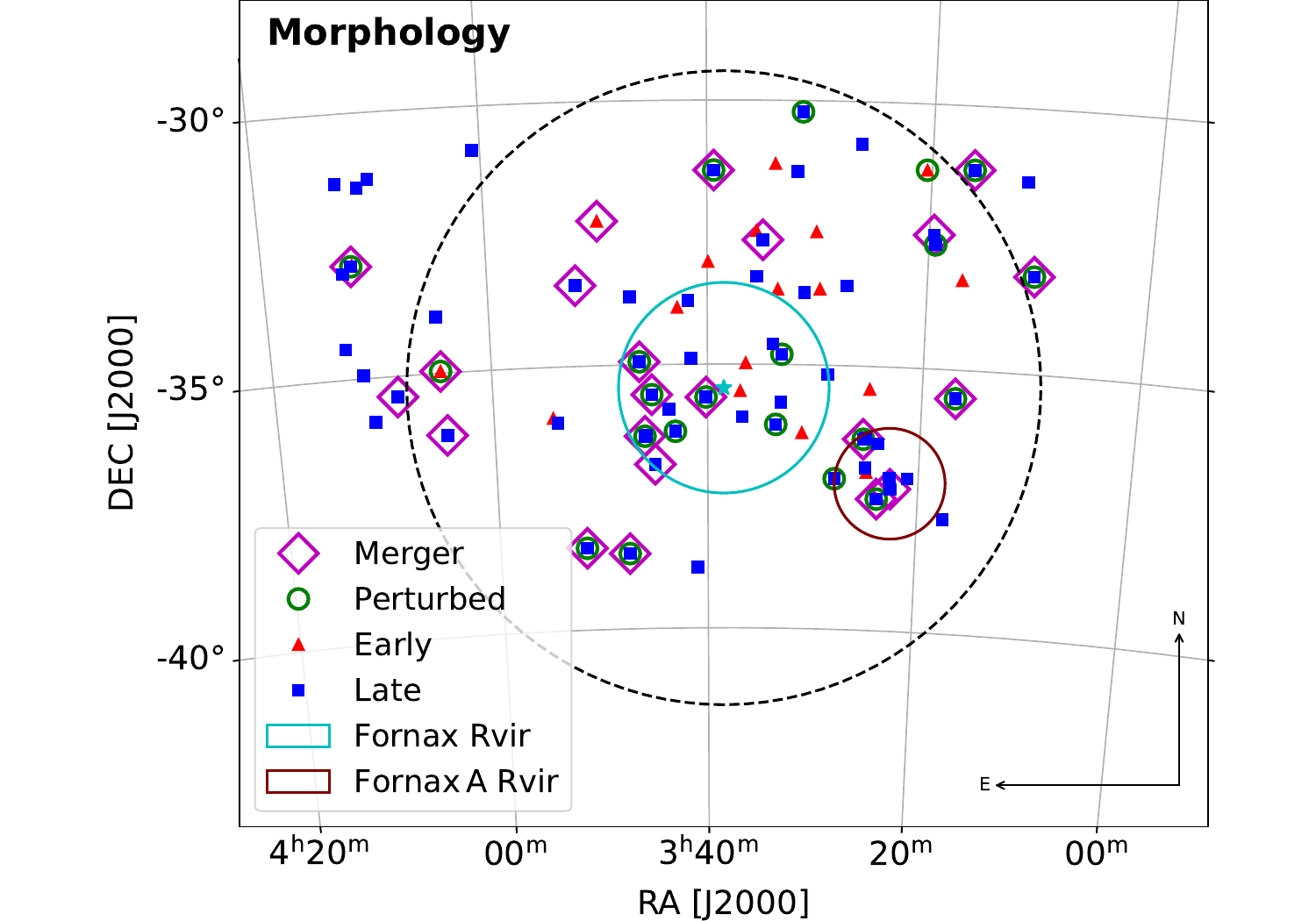}
    \includegraphics[width=0.49\textwidth]{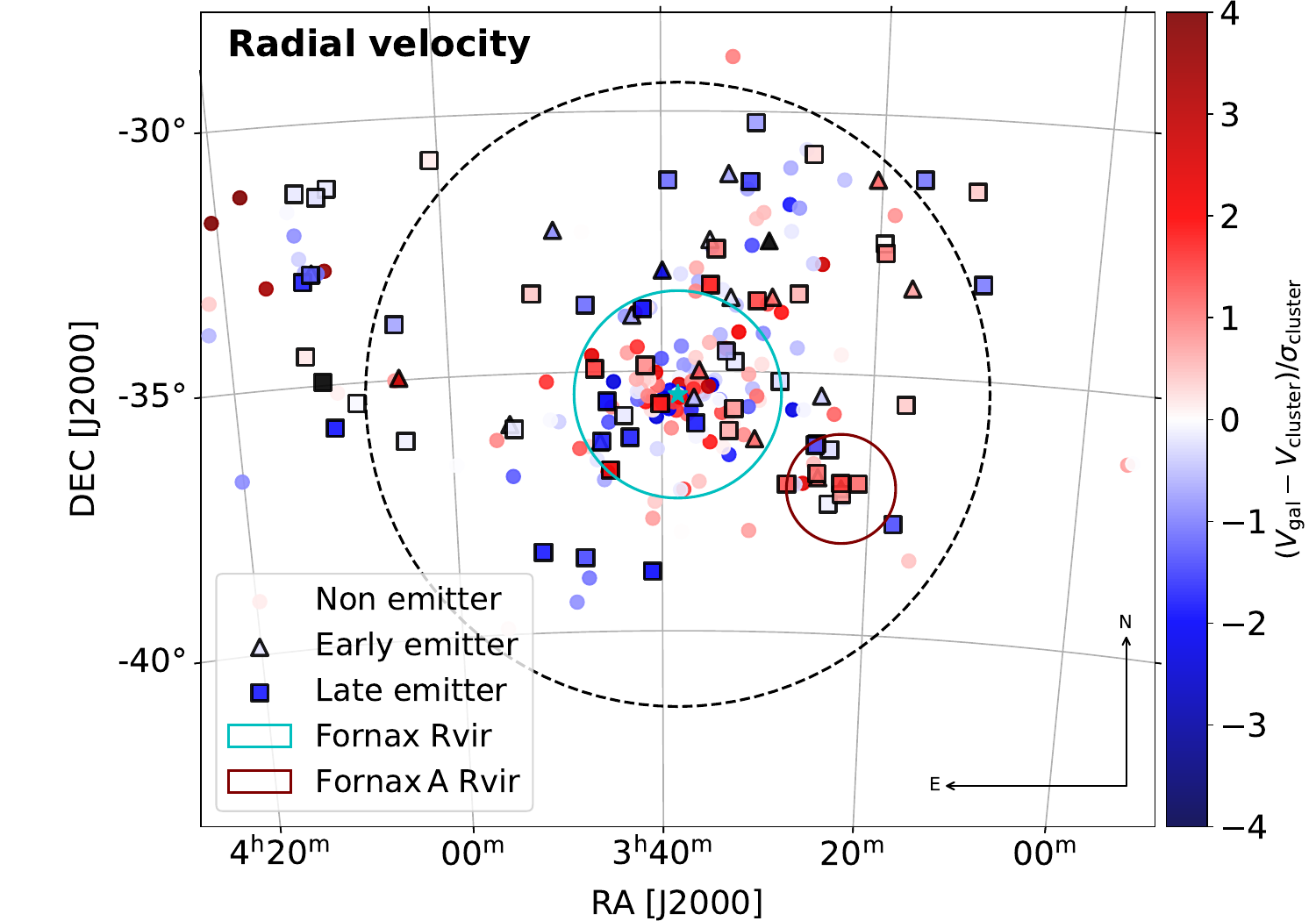}
    \includegraphics[width=0.49\textwidth]{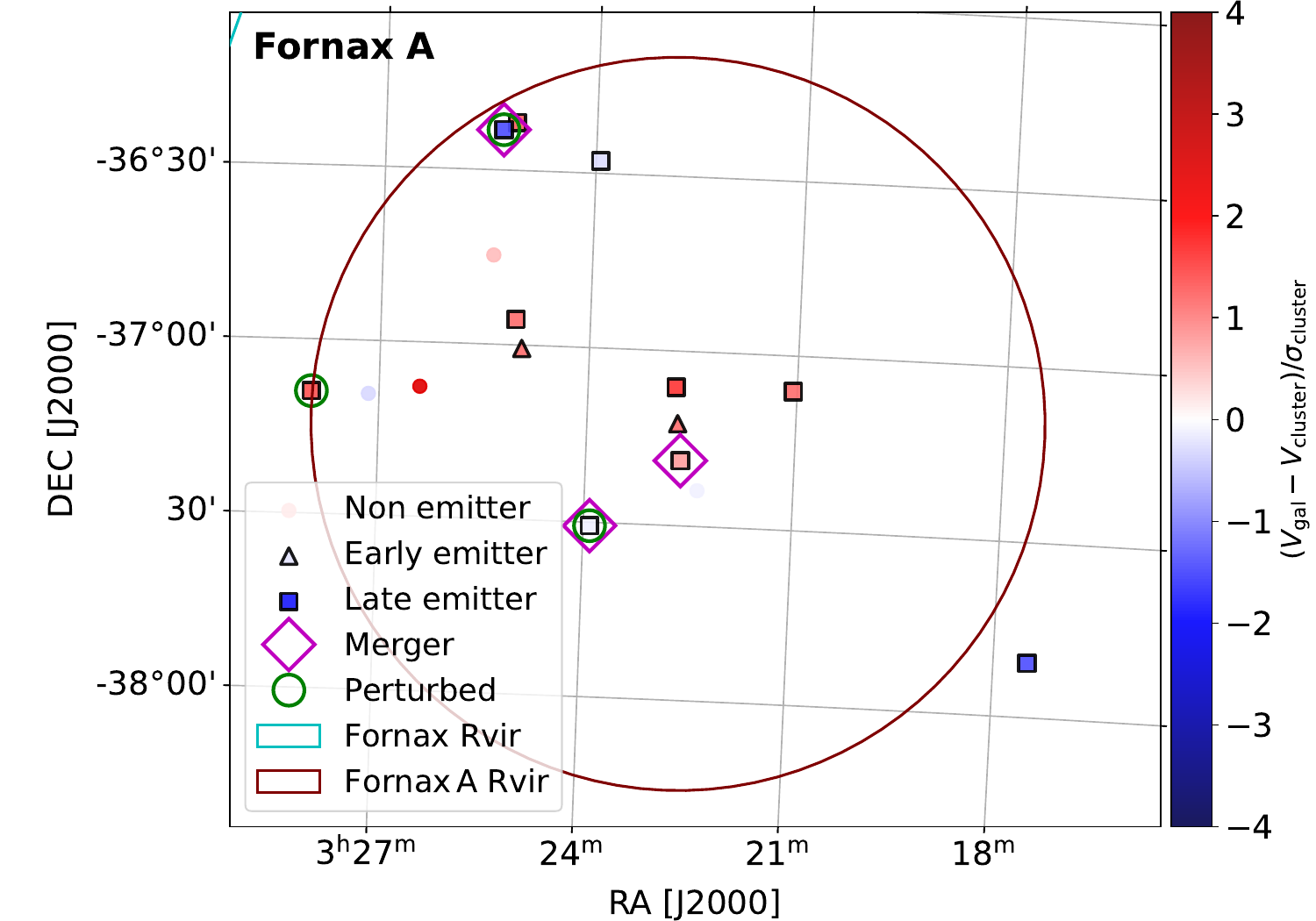}
    \includegraphics[width=0.49\textwidth]{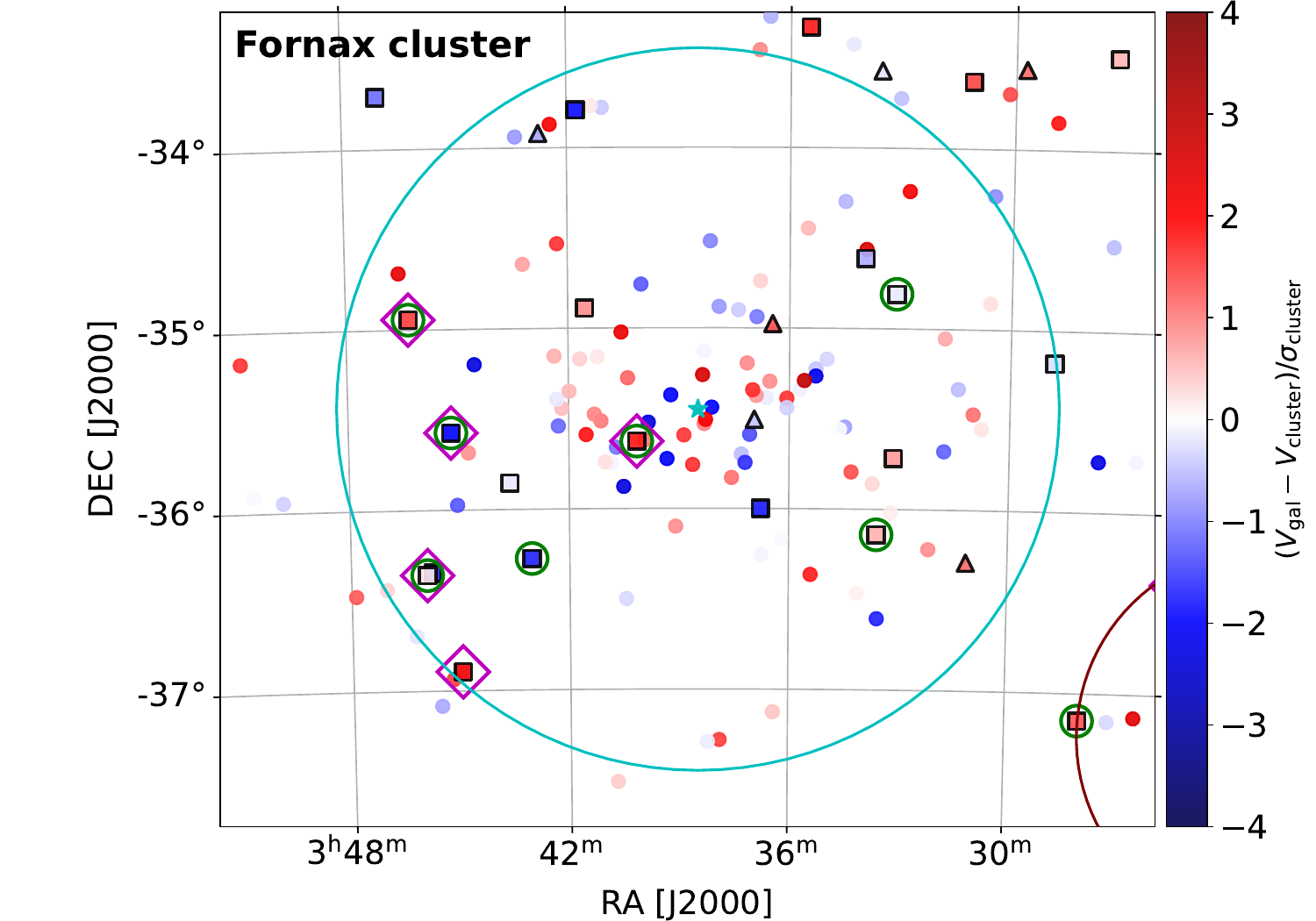}
    \caption{Spatial distribution of the Fornax members  around NGC\,1399 (cyan star) color-coded by radial velocity $V_\mathrm{gal}$ of each galaxy relative to the cluster’s mean velocity, $V_\mathrm{cluster} = 1442$ km s$^{-1}$ and normalized by its velocity dispersion $\sigma_\mathrm{cluster}$=318 km s$^{-1}$ (\citealt{2019/Maddox}) (top right panel) and morphological features for 77 \HalphaNII{} emitters detected by \texttt{PELE} (top left panel). The bottom left provides a close-up view of the spatial distribution of 15 galaxies within the Fornax\,A group, while the bottom right panel focuses on the region within 1 \Rvir{} around the cluster center, both plots use the same symbol conventions. Only galaxies with detected emission are included in the morphological analysis described in Subsection\,\ref{subs:morphology}. LTG are depicted as squares, whereas ETG are shown as triangles. Galaxies identified as perturbed are marked with green open circles, and mergers are denoted by magenta open diamonds. The cyan and brown solid lines represent the $R_\mathrm{vir}$ of the Fornax cluster and the Fornax\,A group, respectively, while the black dashed line indicates 3 $R_\mathrm{vir}$. The black square and triangle in the top right panel correspond to two galaxies without known radial velocities but with detected \HalphaNII{} emission.}
    \label{fig:spatial-dist}
\end{figure*}

Fig.\,\ref{fig:spatial-dist} illustrates the spatial distribution of the 77 identified emitters, highlighting the previously discussed morphological classifications (top left) and the variation in radial velocity of the galaxy relative to the cluster's mean velocity $V_\mathrm{cluster}=1\,442$ km s$^{-1}$ and velocity dispersion $\sigma_\mathrm{cluster} = 318$ km s$^{-1}$ \citep{2019/Maddox}. Additionally, the figure includes a detailed view of the central cluster region (bottom right) and the area encompassing its infalling Fornax\,A group (bottom left). ETG emitters are predominantly concentrated in the north-west region of the cluster, while LTG ones appear more evenly distributed (top left panel). It is noteworthy that the most confirmed Fornax cluster members are concentrated in the northern region of the cluster, as shown in the right top panel. Since these objects serve as the basis for our analysis, there is a naturally lower number of emitters in the southern region, except for those associated with Fornax\,A. 

The top left (morphology) panel highlights another trend, showing a higher concentration of emission-line galaxies with additional features (classified as perturbed, merger, or both) predominantly located on the eastern side of the cluster. This trend is especially prominent within 1 \Rvir{} of NGC\,1399. Notably, LEDA\,707430, LEDA\,655120 and FCC\,B2144, early-type emitters, are identified as perturbed and/or merger candidates and are positioned between 1 \Rvir{} and 3 \Rvir{} in the north. The not perturbed/merged early-type emitters are primarily distributed in the northwestern region within 3 \Rvir{}. In contrast, LTG emitters with single classifications or without any special feature appear more dispersed across the field, without any discernible spatial pattern. Beyond 3 \Rvir{}, only emitters displaying late-type morphologies are found.

In the top right (radial velocity) panel, general patterns can be observed in the emitters following the variations in their radial velocity relative to the cluster, $\Delta V/\sigma =(V_\mathrm{galaxy}-V_\mathrm{cluster})/\sigma_\mathrm{cluster}$. A velocity gradient is observed, increasing from east to west within the emitter sample. On the eastern side of the cluster, many galaxies exhibit comparable $\Delta V/\sigma$ values (shown in blue) extending from the cluster's interior to distances beyond 3\Rvir{}. On the western side, several galaxies displaying similar $\Delta V/\sigma$ values (shown in red) encompassing most of the galaxies in the Fornax\,A group. Within 3\Rvir{}, a southeast-to-northwest gradient is evident, marked by an different in the number of objects and their velocities. A similar result was obtained by \cite{2002/Waugh} for 110 galaxies within an area of $\sim 620$ deg$^{2}$
and a H I mass limit of $1.4 \times 10^{8} M_\odot$ using H I Parkes All Sky Survey (HIPASS) observations.

A closer examination of the Fornax\,A group (bottom left panel) and the cluster center (bottom right panel) provides a clearer comparison of the galaxies in both regions. This analysis is valuable for understanding the differences in galaxy behavior between a more virialized environment and a region where pre-processing is still occurring. The number of emitters around NGC\,1316 extending to 1.05 deg (Fornax\,A \Rvir) is 11 — about half the number found within 1 \Rvir{} of NGC\,1399, which totals 20. This difference in the quantity of detections coincidentally scales with the size disparity between the two regions. However, when considering the objects with known radial velocities (233 in total), we observe a higher relative percentage of emitters within Fornax\,A, 73\% (11 out of 15), compared to only 19\% (20 out of 107) in the Fornax cluster. Unfortunately, the spectroscopic sampling of the Fornax cluster and the Fornax A group is not complete, so we cannot infer much more from these results. 

\begin{figure}
\centering
\includegraphics[width=0.49\textwidth]{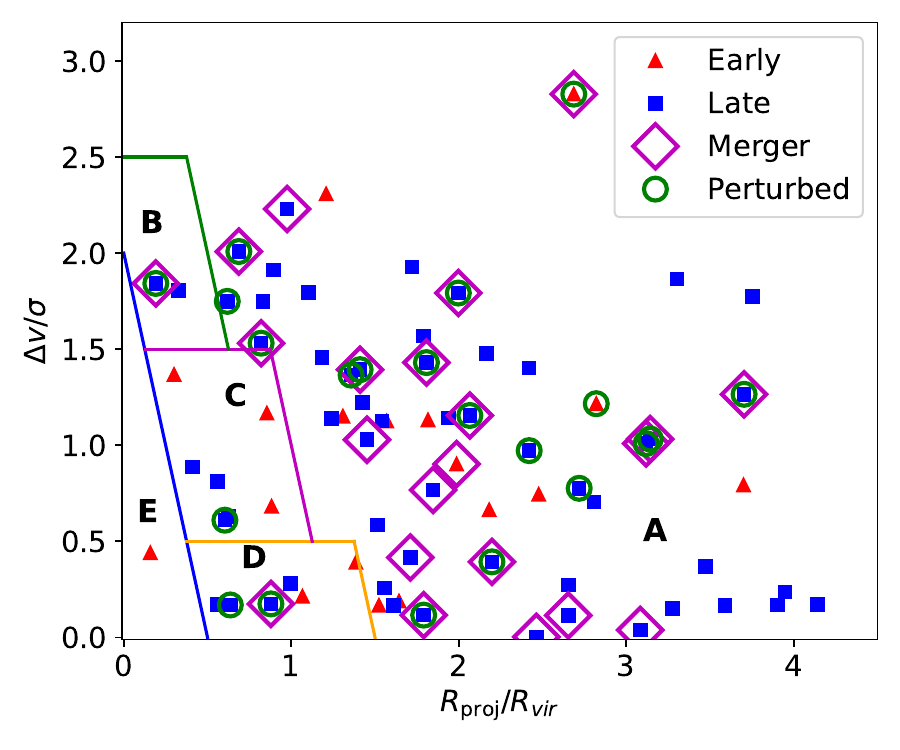}
\caption{Projected Phase Space (PPS) diagram of the 77 \HalphaNII{} emitters found using \texttt{PELE}. \citet{2017/Rhee} defined different areas separated by solid lines where E is dominated by ancient infallers ($t_{\mathrm{infall}} > 8$ Gyr), D contains a mixed population with $t_{\mathrm{infall}}<4-7$ Gyr, and regions A, B and C correspond to galaxies that have fallen into the cluster less than 1 Gyr ago (recent infallers).}
\label{fig:PPS-diagram}
\end{figure}

\subsection{Projected Phase Space Diagrams}
\label{subsec:PPS}
When a field galaxy enters a cluster or group, it takes time for the denser environment to influence its properties. Different studies suggest that quenching processes driven by the environment occur over several gigayears \citep[e.g.][]{2012/DeLucia,2013/Wetzel,2014/Hirschmann,2016/Oman}. However, measuring local environmental density alone may not fully capture galaxy evolution, as properties also depend on how long a galaxy has been in its current environment. Some galaxies may still reflect the characteristics of their previous environment, especially in regions of rapid growth with many recent infallers. Understanding the effect of infall time ($t_\mathrm{infall}$) would provide valuable insights into environmental influences on galaxy evolution.

Assuming a $\Lambda$ cold dark matter cosmology, the infalling substructures have eccentric, descending orbits \citep{2005/Gill,2011/Wetzel}. As a result, at any given radius, there is a broad range of infall times due to the mixture of objects entering the cluster for the first time and those that have been orbiting for several gigayears. However, incorporating both distance and velocity in phase-space diagrams reveals that recent infallers generally exhibit higher velocities, allowing for more accurate estimates of their infall times. In fact, \cite{2017/Rhee} demonstrated that using velocities and distances in the projected phase-space (PPS) diagrams helps separate populations with different infall times more clearly. 

Recent studies have increasingly highlighted the connection between a galaxy's phase-space position and its star formation activity and stellar mass buildup \citep[e.g.][]{2011/Mahajan, 2014/Muzzin, 2014/Hernandez-Fernandez, 2019/Pasquali, 2024/Sampaio, 2024/Romero}.  These findings enhance our understanding of the morphology-density relation \citep{1980/Dressler} and provide insights into how the environment shapes galaxy evolution within clusters. In this subsection, we apply the phase-space analysis of our emitter sample, combining this  with its morphological results to infer the assembly history of Fornax cluster.

Assuming the cluster’s mean velocity $V_\mathrm{cluster} = 1442$ km s$^{-1}$ and velocity dispersion $\sigma_\mathrm{cluster}=318$ km s$^{-1}$ \citep{2019/Maddox} as reference, we constructed the PPS diagram for our sample of emitters, as presented in Fig.\,\ref{fig:PPS-diagram}. Following the division proposed by \cite{2017/Rhee}, the PPS can be split up into five regions ranging from ancient infallers ($t_{\mathrm{infall}} < 8$ Gyr) to recent ones ($t_{\mathrm{infall}} < 1$ Gyr). We find that 91\% of the \HalphaNII{} emitters are located in regions (A, B and C) indicative of recent infall into the cluster. Of these, $\sim77\%$ are LTG, with $40\%$ showing signs of perturbation or merger activity. The interaction between these infalling galaxies and the intracluster medium or nearby cluster galaxies may have triggered star formation, causing the observed \HalphaNII{} emission associated with disturbed optical morphologies. Among the 23\% of ETG in the recent infall regions, only three show evidence of disturbance or merger features.

The only emitter located in the regime of ancient infallers (region E in Fig. \ref{fig:PPS-diagram}) is the early-type galaxy NGC\,1387. In the intermediate infaller regime (region D), which corresponds to galaxies that entered the cluster between 4 and 8 Gyr ago, six galaxies are found: FCC\,32, FCC\,119,FCC\,113, NGC\,1436, NGC\,1437B, and NGC\,1351A. The first two are ETG, and the last four are LTG, with two presenting perturbed and/or merger classification.   

Fig. \ref{fig:PPS-diagram} may differ from other PPS diagrams in the literature due to variations in the reference cluster mean velocity and velocity dispersion. As an example, \cite{2019/Iodice} adopts a line-of-sight radial velocity of $V_\mathrm{rad}=1425$ km s$^{-1}$ for the brightest cluster galaxy, NGC\,1399, and a cluster velocity dispersion of $\sigma_\mathrm{cluster}=300$ km s$^{-1}$, as reported by \cite{2001/Drinkwater}, instead of the values derived by \cite{2019/Maddox} based on the distribution of galaxies. Such differences in the chosen parameters can cause some galaxies to shift their positions within the PPS diagram. For instance, while NGC\,1380 is classified as an ancient infaller (region E) in \cite{2019/Iodice}, its location in Fig. \ref{fig:PPS-diagram} is the region C, near the boundary with region E. All other galaxies common to both studies occupy consistent locations. 

This study expands the analysis of recent infallers in the outskirts of the Fornax cluster, identifying emission line galaxies at clustercentric distances out to 4\Rvir{}. 

\begin{figure}
\centering
\includegraphics[width=0.48\textwidth]{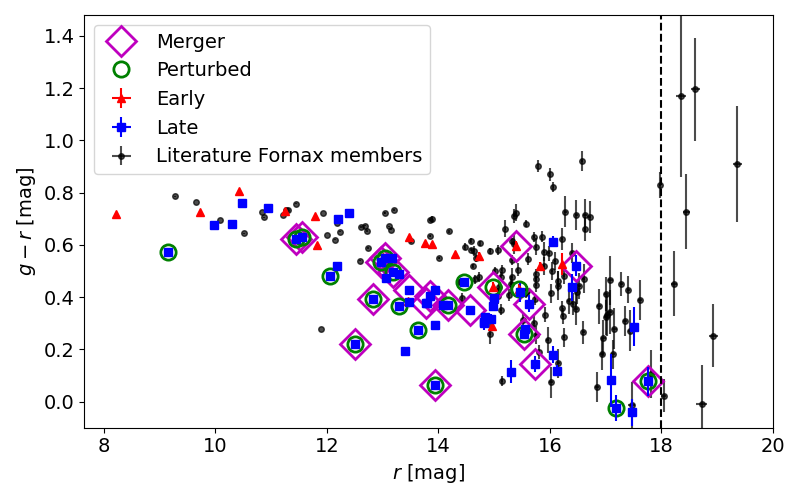}
\caption{Color magnitude relation for Fornax members, with the emitters highlighted according to their respective morphologies, as described in the legend. The dashed black line represents the magnitude limit of our sample.}
\label{fig:CMR_emitters}
\end{figure}

\subsection{Color magnitude relation and stellar mass}

We used the catalog available in the S+FP collaboration \citep[see ][for more details]{PaperI} to retrieve the location in the color magnitude diagram and the stellar mass of our working sample. The magnitude definition is AUTO, which is based on an estimate of the total flux given by the integration pixel values within an adaptively scale aperture \citep[automatic radius derived from Kron's first moment; ][]{1980/Kron}. These data were designed to properly recover the photometry of the larger galaxies of Fornax \citep{2024/Haack}.

\begin{figure}
\centering
\includegraphics[width=0.48\textwidth]{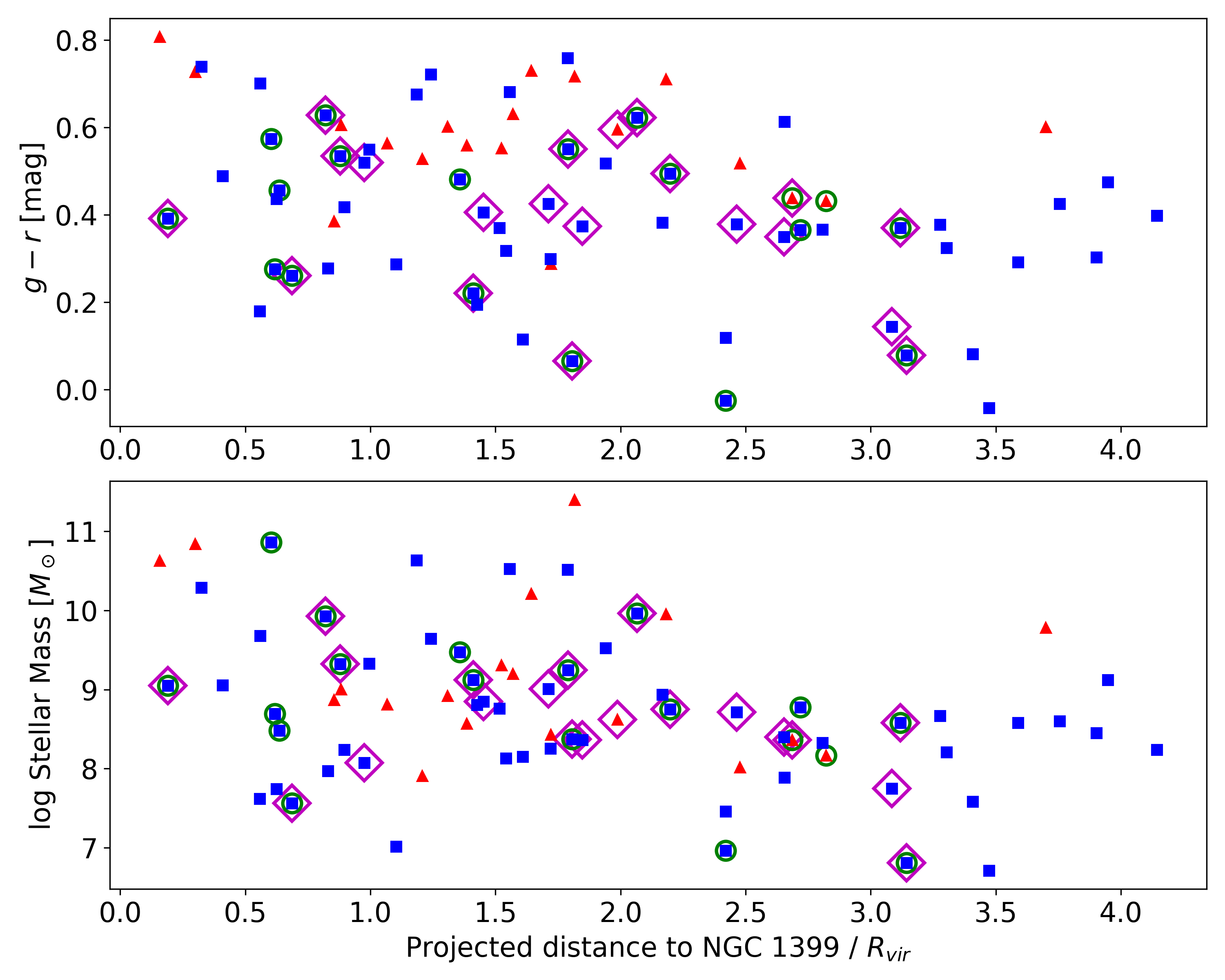}
\caption{Color $(g-r)$ and stellar mass as function of projected distance from NGC 1399 for \HalphaNII{} emitters. The galaxy with higher stellar mass located at $\sim 1.8$ \Rvir{} is the central galaxy of the group Fornax\,A. LTG are represented as squares, whereas ETG  are denoted as triangles. Galaxies identified as perturbed are shown with green open circles, and mergers are marked by magenta open diamonds.}
\label{fig:mass_projdist}
\end{figure}

Fig.\,\ref{fig:CMR_emitters} shows the color-magnitude diagram for all spectroscopic members of the cluster. The brightest galaxy presented here is the central galaxy of Fornax\,A group, NGC\,1316. We exclude NGC\,1532 from this analysis as it is not fully captured in the images, making its magnitudes unreliable. The errors become relevant at $r>16$ mag. The majority (72\%) of the emitters are located below the red sequence. However, it is noticeable that 90\% of the disturbed emitters are found in the blue cloud and the green valley, whereas the red sequence is almost exclusively followed by unperturbed emitters. According to the HyperLeda database \citep{2014/Makarov}\footnote{http://leda.univ-lyon1.fr/}, the few LTG with red color comparable to ELG are all galaxies with identified bars (NGC\,1350, NGC\,1436, NGC\,1317, NGC\,1386, NGC\,1326, FCC\,315).

We derive the stellar mass from the average $(g - i)$ color, using the empirical relation introduced in \cite{2011/Taylor}\footnote{It assumes \cite{2003/Chabrier} initial mass function.},
\begin{equation}
    \log_{10} (M_\mathrm{stellar}/M_{\odot}) = 1.15 + 0.7 \, (g - i) - 0.4 \, M_{i},
\end{equation}
where $M_{i}$ is the absolute magnitude in the $i$-band. In Fig. \ref{fig:mass_projdist}, the upper panel shows a general descending gradient in the color $(g-r)$ for ETG with the clustercentric distance, with a bump of redder galaxies at the projected distance of Fornax\,A. In comparison with the findings of \cite{2019a/Iodice}, who found that the average $(g-r)$ color of bright ETG ($m_\mathrm{B}<15$) within 1\Rvir{} of Fornax tends to decrease with increasing distance from cluster center, our sample of ETG emitters extends this trend out to 4\Rvir. However, the majority of our ETG emitters exhibit $(g-r)>0.4$, which is redder than the ETGs discussed in \cite{2019a/Iodice} within 0.4\Rvir{} and 1\Rvir. This suggests that the presence of emission does not necessarily lead to shift the overall $(g-r)$ color of ETGs toward bluer values. In the bottom panel, we observe a broad range of stellar masses for galaxies located within projected distances ($R_\mathrm{proj}$) of less than 2.5 $R_\mathrm{vir}$. However, beyond this distance, the sample predominantly consists of low-mass galaxies (<$10^{9} M_{\odot}$). Fornax\,A galaxies are distributed at $1.5< R_\mathrm{proj}/R_\mathrm{vir} <2.1$. 

Combining the mass distribution with the PPS diagram and spatial data reveals that at projected distances beyond 2.5 $R_\mathrm{vir}$ from the cluster center, \HalphaNII{} emission line galaxies are predominantly recent infallers. These galaxies tend to be low-mass, blue, and late-type in nature, and those showing evidence of mergers and/or perturbations are preferentially located in the blue cloud and the green valley.

\begin{table*}
\caption {Gas morphology for galaxies around NGC\,1399 reaching 1\Rvir{}.}
\label{table:Fornax} 
\centering
\begin{tabular}{lcccc}
\hline\hline             
Galaxy &  \HalphaNII{} & H I & CO & Infall time \\
\hline\hline
FCC\,90 & extended & deficient \& disturbed  & disturbed & recent \\
NGC\,1437B & extended & deficient \& disturbed & disturbed & intermediate \\
NGC\,1365 & extended with knots in spiral arms  & disturbed  & regular & recent\\
FCC\,312 & extended with knots in spiral arms & disturbed  & regular & recent\\
NGC\,1351A & extended with knots in spiral arms & disturbed  & regular & recent\\ 
FCC\,263 & extended  & deficient & disturbed & recent\\
NGC\,1436 & knots in spiral arms & deficient \& truncated disk & regular & intermediate\\
\hline
  \multicolumn{5}{c}{H I undetected}\\ 
\hline
NGC\,1387 & center concentrated & $-$ & regular & ancient \\
NGC\,1380 & center concentrated & $-$ & regular & (ancient)-recent\\
NGC\,1386 & extended with a central peak & $-$ & regular & recent\\
FCC\,282 & extended & $-$ & disturbed & recent\\
\hline
  \multicolumn{5}{c}{CO not available}\\ 
\hline
FCC\,306 & knots & deficient & $-$ & recent\\
FCC\,113 & knots & deficient & $-$ & intermediate\\
FCC\,115 & knots & deficient & $-$ & recent\\
NGC\,1437A & knots in spiral arms & truncated disk & $-$ & recent\\ 
NGC\,1427A & asymmetric knots & disturbed & $-$ & recent\\ 
\hline
\hline
\end{tabular}
\end{table*}

\subsection{Comparison with radio results}
\label{sec:discussion}
In this subsection, we examine our findings on \HalphaNII{} emission morphology in the context of previous studies that include radio data. The main categories for the emission classification include knots, extended structures, and centrally concentrated regions. This is based on visual inspection, and the categorization is done on the identification of the most prominent or defining characteristics. In some cases, knot structures may exhibit central concentrations, while extended emissions can also include distinct emission knots. We eventually will mention an offset concentration to account for a subset of galaxies that display an emission nucleus or dominating giant H II region displaced from their center. All the maps discussed are included in the Appendix \ref{app:maps}, with Table \ref{tab:data_morph1} presenting their main properties.

\subsubsection{Fornax cluster}
Inside 1\Rvir{} around NGC\,1399, we have 20 \HalphaNII{} maps generated by \texttt{PELE} with 80\% LTG and 20\% ETG. Among the early-types, NGC\,1387 and NGC\,1380 exhibit centrally concentrated emission with no signs of perturbation or mergers in the optical images. Both are located near the cluster center. NGC\,1386 is a recent infaller with extended emission with a distinct central peak. The irregular galaxies FCC\,B905 and FCC\,302 exhibit a prominent offset knot. Seven galaxies—NGC\,1427A, FCC\,115, FCC\,113, FCC\,306, NGC\,1437A, FCC\,299, and FCC\,267—show emission concentrated in knots. The remaining eight—FCC\,90, FCC\,263, FCC\,282, FCC\,312, NGC\,1351A, NGC\,1365, NGC\,1436, and NGC\,1437B—exhibit extended emission. Typically, ETG display centrally concentrated or extended emission without distinct knots, whereas irregulars often feature large, offset emission knots, and spirals show extended or knot-like emission along their arms.

Following results from \cite{2021/Loni} we correlated emission data with cold gas morphology. They first conducted a blind H I survey using ATCA around NGC 1399, covering a region extending to approximately 1\Rvir{}. 
This survey has a spatial resolution of $\sim6 \times 9$ kpc at the Fornax cluster distance and a $3\sigma$ sensitivity in H I mass of $\sim 2 \times 10^{7} M_{\odot}$. These authors detected 16 H I sources, which avoided the central cluster region in both sky distribution and projected phase-space (PPS) analysis, with several showing disturbed H I morphologies, particularly among LTG. These findings align with results from the F3D data \citep[e.g.,][]{2019/Iodice}, which indicates that these galaxies are actively forming stars and are located in the cluster’s low-density regions, where X-ray emission is faint or absent. Similarly, our sample of \HalphaNII{} emitters predominantly consists of recent infallers positioned farther from the cluster center, with many displaying disturbed or merger-like features in optical images. This suggests that several galaxies have recently entered the cluster and are experiencing interactions with other galaxies or the intergalactic medium. Since our selected emitters extend to distances beyond 1\Rvir{}, we observe several galaxies with disturbed classifications but without available H I maps. Obtaining H I measurements for these outskirts objects in the future would be valuable, as their H I morphology could help determine whether these galaxies are already experiencing cluster-induced effects at such distances.

Additionally, \cite{2021/Loni} analyzes the ratio of molecular to atomic gas mass as an indicator of environmental influence on galaxies. For instance, galaxies exhibiting disturbed H I but regular H$_{2}$ suggest that only the atomic gas is impacted by the dense environment. Using the H$_{2}$ masses derived by the conversion from CO given by \cite{2019/Zabel}, they found that the fast process of H I removal can be seen in the large scatter in the H$_{2}$-to-H I mass ratio and in the high number of H I-undetected but H$_{2}$-detected star-forming galaxies. Comparing our sample with the list of objects analyzed by \cite{2021/Loni} and \cite{2019/Zabel}, we have 5 galaxies that overlap only with the first and 4 with the second, plus 7 objects that are common to all three studies. A summary of the gas morphology for these galaxies is shown in Table \ref{table:Fornax}. 

Three galaxies for which we possess \HalphaNII{} maps and CO emission data (indicating the presence of H$_{2}$) but lack H I detection are early-type systems, NGC\,1387, NGC\,1380 and FCC\,282, while NGC\,1386 is a late-type. The first two galaxies exhibit centrally concentrated emission accompanied by regular CO profiles. They are in the high-density region of the cluster \citep[$<0.4$\Rvir{}, e.g.][]{2019a/Iodice}, within the area of detected X-ray emission \citep[e.g.][]{2013/Frank}. NGC\,1386 and FCC\,282 display extended emission, with the latter showing evidence of disturbed CO distribution. Both galaxies are recent infallers. 

From our sample, four objects (FCC\,306, FCC\,113, FCC\,115, and NGC\,1437A) exhibit H I deficiency and display either extended structures or emission knots in their \HalphaNII{} maps. NGC\,1427A shows disturbed H I morphology, accompanied by asymmetric emission line knots and perturbed optical features. NGC\,1437A has a truncated H I disk, characterized by \HalphaNII{} emission knots and a disrupted optical structure.

FCC\,90 and NGC\,1437B are H I deficient with disturbed H I and CO profiles, and display extended emission. NGC\,1365, FCC\,312, and NGC\,1351A present disturbed H I distributions but maintain regular CO profiles, with \HalphaNII{} emission showing both extended regions and knots. FCC\,263 is H I deficient, exhibits disturbed CO, and shows extended \HalphaNII{} emission. NGC\,1436 also displays H I deficiency with a truncated H I disk, regular CO, and emission appearing in knots.

\subsubsection{Fornax A group}
\label{sec:FornaxA}
The Fornax A group is an ideal target for studying pre-processing effects on galaxies, as it is currently infalling toward the Fornax cluster. This group, rich in gas, predominantly consists of LTG. We obtained \HalphaNII{} maps for 11 galaxies within \Rvir{} of Fornax A centered around NGC\,1316. From this sample, 9 sources were matched with neutral hydrogen detections from Meerkat observations presented in \cite{2021/Kleiner}. Based on the author's classification of pre-processing stages, which uses H I morphology along with H$_{2}$ and CO measurements \citep{2019/Zabel,2019/Morokuma-Matsui,2022/Morokuma-Matsui}, the galaxies are separated in three categories:
\begin{enumerate}
    \item Early: H I-rich with extended H I disks, H$_{2}$-to-H I ratios lower than the median for their stellar mass (NGC\,1326A, NGC\,1326B);
    \item Ongoing: H I tails and truncated H I disks, with typical H$_{2}$-to-H I ratios (FCC\,28, FCC\,35, NGC\,1310, NGC\,1316, NGC\,1326);
    \item Advanced: H I deficient with no H I outer disk, H$_{2}$-to-H I ratios higher than the median for their stellar mass (NGC\,1316C, NGC\,1317).
\end{enumerate}

The galaxies NGC\,1326A and NGC\,1326B, categorized as being in the early stages of pre-processing present prominent knots in their emission line maps. Although they may appear to form an interacting pair due to their proximity in projection in the sky, their velocity difference of $\sim 832$ km s$^{-1}$ indicates they are not a close interacting system \citep{2018Ap&SS.363..131R}. This distinction was also addressed by \cite{2020/Raj}, who noted that the two galaxies occupy different positions in the phase-space diagram of Fornax\,A, with NGC\,1326A classified as an intermediate infaller and NGC\,1326B as a recent infaller to the group. (see Figure 7 therein).

The galaxies in advanced pre-processing, NGC\,1316C and NGC\,1317, present extended emission line maps. As can be seen in the panel located at the fourth line of the second column in Fig. \ref{fig:1.6Rvir_1.99Rvir}, NGC\,1317 has a ring of emission located around the center of the galaxy. No emission was detected in the disk, which coincides with the lack of an H I disk in these galaxies.

The list of galaxies undergoing active pre-processing include several fascinating objects. Among them is FCC\,28, which exhibits asymmetrically distributed emission knots, almost forming a ring-like structure. Two sources, NGC\,1316 and NGC\,1326, are classified as AGNs based on X-ray detections \citep[e.g.][]{2020/Maccagni,2024/Hou}. NGC\,1326 displays centrally concentrated emission with a few knots in its outer regions, while NGC\,1316 features central emission accompanied by extended structures reminiscent of X-ray lobes, albeit on a smaller scale. NGC\,1310 presents a more organized extended emission pattern, with knots aligning along its spiral arms. Lastly, FCC\,35 exhibits distinctly extended emission.

\begin{figure*}
    \centering
    \includegraphics[width=\textwidth]{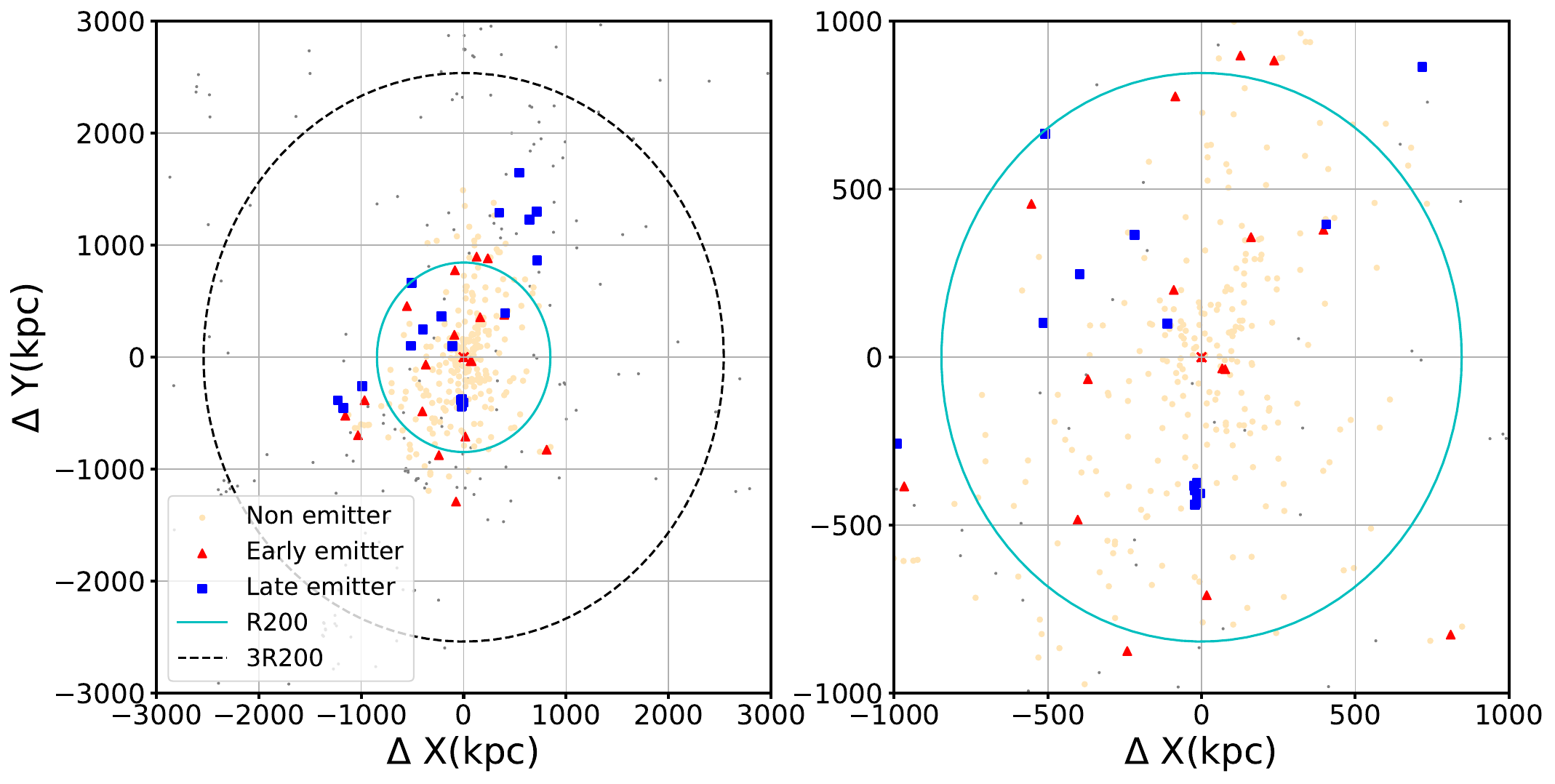}
    \caption{Spatial distribution of simulated galaxies in the Fornax-like system present in IllustrisTNG-50. \textit{Left:} Distribution of galaxies up to $1R_{200}$ (cyan continuous line) and $3R_{200}$ (black dashed line). Late, early, and non-emitter galaxies bound to the cluster are shown in squares (blue), triangles (red), and full circles (maroon), respectively. Additionally, gray dots show galaxies within $5R_{200}$ of the cluster that are not gravitationally bound to the cluster.
    \textit{Right:} Zoom region showing the galaxy distribution within $1R_{200}$ of the Fornax-like system.}
    \label{fig:TNG_FornaxDist}
\end{figure*}

\subsection{Beyond \Rvir{} and assembly history of the Fornax cluster}
Forty eight of our emitters are galaxies distributed beyond 1\Rvir{} of the Fornax cluster, but not belonging to the Fornax A group. From these objects, we find an extended emission in 24 (58\%) extended, knots in 20 (42\%), and 4 (6\%) with central concentration. They are mostly LTG (73\%) and 40\% present at least one sign of perturbations and/or merger.

Studies such as F3D \citep[e.g.][]{2019/Iodice} and FDS \citep[e.g.][]{2019a/Iodice, 2020/Spavone}, that examine bright ($m_\mathrm{B}\le 15$ mag) galaxies within the virial radius of the Fornax system have identified three main groups with galaxies with similar properties: the core, the north-south (N-S) clump and the infalling galaxies, which are added to the southwest Fornax\,A group centered on NGC\,1316. The core is dominated by the massive luminous NGC\,1399, which coincides with the X-ray emission detection \citep[e.g.][]{2013/Frank}. Still inside the X-ray halo, there is a high density region until 0.4\Rvir{}, where the N-S clump is found. The galaxies located at the clump on the north-northwest side of the cluster are redder, fast rotators, and metal-rich. The third group is composed of intermediate and recent infallers positioned in an approximately symmetrical pattern around the core within the low-density region (0.4 < $R_\mathrm{proj}$ < 1). These galaxies are predominantly late-type, exhibiting ongoing star formation and evidence of structural transformations. For instance, FCC\,263 displays disturbed ionized gas (see Fig. \ref{fig:musexsplus}), which is likely driven by environmental interactions or merger events.

\begin{figure*}
    \centering
    \includegraphics[width=\textwidth]{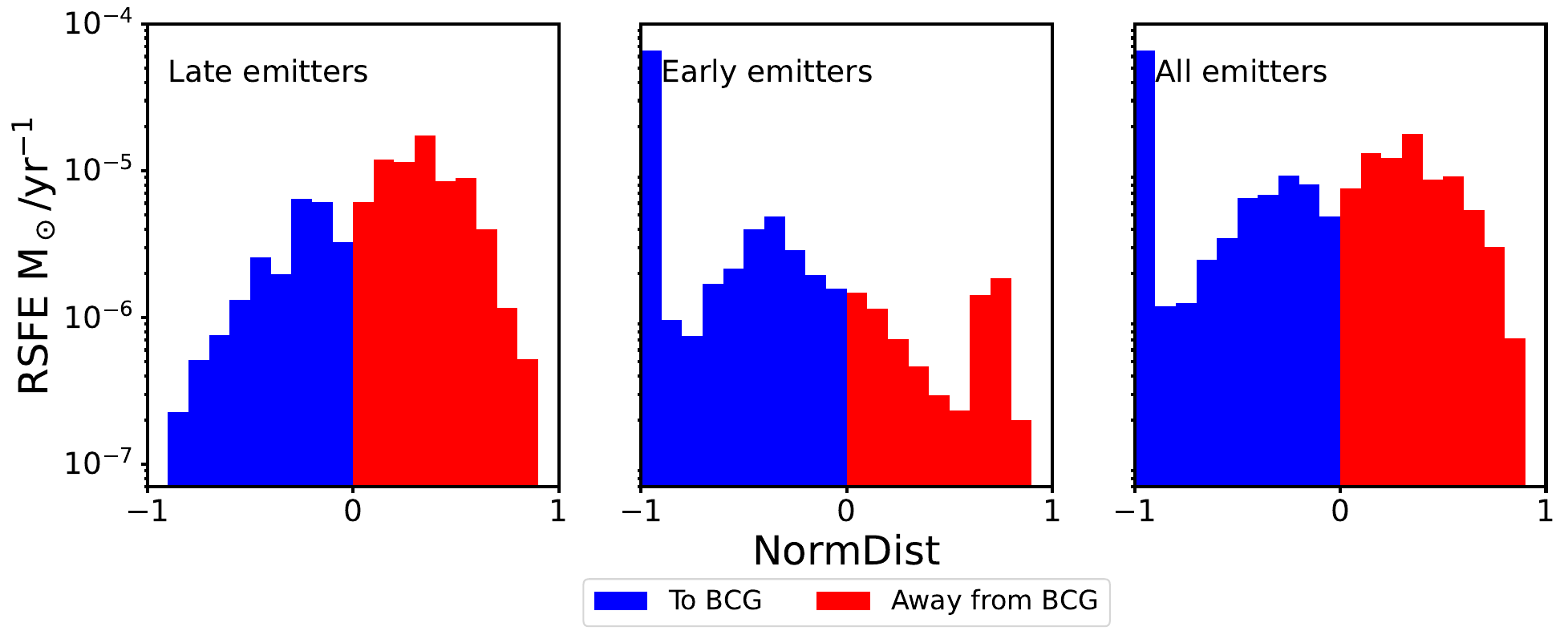}
    \caption{Evidence of misaligned star formation rate for satellite galaxies from the Fornax-like cluster in IllustrisTNG-50. The left, central, and rightmost panels show the resolved star formation efficiency spatial distribution for late-, early- and all-emitters, respectively.
    The zero value split the galaxy into halves, with negative values (between [-1,0); blue bars) being the half of the galaxy closer to the cluster's center, and positive values (between (0,1]; red bars), the half of the galaxy to the outskirts of the cluster. Late emitters show a clear signal of a higher star formation rate toward the cluster's outskirts, while early emitters show the opposite trend.} 
    \label{fig:TNG_SFE}
\end{figure*}

Our results agree with previous findings with a low number of emitters being detected in the N-S clump. Agreeing with the conclusion that the star formation in these galaxies stopped earlier than for those that entered later into the cluster potential \citep{2019/Iodice}. As our focus is on  \HalphaNII{} emitters, the patterns seen in galaxies without signs of star formation or AGN activity, particularly ETG within $R_\mathrm{proj}<0.4$\Rvir, are not reproduced here. For instance, within 1\Rvir{}, we successfully mapped \HalphaNII{} emission in three ETG, NGC\,1387 and NGC\,1380, both situated in the high density core of the cluster, and FCC\,90, located in the low density region. None of these objects present any disturbance or merger feature, agreeing with previous results from ETG \citep{2020/Spavone}. Indeed, only three ETG presented indications of perturbed or merger features, and they are located at projected distances larger than 2\Rvir.

We also expand the previous analysis to emission line galaxies located in regions outside the virial radius and $r<18$. We found the majority of emitters in the low density region of Fornax. A high number of emitters are located in the north-northwest side of the cluster between 1\Rvir{} and 3\Rvir. These galaxies seem be the ones belonging to the filament that connects the Fornax-Eridanus large scale structure, as suggested by \cite{2011/Nasonova}. In the west-southwest, the Fornax A group is the main location of the emitters. 

In the context of the pre-processing scenario, \cite{2009/McGee} found through hydrodynamical simulations that galaxies that were accreted recently into the cluster are more likely to have been in a group and that low-stellar-mass galaxies are effective tracers of the recent mass accretion history of the cluster. Within this framework, the spatial distribution of ETG and LTG presented in this paper traces the assembly history of Fornax. On one hand, most of our ETG emitters are first infallers and exhibit no signs of morphological disturbance. This suggests  that these galaxies had already evolved prior to falling into the cluster environment. In fact, the ELG population may have been quenched prior to infall (the time when a galaxy enters for the first time the virial volume), as already indicated by \cite{2020/Spavone}. This interpretation is consistent with results from studies of high-redshift clusters $(z\gtrsim1)$, such as those conducted by the Gemini Observations of Galaxies in Rich Early ENvironments \citep[GOGREEN,][]{2021/Balogh} collaboration \citep[see e.g.,][]{2020/vanderBurg,2021/McNab,2022/Baxter}. On the other hand, the LTG population seems to be tracing the more recent mass-assembly from the field, with galaxies coming from different directions, including along filaments, as suggested by their spatial distribution (Figure \ref{fig:spatial-dist}). Evidence of morphological transformation during the cluster assembly process is more apparent among these star-forming systems. Notably, 44\% of the late-type emitters show morphological signs of perturbation or merging process. In particular, LTG located at larger cluster-centric distances are low-mass, $\log(M_\mathrm{stellar}) < 10^{9} M_\odot$, which is consistent with the scenario in which low-mass galaxies are more susceptible to environmental effects, such as galaxy-galaxy interactions in groups and filaments. The presence of star-forming, low-mass galaxies with disturbed morphologies at $R_\mathrm{proj} > 2.5$\Rvir{} strongly supports a  pre-processing scenario, i.e., galaxies start the transform processes prior to their accretion into the main cluster. 

\subsection{Fornax-like systems in IllustrisTNG-50}
Hydrodynamical simulations have proven to be a key tool in understanding the role that the environment plays in galaxy evolution \citep[e.g.][]{Bahe13, Wright18, Donnari20, Pallero19, Pallero22}. However, due to the lack of resolution in large cosmological hydrodynamical simulations, studying the evolution of dwarf galaxies in complex systems like Fornax has been a difficult task. To shed light on the matter, we will use the high-resolution box from the IllustrisTNG suite \citep{Marinacci18, Nelson18, Pillepich18, Springel18} to look for the distribution of emitters in Fornax-like systems. The simulations from IllustrisTNG, count with a full-physics galaxy formation model, which includes magneto-hydrodynamic physics to model several processes affecting galaxy evolution, such as star formation, chemical enrichment, supernova feedback, AGN feedback, photoionization, etc. From this suite, IllustrisTNG-50 has the smallest volume (51.6$^3$cMpc) and the highest resolution with a baryonic particle mass of $8.5\times 10^{4} M_\odot$ and dark matter of $4.5\times10^{5} M_\odot$ \citep{Nelson19, Pillepich19}.
This allows us to resolve satellites  $\sim 10^7M_\star / M_\odot$ with $>100$ stellar particles and $\sim 1000$ dark matter particles. Moreover, despite the simulation's small box size, IllustrisTNG-50 counts with one Fornax-like system with a virial mass $M_{200} = 6.7\times10^{13}M_\odot$  and virial radii $R_{200}$ of 846 kpc, with 303 satellite galaxies $M_\star > 10^{7} M_\odot$ bound to the cluster. 

Fig. \ref{fig:TNG_FornaxDist} shows the galaxy distribution of the Fornax-like system and its surroundings up to $3R_{200}$ (left panel) and $1R_{200}$ (right panel). 
For visualization and to make a comparison with observations more clear, we split our satellite population into three samples based on its sSFR:
\begin{itemize}
    \item Late-emitters: Galaxies with high sSFR, considered to be on the main sequence (log$_{10}$sSFR > 10.5)
    \item  Early-emitters: Galaxies with low sSFR, considered below the star formation main sequence (log$_{10}$sSFR < 10.5), but with some emission.
    \item Non-emitters: Galaxies with no star formation rate.

\end{itemize}
Squares (blue), triangles (red), and filled circles (maroon) show the distribution of galaxies bound to the Fornax-like system. The continuous (cyan) and dashed (black) lines correspond to 1 and 3$R_{200}$, respectively. Gray dots show galaxies that are within $3R_{200}$ of the system but are not bound to the cluster, according to a FoF algorithm \citep{Springel01, Dolag09}.
Of these 303 galaxies, only 40 galaxies are emitters, being 19 and 21 late- and early-emitters, respectively.
Nonetheless, when counting all emitters within $3R_{200}$, we found 74 emitters around the Fornax-like system, similar to what we report in this work.

To analyze the distribution of the star-forming regions from satellites in our Fornax-like system, we measure, per galaxy, the distance of each star-forming region to the center of the cluster and split the distance between 20 radial bins, normalized by the galaxy size. In each bin, we measure the resolved star-formation efficiency, defined as 
\begin{equation}
    RSFE_i = SFR_{i}/M_{gas, i}
\end{equation}
where $SFR_i$ is the star formation, per radial bin and $M_{gas ,i}$ is the gas mass enclosed in the $i^{th}$ radial bin.
Then, we stacked the results for all our emitters as shown in Fig. \ref{fig:TNG_SFE}. For visualization purposes, we fixed the x-axis to be from -1 to 1, where -1 is the bin closer to the center of the cluster, and 1 is the bin farthest to the center.
The zero value marks the center of each galaxy. 
The left, middle, and rightmost panels show the distribution of RSFE for late- early and all-emitters, respectively. As can be seen, late-emitters increase their RSFE toward the outskirts of clusters, while early-emitters have this trend reversed. This change in the distribution is related to the dynamical state of early- and late-emitters regarding the cluster. Typically in our sample, late-emitters correspond to galaxies undergoing their first infall, while early-emitters have spent more time within the cluster. This distribution of misaligned RSFE could help us in the future to better constrain the direction of motion of galaxies during their infall into structures and will be further explored in a follow-up paper. 

\section{Conclusions}
\label{sec:conclusions}
In this paper we present an approach to create \HalphaNII{} maps for S-PLUS images applying the Three Filter Method in the region of the Fornax cluster. We successfully map the \HalphaNII{} emission in 77 galaxies within the four virial radius of the Fornax cluster and with $r < 18$ mag, with only two not being confirmed spectroscopic members. Our method reaches a limit of H$\alpha$ flux $\sim2\times 10^{-17}$ erg s$^{-1}$ cm$^{-2}$, when compared to IFS results from MUSE/F3D. The main findings of this study are summarized as follows:
\begin{itemize}
    \item 25\% are ETG (elliptical or lenticular), while 75\% are LTG (spiral or irregular). The ETG are predominantly concentrated in the north-west region of the cluster, whereas the LTG are more evenly distributed.
   \item  The analysis of the morphologies using the \texttt{ASTROMORPHLIB} code applied to DESI Legacy $g$-band images of the galaxies revealed that about 38\% of the \HalphaNII{} emitters are experiencing morphological transformation, as determined by considering both the region defined as perturbed in the Concentration versus Asymmetry diagram from \cite{2024/Krabbe} and the merger region in the Gini versus $M_{20}$ diagram from \cite{2004/Lotz}, where galaxies that fall within these regions are considered to have disturbed morphologies.
    \item Evidence of perturbation and/or merging is observed in 15\% of the early-type emitters, while this fraction increases to 44\% among late-type emitters.
    \item 91\% of them are recent infallers, with $\sim77\%$ being LTG and 40\% showing signs of perturbation or merger activity. In contrast, only two ETG, located near 3\Rvir{}, show evidence of disturbance or merger features.
    \item Only one emitter has entered the cluster at > 8 Gyr, NGC\,1387. In the intermediate regime, between 4 and 8 Gyr, we have 6 emitters, 2 displaying early and 4 late-type morphologies.
    \item Among the emitters, 8\% display centrally concentrated emission, 40\% exhibit emission in knots, 52\% show extended emission. All galaxies with centrally concentrated emission are unperturbed objects, and the ones with radio data present a regular CO distribution.
    \item At the projected distances beyond 2.5\Rvir{} from the cluster center, the \HalphaNII{} emitters are predominantly low-mass, blue, late-type recent infallers. 
    \item On the northwest outskirts of the cluster, a clump of emitters appears to extend the pattern observed in the north-south clump within the Fornax virial radius \citep{2019a/Iodice,2019/Iodice}. This distribution suggests they may be part of a filament feeding into the cluster, potentially linking Fornax to the Eridanus cluster to the north. Further analysis, extending observations toward Eridanus, is needed to confirm this connection.
    \item  The eastern region of the cluster, although hosting several disturbed galaxies, does not exhibit a distinct substructure.
    \item In a Fornax-like system from the IllustrisTNG-50 simulation, we identified 74 emitters within $3R_{200}$, closely matching our observational results. The simulated galaxies exhibit a clear trend: ETG show enhanced star formation toward the cluster center, while LTG display the opposite behavior. The late-type emitters are typically first-infallers, whereas early-type emitters have resided in the cluster for longer. Future observational studies of such star formation rate (SFR) misalignments could provide valuable insights into the dynamics of galaxy infall through substructures toward the cluster center.  
\end{itemize}

It is well established that galaxy clusters grow by hierarchical assembling of structures. Therefore the analysis of galaxy properties in the outer regions of the cluster (beyond \Rvir) provides valuable insight into the cluster's assembly history and its connection with galaxy evolution. Indeed, our results reflect the assembly history of the Fornax cluster, with LTG appearing to trace the recent mass assembly of the cluster from the field, with galaxies infalling from multiple directions (see Figure \ref{fig:spatial-dist}), including a filament in the northwest of the cluster that possibly connects to Eridanus system. While ETG could have already evolved prior to falling into the cluster environment. Additionally, the presence of disturbed, low-mass galaxies beyond 2.5\Rvir{} provides evidence of pre-processing, i.e. these galaxies begin undergoing transformation before entering into the main cluster.

It is important to note that while our photometric approach does not replace IFS spectroscopic analysis, it can be valuable for a broad assessment of a given emission line across larger areas, such as in galaxy clusters and groups. Some limitations of the results presented here will be addressed in forthcoming papers, including the absence of internal dust extinction and \NII{} emission corrections, and the assessment of the equivalent width of H$\alpha$ using S-PLUS filters. For such, we will evaluate current known recipes to correct for internal dust extinction and to remove the \NII{} contamination using photometry, and potentially develop new ones. The absence of \NII{} correction can lead to an overestimation of the H$\alpha$ emission flux and, consequently, of the derived star formation rate (SFR). Correction factors based on nuclear measurements (e.g., slit spectroscopy) may also underestimate the true H$\alpha$ flux in star-forming regions, particularly in spiral galaxies \citep{2005/James}. Additionally, in cases such as active galactic nuclei (AGNs), \NII{} emission may dominate the narrow-band filter response, further biasing the results.

Several galaxies in our sample are located in the outskirts of the cluster, where no radio data is currently available. Although galaxies within 1\Rvir{} of the Fornax cluster and the Fornax\,A group are being studied across multiple wavelengths, the outer regions still lack sufficient radio data coverage. The combination of cold and molecular gas is crucial for understanding how the environment impacts and transforms galaxies, providing valuable pre-processing diagnostics and revealing physical processes such as ram pressure stripping and tidal stripping during interactions with the intracluster medium or other galaxies. Targeting these emitters for future radio observations would be highly beneficial.

Our results confirm that the Fornax cluster is still evolving, with numerous indications of morphological transformations occurring at large clustercentric distances. The distribution of \HalphaNII{} emitters suggests a connection to the large-scale Fornax-Eridanus structure, emphasizing their potential for tracing cosmic filaments and unveiling the underlying structure of the cosmic web. 

This study highlights the effectiveness of S-PLUS in detecting and characterizing emission-line features in galaxy clusters, providing a valuable tool for large-scale investigations of galaxy evolution. The observed trends reinforce the significant influence of environmental factors on star formation history and morphological transformations in galaxies. The methodology and findings presented here can be extended to other clusters, offering broader insights into the role of cluster environments in galaxy evolution across cosmic scales.

\begin{acknowledgements}
We would like to thank the anonymous referee for their suggestions, which significantly helped us improve the paper. A.R.L. and A.V.S.C. acknowledge financial support from Consejo Nacional de Investigaciones Científicas y Técnicas (CONICET) (PIP 1504), Agencia I+D+i (PICT 2019–03299) and Universidad Nacional de La Plata (Argentina). A.C.K thanks Fundação de Amparo à Pesquisa do Estado de São Paulo (FAPESP) for the support grant 2024/05467-9  and the Conselho Nacional de Desenvolvimento Científico e Tecnológico (CNPq). A.R.L. also acknowledges the financial support from Conselho Nacional de Pesquisa (CNPq). J.T.B. acknowledges CAPES 88881.892595/2023-01, FAPESC (CP 48/2021). M.S.C acknowledges support from São Paulo Research Foundation (FAPESP), grant 2023/10774-5. D.P. acknowledges financial support from ANID through FONDECYT Postdoctrorado Project 3230379. C.L. acknowledges support by Fundação para a Ciência e a Tecnologia (FCT) through the research grants UIDB/04434/2020 and UIDP/04434/2020. DOI: 10.54499/UIDB/04434/2020 and DOI: 10.54499/UIDP/04434/2020. F.R.H. acknowledges support from FAPESP grants 2018/21661-9 and 2021/11345-5. R.D. gratefully acknowledges support by the ANID BASAL project FB210003. P.K.H. gratefully acknowledges the Fundação de Amparo à Pesquisa do Estado de São Paulo (FAPESP) for the support grant 2023/14272-4. L.L.N. thanks Funda\c{c}\~ao de Amparo \`a Pesquisa do Estado do Rio de Janeiro (FAPERJ) for granting the postdoctoral research fellowship E-40/2021(280692). C.L-D acknowledges financial support from the ESO Comite Mixto 2022. D.E.O-R acknowledges the financial support from the Chilean National Agency for Research and Development (ANID) through the InESG\'{e}nero project INGE210025, and funding from the Fondo Aporte para el Desarrollo de Actividades de Inter\'{e}s Nacional from the Ministerio de Educaci\'{o}n through the project TAL2293.
S.V.W. is supported by the United Kingdom Research and Innovation (UKRI) Future Leaders Fellowship `Using Cosmic Beasts to uncover the Nature of Dark Matter' (grant number MR/X006069/1).

The S-PLUS project, including the T80-South robotic telescope and the S-PLUS scientific survey, was founded as a partnership between the Fundação de Amparo à Pesquisa do Estado de São Paulo (FAPESP), the Observatório Nacional (ON), the Federal University of Sergipe (UFS), and the Federal University of Santa Catarina (UFSC), with important financial and practical contributions from other collaborating institutes in Brazil, Chile (Universidad de La Serena), and Spain (Centro de Estudios de Física del Cosmos de Aragón, CEFCA). We further acknowledge financial support from the São Paulo Research Foundation (FAPESP), Fundação de Amparo à Pesquisa do Estado do RS (FAPERGS), the Brazilian National Research Council (CNPq), the Coordination for the Improvement of Higher Education Personnel (CAPES), the Carlos Chagas Filho Rio de Janeiro State Research Foundation (FAPERJ), and the Brazilian Innovation Agency (FINEP). The authors who are members of the S-PLUS collaboration are grateful for the contributions from CTIO staff in helping in the construction, commissioning and maintenance of the T80-South telescope and camera. We are also indebted to Rene Laporte and INPE, as well as Keith Taylor, for their important contributions to the project. From CEFCA, we particularly would like to thank Antonio Marín-Franch for his invaluable contributions in the early phases of the project, David Cristóbal-Hornillos and his team for their help with the installation of the data reduction package jype version 0.9.9, César Íñiguez for providing 2D measurements of the filter transmissions, and all other staff members for their support with various aspects of the project. 

The Legacy Surveys consist of three individual and complementary projects: the Dark Energy Camera Legacy Survey (DECaLS; Proposal ID 2014B-0404; PIs: David Schlegel and Arjun Dey), the Beijing-Arizona Sky Survey (BASS; NOAO Prop. ID 2015A-0801; PIs: Zhou Xu and Xiaohui Fan), and the Mayall $z$-band Legacy Survey (MzLS; Prop. ID 2016A-0453; PI: Arjun Dey). DECaLS, BASS and MzLS together include data obtained, respectively, at the Blanco telescope, Cerro Tololo Inter-American Observatory, NSF’s NOIRLab; the Bok telescope, Steward Observatory, University of Arizona; and the Mayall telescope, Kitt Peak National Observatory, NOIRLab. Pipeline processing and analyses of the data were supported by NOIRLab and the Lawrence Berkeley National Laboratory (LBNL). The Legacy Surveys project is honored to be permitted to conduct astronomical research on Iolkam Du’ag (Kitt Peak), a mountain with particular significance to the Tohono O’odham Nation.

This project used data obtained with the Dark Energy Camera (DECam), which was constructed by the Dark Energy Survey (DES) collaboration. Funding for the DES Projects has been provided by the U.S. Department of Energy, the U.S. National Science Foundation, the Ministry of Science and Education of Spain, the Science and Technology Facilities Council of the United Kingdom, the Higher Education Funding Council for England, the National Center for Supercomputing Applications at the University of Illinois at Urbana-Champaign, the Kavli Institute of Cosmological Physics at the University of Chicago, Center for Cosmology and Astro-Particle Physics at the Ohio State University, the Mitchell Institute for Fundamental Physics and Astronomy at Texas A\&M University, Financiadora de Estudos e Projetos, Fundacao Carlos Chagas Filho de Amparo, Financiadora de Estudos e Projetos, Fundacao Carlos Chagas Filho de Amparo a Pesquisa do Estado do Rio de Janeiro, Conselho Nacional de Desenvolvimento Cientifico e Tecnologico and the Ministerio da Ciencia, Tecnologia e Inovacao, the Deutsche Forschungsgemeinschaft and the Collaborating Institutions in the Dark Energy Survey. The Collaborating Institutions are Argonne National Laboratory, the University of California at Santa Cruz, the University of Cambridge, Centro de Investigaciones Energeticas, Medioambientales y Tecnologicas-Madrid, the University of Chicago, University College London, the DES-Brazil Consortium, the University of Edinburgh, the Eidgenossische Technische Hochschule (ETH) Zurich, Fermi National Accelerator Laboratory, the University of Illinois at Urbana-Champaign, the Institut de Ciencies de l’Espai (IEEC/CSIC), the Institut de Fisica d’Altes Energies, Lawrence Berkeley National Laboratory, the Ludwig Maximilians Universitat Munchen and the associated Excellence Cluster Universe, the University of Michigan, NSF’s NOIRLab, the University of Nottingham, the Ohio State University, the University of Pennsylvania, the University of Portsmouth, SLAC National Accelerator Laboratory, Stanford University, the University of Sussex, and Texas A\&M University.

BASS is a key project of the Telescope Access Program (TAP), which has been funded by the National Astronomical Observatories of China, the Chinese Academy of Sciences (the Strategic Priority Research Program “The Emergence of Cosmological Structures” Grant XDB09000000), and the Special Fund for Astronomy from the Ministry of Finance. The BASS is also supported by the External Cooperation Program of Chinese Academy of Sciences (Grant  114A11KYSB20160057), and Chinese National Natural Science Foundation (Grant  12120101003,  11433005).

The Legacy Survey team makes use of data products from the Near-Earth Object Wide-field Infrared Survey Explorer (NEOWISE), which is a project of the Jet Propulsion Laboratory/California Institute of Technology. NEOWISE is funded by the National Aeronautics and Space Administration.

The Legacy Surveys imaging of the DESI footprint is supported by the Director, Office of Science, Office of High Energy Physics of the U.S. Department of Energy under Contract No. DE-AC02-05CH1123, by the National Energy Research Scientific Computing Center, a DOE Office of Science User Facility under the same contract; and by the U.S. National Science Foundation, Division of Astronomical Sciences under Contract No. AST-0950945 to NOAO.

This research has made use of NASA’s Astrophysics Data System (ADS) Bibliographic Services. The ADS is operated by the Smithsonian Astrophysical Observatory under NASA Cooperative Agreement 80NSSC21M0056.
\end{acknowledgements}

\bibliographystyle{aa} 
\bibliography{S+FP_Halpha} 

\begin{thebibliography}{130}
\expandafter\ifx\csname natexlab\endcsname\relax\def\natexlab#1{#1}\fi

\bibitem[{{Bacon} {et~al.}(2010){Bacon}, {Accardo}, {Adjali}, {Anwand},
  {Bauer}, {Biswas}, {Blaizot}, {Boudon}, {Brau-Nogue}, {Brinchmann},
  {Caillier}, {Capoani}, {Carollo}, {Contini}, {Couderc}, {Daguis{\'e}},
  {Deiries}, {Delabre}, {Dreizler}, {Dubois}, {Dupieux}, {Dupuy}, {Emsellem},
  {Fechner}, {Fleischmann}, {Fran{\c{c}}ois}, {Gallou}, {Gharsa}, {Glindemann},
  {Gojak}, {Guiderdoni}, {Hansali}, {Hahn}, {Jarno}, {Kelz}, {Koehler},
  {Kosmalski}, {Laurent}, {Le Floch}, {Lilly}, {Lizon}, {Loupias}, {Manescau},
  {Monstein}, {Nicklas}, {Olaya}, {Pares}, {Pasquini}, {P{\'e}contal-Rousset},
  {Pell{\'o}}, {Petit}, {Popow}, {Reiss}, {Remillieux}, {Renault}, {Roth},
  {Rupprecht}, {Serre}, {Schaye}, {Soucail}, {Steinmetz}, {Streicher}, {Stuik},
  {Valentin}, {Vernet}, {Weilbacher}, {Wisotzki}, \& {Yerle}}]{2010/Bacon}
{Bacon}, R., {Accardo}, M., {Adjali}, L., {et~al.} 2010, in Society of
  Photo-Optical Instrumentation Engineers (SPIE) Conference Series, Vol. 7735,
  Ground-based and Airborne Instrumentation for Astronomy III, ed. I.~S.
  {McLean}, S.~K. {Ramsay}, \& H.~{Takami}, 773508

\bibitem[{{Bah{\'e}} {et~al.}(2013){Bah{\'e}}, {McCarthy}, {Balogh}, \&
  {Font}}]{Bahe13}
{Bah{\'e}}, Y.~M., {McCarthy}, I.~G., {Balogh}, M.~L., \& {Font}, A.~S. 2013,
  \mnras, 430, 3017

\bibitem[{{Balogh} {et~al.}(1999){Balogh}, {Morris}, {Yee}, {Carlberg}, \&
  {Ellingson}}]{1999/Balogh}
{Balogh}, M.~L., {Morris}, S.~L., {Yee}, H.~K.~C., {Carlberg}, R.~G., \&
  {Ellingson}, E. 1999, \apj, 527, 54

\bibitem[{{Balogh} {et~al.}(2021){Balogh}, {van der Burg}, {Muzzin}, {Rudnick},
  {Wilson}, {Webb}, {Biviano}, {Boak}, {Cerulo}, {Chan}, {Cooper}, {Gilbank},
  {Gwyn}, {Lidman}, {Matharu}, {McGee}, {Old}, {Pintos-Castro}, {Reeves},
  {Shipley}, {Vulcani}, {Yee}, {Alonso}, {Bellhouse}, {Cooke}, {Davidson}, {De
  Lucia}, {Demarco}, {Drakos}, {Fillingham}, {Finoguenov}, {Forrest},
  {Golledge}, {Jablonka}, {Lambas Garcia}, {McNab}, {Muriel}, {Nantais},
  {Noble}, {Parker}, {Petter}, {Poggianti}, {Townsend}, {Valotto}, {Webb}, \&
  {Zaritsky}}]{2021/Balogh}
{Balogh}, M.~L., {van der Burg}, R. F.~J., {Muzzin}, A., {et~al.} 2021, \mnras,
  500, 358

\bibitem[{{Baxter} {et~al.}(2022){Baxter}, {Cooper}, {Balogh}, {Carleton},
  {Cerulo}, {De Lucia}, {Demarco}, {McGee}, {Muzzin}, {Nantais},
  {Pintos-Castro}, {Reeves}, {Rudnick}, {Sarron}, {van der Burg}, {Vulcani},
  {Wilson}, \& {Zaritsky}}]{2022/Baxter}
{Baxter}, D.~C., {Cooper}, M.~C., {Balogh}, M.~L., {et~al.} 2022, \mnras, 515,
  5479

\bibitem[{{Benitez} {et~al.}(2014){Benitez}, {Dupke}, {Moles}, {Sodre},
  {Cenarro}, {Marin-Franch}, {Taylor}, {Cristobal}, {Fernandez-Soto}, {Mendes
  de Oliveira}, {Cepa-Nogue}, {Abramo}, {Alcaniz}, {Overzier},
  {Hernandez-Monteagudo}, {Alfaro}, {Kanaan}, {Carvano}, {Reis}, {Martinez
  Gonzalez}, {Ascaso}, {Ballesteros}, {Xavier}, {Varela}, {Ederoclite},
  {Vazquez Ramio}, {Broadhurst}, {Cypriano}, {Angulo}, {Diego}, {Zandivarez},
  {Diaz}, {Melchior}, {Umetsu}, {Spinelli}, {Zitrin}, {Coe}, {Yepes}, {Vielva},
  {Sahni}, {Marcos-Caballero}, {Kitaura}, {Maroto}, {Masip}, {Tsujikawa},
  {Carneiro}, {Gonzalez Nuevo}, {Carvalho}, {Reboucas}, {Carvalho}, {Abdalla},
  {Bernui}, {Pigozzo}, {Ferreira}, {Chandrachani Devi}, {Bengaly}, {Campista},
  {Amorim}, {Asari}, {Bongiovanni}, {Bonoli}, {Bruzual}, {Cardiel}, {Cava},
  {Cid Fernandes}, {Coelho}, {Cortesi}, {Delgado}, {Diaz Garcia}, {Espinosa},
  {Galliano}, {Gonzalez-Serrano}, {Falcon-Barroso}, {Fritz}, {Fernandes},
  {Gorgas}, {Hoyos}, {Jimenez-Teja}, {Lopez-Aguerri}, {Lopez-San Juan},
  {Mateus}, {Molino}, {Novais}, {OMill}, {Oteo}, {Perez-Gonzalez}, {Poggianti},
  {Proctor}, {Ricciardelli}, {Sanchez-Blazquez}, {Storchi-Bergmann}, {Telles},
  {Schoennell}, {Trujillo}, {Vazdekis}, {Viironen}, {Daflon},
  {Aparicio-Villegas}, {Rocha}, {Ribeiro}, {Borges}, {Martins}, {Marcolino},
  {Martinez-Delgado}, {Perez-Torres}, {Siffert}, {Calvao}, {Sako}, {Kessler},
  {Alvarez-Candal}, {De Pra}, {Roig}, {Lazzaro}, {Gorosabel}, {Lopes de
  Oliveira}, {Lima-Neto}, {Irwin}, {Liu}, {Alvarez}, {Balmes}, {Chueca},
  {Costa-Duarte}, {da Costa}, {Dantas}, {Diaz}, {Fabregat}, {Ferrari},
  {Gavela}, {Gracia}, {Gruel}, {Gutierrez}, {Guzman}, {Hernandez-Fernandez},
  {Herranz}, {Hurtado-Gil}, {Jablonsky}, {Laporte}, {Le Tiran}, {Licandro},
  {Lima}, {Martin}, {Martinez}, {Montero}, {Penteado}, {Pereira}, {Peris},
  {Quilis}, {Sanchez-Portal}, {Soja}, {Solano}, {Torra}, \&
  {Valdivielso}}]{2014/Benitez}
{Benitez}, N., {Dupke}, R., {Moles}, M., {et~al.} 2014, arXiv e-prints,
  arXiv:1403.5237

\bibitem[{{Blakeslee} {et~al.}(2009){Blakeslee}, {Jord{\'a}n}, {Mei},
  {C{\^o}t{\'e}}, {Ferrarese}, {Infante}, {Peng}, {Tonry}, \&
  {West}}]{2009/Blakeslee}
{Blakeslee}, J.~P., {Jord{\'a}n}, A., {Mei}, S., {et~al.} 2009, \apj, 694, 556

\bibitem[{{Boselli} {et~al.}(2014){Boselli}, {Voyer}, {Boissier}, {Cucciati},
  {Consolandi}, {Cortese}, {Fumagalli}, {Gavazzi}, {Heinis}, {Roehlly}, \&
  {Toloba}}]{2014/Boselli}
{Boselli}, A., {Voyer}, E., {Boissier}, S., {et~al.} 2014, \aap, 570, A69

\bibitem[{{Cantiello} {et~al.}(2020){Cantiello}, {Venhola}, {Grado},
  {Paolillo}, {D'Abrusco}, {Raimondo}, {Quintini}, {Hilker}, {Mieske},
  {Tortora}, {Spavone}, {Capaccioli}, {Iodice}, {Peletier}, {Barroso},
  {Limatola}, {Napolitano}, {Schipani}, {van de Ven}, {Gentile}, \&
  {Covone}}]{2020/Cantiello}
{Cantiello}, M., {Venhola}, A., {Grado}, A., {et~al.} 2020, \aap, 639, A136

\bibitem[{{Cappellari} \& {Copin}(2003)}]{2003/Cappellari}
{Cappellari}, M. \& {Copin}, Y. 2003, \mnras, 342, 345

\bibitem[{{Cardamone} {et~al.}(2010){Cardamone}, {van Dokkum}, {Urry},
  {Taniguchi}, {Gawiser}, {Brammer}, {Taylor}, {Damen}, {Treister}, {Cobb},
  {Bond}, {Schawinski}, {Lira}, {Murayama}, {Saito}, \&
  {Sumikawa}}]{2010/Cardamone}
{Cardamone}, C.~N., {van Dokkum}, P.~G., {Urry}, C.~M., {et~al.} 2010, \apjs,
  189, 270

\bibitem[{{Cenarro} {et~al.}(2019){Cenarro}, {Moles},
  {Crist{\'o}bal-Hornillos}, {Mar{\'\i}n-Franch}, {Ederoclite}, {Varela},
  {L{\'o}pez-Sanjuan}, {Hern{\'a}ndez-Monteagudo}, {Angulo}, {V{\'a}zquez
  Rami{\'o}}, {Viironen}, {Bonoli}, {Orsi}, {Hurier}, {San Roman}, {Greisel},
  {Vilella-Rojo}, {D{\'\i}az-Garc{\'\i}a}, {Logro{\~n}o-Garc{\'\i}a},
  {Gurung-L{\'o}pez}, {Spinoso}, {Izquierdo-Villalba}, {Aguerri}, {Allende
  Prieto}, {Bonatto}, {Carvano}, {Chies-Santos}, {Daflon}, {Dupke},
  {Falc{\'o}n-Barroso}, {Gon{\c{c}}alves}, {Jim{\'e}nez-Teja}, {Molino},
  {Placco}, {Solano}, {Whitten}, {Abril}, {Ant{\'o}n}, {Bello}, {Bielsa de
  Toledo}, {Castillo-Ram{\'\i}rez}, {Chueca}, {Civera},
  {D{\'\i}az-Mart{\'\i}n}, {Dom{\'\i}nguez-Mart{\'\i}nez},
  {Garzar{\'a}n-Calderaro}, {Hern{\'a}ndez-Fuertes}, {Iglesias-Marzoa},
  {I{\~n}iguez}, {Jim{\'e}nez Ruiz}, {Kruuse}, {Lamadrid}, {Lasso-Cabrera},
  {L{\'o}pez-Alegre}, {L{\'o}pez-Sainz}, {Ma{\'\i}cas}, {Moreno-Signes},
  {Muniesa}, {Rodr{\'\i}guez-Llano}, {Rueda-Teruel}, {Rueda-Teruel},
  {Soriano-Lagu{\'\i}a}, {Tilve}, {Valdivielso}, {Yanes-D{\'\i}az}, {Alcaniz},
  {Mendes de Oliveira}, {Sodr{\'e}}, {Coelho}, {Lopes de Oliveira}, {Tamm},
  {Xavier}, {Abramo}, {Akras}, {Alfaro}, {Alvarez-Candal}, {Ascaso}, {Beasley},
  {Beers}, {Borges Fernandes}, {Bruzual}, {Buzzo}, {Carrasco}, {Cepa},
  {Cortesi}, {Costa-Duarte}, {De Pr{\'a}}, {Favole}, {Galarza}, {Galbany},
  {Garcia}, {Gonz{\'a}lez Delgado}, {Gonz{\'a}lez-Serrano},
  {Guti{\'e}rrez-Soto}, {Hernandez-Jimenez}, {Kanaan}, {Kuncarayakti},
  {Landim}, {Laur}, {Licandro}, {Lima Neto}, {Lyman}, {Ma{\'\i}z
  Apell{\'a}niz}, {Miralda-Escud{\'e}}, {Morate}, {Nogueira-Cavalcante},
  {Novais}, {Oncins}, {Oteo}, {Overzier}, {Pereira}, {Rebassa-Mansergas},
  {Reis}, {Roig}, {Sako}, {Salvador-Rusi{\~n}ol}, {Sampedro},
  {S{\'a}nchez-Bl{\'a}zquez}, {Santos}, {Schmidtobreick}, {Siffert}, {Telles},
  \& {Vilchez}}]{2019/Cenarro}
{Cenarro}, A.~J., {Moles}, M., {Crist{\'o}bal-Hornillos}, D., {et~al.} 2019,
  \aap, 622, A176

\bibitem[{{Chabrier}(2003)}]{2003/Chabrier}
{Chabrier}, G. 2003, \pasp, 115, 763

\bibitem[{{Chevance} {et~al.}(2020){Chevance}, {Kruijssen}, {Hygate},
  {Schruba}, {Longmore}, {Groves}, {Henshaw}, {Herrera}, {Hughes}, {Jeffreson},
  {Lang}, {Leroy}, {Meidt}, {Pety}, {Razza}, {Rosolowsky}, {Schinnerer},
  {Bigiel}, {Blanc}, {Emsellem}, {Faesi}, {Glover}, {Haydon}, {Ho}, {Kreckel},
  {Lee}, {Liu}, {Querejeta}, {Saito}, {Sun}, {Usero}, \&
  {Utomo}}]{2020/Chevance}
{Chevance}, M., {Kruijssen}, J.~M.~D., {Hygate}, A. P.~S., {et~al.} 2020,
  \mnras, 493, 2872

\bibitem[{{Combes} {et~al.}(1988){Combes}, {Dupraz}, {Casoli}, \&
  {Pagani}}]{1988/Combes}
{Combes}, F., {Dupraz}, C., {Casoli}, F., \& {Pagani}, L. 1988, \aap, 203, L9

\bibitem[{{Conselice}(2003)}]{2003/Conselice}
{Conselice}, C.~J. 2003, \apjs, 147, 1

\bibitem[{{Cook} {et~al.}(2019){Cook}, {Kasliwal}, {Van Sistine}, {Kaplan},
  {Sutter}, {Kupfer}, {Shupe}, {Laher}, {Masci}, {Dale}, {Sesar}, {Brady},
  {Yan}, {Ofek}, {Reitze}, \& {Kulkarni}}]{2019/Cook}
{Cook}, D.~O., {Kasliwal}, M.~M., {Van Sistine}, A., {et~al.} 2019, \apj, 880,
  7

\bibitem[{{De Lucia} {et~al.}(2012){De Lucia}, {Weinmann}, {Poggianti},
  {Arag{\'o}n-Salamanca}, \& {Zaritsky}}]{2012/DeLucia}
{De Lucia}, G., {Weinmann}, S., {Poggianti}, B.~M., {Arag{\'o}n-Salamanca}, A.,
  \& {Zaritsky}, D. 2012, \mnras, 423, 1277

\bibitem[{{Dey} {et~al.}(2019){Dey}, {Schlegel}, {Lang}, {Blum}, {Burleigh},
  {Fan}, {Findlay}, {Finkbeiner}, {Herrera}, {Juneau}, {Landriau}, {Levi},
  {McGreer}, {Meisner}, {Myers}, {Moustakas}, {Nugent}, {Patej}, {Schlafly},
  {Walker}, {Valdes}, {Weaver}, {Y{\`e}che}, {Zou}, {Zhou}, {Abareshi},
  {Abbott}, {Abolfathi}, {Aguilera}, {Alam}, {Allen}, {Alvarez}, {Annis},
  {Ansarinejad}, {Aubert}, {Beechert}, {Bell}, {BenZvi}, {Beutler}, {Bielby},
  {Bolton}, {Brice{\~n}o}, {Buckley-Geer}, {Butler}, {Calamida}, {Carlberg},
  {Carter}, {Casas}, {Castander}, {Choi}, {Comparat}, {Cukanovaite}, {Delubac},
  {DeVries}, {Dey}, {Dhungana}, {Dickinson}, {Ding}, {Donaldson}, {Duan},
  {Duckworth}, {Eftekharzadeh}, {Eisenstein}, {Etourneau}, {Fagrelius},
  {Farihi}, {Fitzpatrick}, {Font-Ribera}, {Fulmer}, {G{\"a}nsicke},
  {Gaztanaga}, {George}, {Gerdes}, {Gontcho}, {Gorgoni}, {Green}, {Guy},
  {Harmer}, {Hernandez}, {Honscheid}, {Huang}, {James}, {Jannuzi}, {Jiang},
  {Joyce}, {Karcher}, {Karkar}, {Kehoe}, {Kneib}, {Kueter-Young}, {Lan},
  {Lauer}, {Le Guillou}, {Le Van Suu}, {Lee}, {Lesser}, {Perreault Levasseur},
  {Li}, {Mann}, {Marshall}, {Mart{\'\i}nez-V{\'a}zquez}, {Martini}, {du Mas des
  Bourboux}, {McManus}, {Meier}, {M{\'e}nard}, {Metcalfe},
  {Mu{\~n}oz-Guti{\'e}rrez}, {Najita}, {Napier}, {Narayan}, {Newman}, {Nie},
  {Nord}, {Norman}, {Olsen}, {Paat}, {Palanque-Delabrouille}, {Peng},
  {Poppett}, {Poremba}, {Prakash}, {Rabinowitz}, {Raichoor}, {Rezaie},
  {Robertson}, {Roe}, {Ross}, {Ross}, {Rudnick}, {Safonova}, {Saha},
  {S{\'a}nchez}, {Savary}, {Schweiker}, {Scott}, {Seo}, {Shan}, {Silva},
  {Slepian}, {Soto}, {Sprayberry}, {Staten}, {Stillman}, {Stupak}, {Summers},
  {Sien Tie}, {Tirado}, {Vargas-Maga{\~n}a}, {Vivas}, {Wechsler}, {Williams},
  {Yang}, {Yang}, {Yapici}, {Zaritsky}, {Zenteno}, {Zhang}, {Zhang}, {Zhou}, \&
  {Zhou}}]{2019/Dey}
{Dey}, A., {Schlegel}, D.~J., {Lang}, D., {et~al.} 2019, \aj, 157, 168

\bibitem[{{Dolag} {et~al.}(2009){Dolag}, {Borgani}, {Murante}, \&
  {Springel}}]{Dolag09}
{Dolag}, K., {Borgani}, S., {Murante}, G., \& {Springel}, V. 2009, \mnras, 399,
  497

\bibitem[{{Donnari} {et~al.}(2020){Donnari}, {Pillepich}, {Nelson},
  {Marinacci}, {Vogelsberger}, \& {Hernquist}}]{Donnari20}
{Donnari}, M., {Pillepich}, A., {Nelson}, D., {et~al.} 2020, arXiv e-prints,
  arXiv:2008.00004

\bibitem[{{Dressler}(1980)}]{1980/Dressler}
{Dressler}, A. 1980, \apj, 236, 351

\bibitem[{{Drinkwater} {et~al.}(2001){Drinkwater}, {Gregg}, \&
  {Colless}}]{2001/Drinkwater}
{Drinkwater}, M.~J., {Gregg}, M.~D., \& {Colless}, M. 2001, \apjl, 548, L139

\bibitem[{{Fahrion} {et~al.}(2020{\natexlab{a}}){Fahrion}, {Lyubenova},
  {Hilker}, {van de Ven}, {Falc{\'o}n-Barroso}, {Leaman},
  {Mart{\'\i}n-Navarro}, {Bittner}, {Coccato}, {Corsini}, {Gadotti}, {Iodice},
  {McDermid}, {Pinna}, {Sarzi}, {Viaene}, {de Zeeuw}, \& {Zhu}}]{2020/Fahrion}
{Fahrion}, K., {Lyubenova}, M., {Hilker}, M., {et~al.} 2020{\natexlab{a}},
  \aap, 637, A26

\bibitem[{{Fahrion} {et~al.}(2020{\natexlab{b}}){Fahrion}, {Lyubenova},
  {Hilker}, {van de Ven}, {Falc{\'o}n-Barroso}, {Leaman},
  {Mart{\'\i}n-Navarro}, {Bittner}, {Coccato}, {Corsini}, {Gadotti}, {Iodice},
  {McDermid}, {Pinna}, {Sarzi}, {Viaene}, {de Zeeuw}, \& {Zhu}}]{2020B/Fahrion}
{Fahrion}, K., {Lyubenova}, M., {Hilker}, M., {et~al.} 2020{\natexlab{b}},
  \aap, 637, A27

\bibitem[{{Fossati} {et~al.}(2019){Fossati}, {Fumagalli}, {Gavazzi},
  {Consolandi}, {Boselli}, {Yagi}, {Sun}, \& {Wilman}}]{2019/Fossati}
{Fossati}, M., {Fumagalli}, M., {Gavazzi}, G., {et~al.} 2019, \mnras, 484, 2212

\bibitem[{{Frank} {et~al.}(2013){Frank}, {Peterson}, {Andersson}, {Fabian}, \&
  {Sanders}}]{2013/Frank}
{Frank}, K.~A., {Peterson}, J.~R., {Andersson}, K., {Fabian}, A.~C., \&
  {Sanders}, J.~S. 2013, \apj, 764, 46

\bibitem[{{Fujita} {et~al.}(2003){Fujita}, {Ajiki}, {Shioya}, {Nagao},
  {Murayama}, {Taniguchi}, {Okamura}, {Ouchi}, {Shimasaku}, {Doi}, {Furusawa},
  {Hamabe}, {Kimura}, {Komiyama}, {Miyazaki}, {Miyazaki}, {Nakata},
  {Sekiguchi}, {Yagi}, {Yasuda}, {Matsuda}, {Tamura}, {Hayashino}, {Kodaira},
  {Karoji}, {Yamada}, {Ohta}, \& {Umemura}}]{2003/Fujita}
{Fujita}, S.~S., {Ajiki}, M., {Shioya}, Y., {et~al.} 2003, \aj, 125, 13

\bibitem[{{Geach} {et~al.}(2008){Geach}, {Smail}, {Best}, {Kurk}, {Casali},
  {Ivison}, \& {Coppin}}]{2008/Geach}
{Geach}, J.~E., {Smail}, I., {Best}, P.~N., {et~al.} 2008, \mnras, 388, 1473

\bibitem[{{Gill} {et~al.}(2005){Gill}, {Knebe}, \& {Gibson}}]{2005/Gill}
{Gill}, S. P.~D., {Knebe}, A., \& {Gibson}, B.~K. 2005, \mnras, 356, 1327

\bibitem[{{Giovanelli} \& {Haynes}(1983)}]{1983/Giovanelli}
{Giovanelli}, R. \& {Haynes}, M.~P. 1983, \aj, 88, 881

\bibitem[{{Glazebrook} {et~al.}(2004){Glazebrook}, {Tober}, {Thomson},
  {Bland-Hawthorn}, \& {Abraham}}]{2004/Glazebrook}
{Glazebrook}, K., {Tober}, J., {Thomson}, S., {Bland-Hawthorn}, J., \&
  {Abraham}, R. 2004, \aj, 128, 2652

\bibitem[{{Gunn} \& {Gott}(1972)}]{1972/Gunn}
{Gunn}, J.~E. \& {Gott}, J.~Richard, I. 1972, \apj, 176, 1

\bibitem[{{Haack} {et~al.}(2024){Haack}, {Smith Castelli}, {Mendes de
  Oliveira}, {Almeida-Fernandes}, {Faifer}, {Lopes}, {Jaffe}, {Demarco},
  {Lima-Dias}, {Lomel{\'\i}-Nu{\~n}ez}, {Montaguth}, {Schoenell}, {Ribeiro}, \&
  {Kanaan}}]{2024/Haack}
{Haack}, R.~F., {Smith Castelli}, A.~V., {Mendes de Oliveira}, C., {et~al.}
  2024, \mnras, 530, 3195

\bibitem[{{Haines} {et~al.}(2015){Haines}, {Pereira}, {Smith}, {Egami},
  {Babul}, {Finoguenov}, {Ziparo}, {McGee}, {Rawle}, {Okabe}, \&
  {Moran}}]{2015/Haines}
{Haines}, C.~P., {Pereira}, M.~J., {Smith}, G.~P., {et~al.} 2015, \apj, 806,
  101

\bibitem[{{Haines} {et~al.}(2013){Haines}, {Pereira}, {Smith}, {Egami},
  {Sanderson}, {Babul}, {Finoguenov}, {Merluzzi}, {Busarello}, {Rawle}, \&
  {Okabe}}]{2013/Haines}
{Haines}, C.~P., {Pereira}, M.~J., {Smith}, G.~P., {et~al.} 2013, \apj, 775,
  126

\bibitem[{{Hern{\'a}ndez-Fern{\'a}ndez}
  {et~al.}(2014){Hern{\'a}ndez-Fern{\'a}ndez}, {Haines}, {Diaferio},
  {Iglesias-P{\'a}ramo}, {Mendes de Oliveira}, \&
  {Vilchez}}]{2014/Hernandez-Fernandez}
{Hern{\'a}ndez-Fern{\'a}ndez}, J.~D., {Haines}, C.~P., {Diaferio}, A., {et~al.}
  2014, \mnras, 438, 2186

\bibitem[{{Hernandez-Jimenez} \& {Krabbe}(2022)}]{2022/Hernandez}
{Hernandez-Jimenez}, J.~A. \& {Krabbe}, A.~C. 2022, {Astromorphlib: Python
  scripts to analyze the morphology of isolated and interacting galaxies}

\bibitem[{{Herpich} {et~al.}(2024){Herpich}, {Almeida-Fernandes}, {Oliveira
  Schwarz}, {Lima}, {Nakazono}, {Alonso-Garc{\'\i}a}, {Fonseca-Faria},
  {Sartori}, {Bolutavicius}, {Fabiano de Souza}, {Hartmann}, {Li}, {Espinosa},
  {Kanaan}, {Schoenell}, {Werle}, {Machado-Pereira}, {Guti{\'e}rrez-Soto},
  {Santos-Silva}, {Smith Castelli}, {Lacerda}, {Barbosa}, {Perottoni},
  {Ferreira Lopes}, {Valen{\c{c}}a}, {Re Martho}, {Bom}, {Bonatto}, {Carvalho},
  {Cernic}, {Cid Fernandes}, {Coelho}, {Cortesi}, {Cubillos Palma}, {Doubrawa},
  {Ferreira Alberice}, {Quispe-Huaynasi}, {Jacob Perin}, {Jaque Arancibia},
  {Krabbe}, {Lima-Dias}, {Lomel{\'\i}-N{\'u}{\~n}ez}, {Lopes de Oliveira},
  {Lopes}, {Luiz Figueiredo}, {L{\"o}sch}, {Navarete}, {Oliveira}, {Overzier},
  {Placco}, {Roig}, {Rubet}, {Santos}, {Sasse}, {Thain{\'a}-Batista},
  {Torres-Flores}, {Beers}, {Alvarez-Candal}, {Akras}, {Panda}, {Limberg},
  {Nilo Castell{\'o}n}, {Telles}, {Lopes}, {Pardo Montaguth}, {Beraldo e
  Silva}, {Humire}, {Borges Fernandes}, {Cordeiro}, {Ribeiro}, \& {Mendes de
  Oliveira}}]{2024/Herpich}
{Herpich}, F.~R., {Almeida-Fernandes}, F., {Oliveira Schwarz}, G.~B., {et~al.}
  2024, \aap, 689, A249

\bibitem[{{Hirschmann} {et~al.}(2014){Hirschmann}, {De Lucia}, {Wilman},
  {Weinmann}, {Iovino}, {Cucciati}, {Zibetti}, \&
  {Villalobos}}]{2014/Hirschmann}
{Hirschmann}, M., {De Lucia}, G., {Wilman}, D., {et~al.} 2014, \mnras, 444,
  2938

\bibitem[{{Horellou} {et~al.}(1995){Horellou}, {Casoli}, \&
  {Dupraz}}]{1995/Horellou}
{Horellou}, C., {Casoli}, F., \& {Dupraz}, C. 1995, \aap, 303, 361

\bibitem[{{Hou} {et~al.}(2024){Hou}, {Hu}, \& {Li}}]{2024/Hou}
{Hou}, M., {Hu}, Z., \& {Li}, Z. 2024, \apjl, 965, L24

\bibitem[{{Iodice} {et~al.}(2016){Iodice}, {Capaccioli}, {Grado}, {Limatola},
  {Spavone}, {Napolitano}, {Paolillo}, {Peletier}, {Cantiello}, {Lisker},
  {Wittmann}, {Venhola}, {Hilker}, {D'Abrusco}, {Pota}, \&
  {Schipani}}]{2016/Iodice}
{Iodice}, E., {Capaccioli}, M., {Grado}, A., {et~al.} 2016, \apj, 820, 42

\bibitem[{{Iodice} {et~al.}(2019{\natexlab{a}}){Iodice}, {Sarzi}, {Bittner},
  {Coccato}, {Costantin}, {Corsini}, {van de Ven}, {de Zeeuw},
  {Falc{\'o}n-Barroso}, {Gadotti}, {Lyubenova}, {Mart{\'\i}n-Navarro},
  {McDermid}, {Nedelchev}, {Pinna}, {Pizzella}, {Spavone}, \&
  {Viaene}}]{2019/Iodice}
{Iodice}, E., {Sarzi}, M., {Bittner}, A., {et~al.} 2019{\natexlab{a}}, \aap,
  627, A136

\bibitem[{{Iodice} {et~al.}(2017){Iodice}, {Spavone}, {Capaccioli}, {Peletier},
  {Richtler}, {Hilker}, {Mieske}, {Limatola}, {Grado}, {Napolitano},
  {Cantiello}, {D'Abrusco}, {Paolillo}, {Venhola}, {Lisker}, {Van de Ven},
  {Falcon-Barroso}, \& {Schipani}}]{2017/Iodice}
{Iodice}, E., {Spavone}, M., {Capaccioli}, M., {et~al.} 2017, \apj, 839, 21

\bibitem[{{Iodice} {et~al.}(2019{\natexlab{b}}){Iodice}, {Spavone},
  {Capaccioli}, {Peletier}, {van de Ven}, {Napolitano}, {Hilker}, {Mieske},
  {Smith}, {Pasquali}, {Limatola}, {Grado}, {Venhola}, {Cantiello}, {Paolillo},
  {Falcon-Barroso}, {D'Abrusco}, \& {Schipani}}]{2019a/Iodice}
{Iodice}, E., {Spavone}, M., {Capaccioli}, M., {et~al.} 2019{\natexlab{b}},
  \aap, 623, A1

\bibitem[{{J{\'a}chym} {et~al.}(2017){J{\'a}chym}, {Sun}, {Kenney}, {Cortese},
  {Combes}, {Yagi}, {Yoshida}, {Palou{\v{s}}}, \& {Roediger}}]{2017/Jachym}
{J{\'a}chym}, P., {Sun}, M., {Kenney}, J. D.~P., {et~al.} 2017, \apj, 839, 114

\bibitem[{{James} {et~al.}(2005){James}, {Shane}, {Knapen}, {Etherton}, \&
  {Percival}}]{2005/James}
{James}, P.~A., {Shane}, N.~S., {Knapen}, J.~H., {Etherton}, J., \& {Percival},
  S.~M. 2005, \aap, 429, 851

\bibitem[{{Jones} {et~al.}(1997){Jones}, {Stern}, {Forman}, {Breen}, {David},
  {Tucker}, \& {Franx}}]{1997/Jones}
{Jones}, C., {Stern}, C., {Forman}, W., {et~al.} 1997, \apj, 482, 143

\bibitem[{{Kauffmann} {et~al.}(2003){Kauffmann}, {Heckman}, {White}, {Charlot},
  {Tremonti}, {Brinchmann}, {Bruzual}, {Peng}, {Seibert}, {Bernardi},
  {Blanton}, {Brinkmann}, {Castander}, {Cs{\'a}bai}, {Fukugita}, {Ivezic},
  {Munn}, {Nichol}, {Padmanabhan}, {Thakar}, {Weinberg}, \&
  {York}}]{2003/Kauffmann}
{Kauffmann}, G., {Heckman}, T.~M., {White}, S. D.~M., {et~al.} 2003, \mnras,
  341, 33

\bibitem[{{Kellar} {et~al.}(2012){Kellar}, {Salzer}, {Wegner}, {Gronwall}, \&
  {Williams}}]{Kellar2012}
{Kellar}, J.~A., {Salzer}, J.~J., {Wegner}, G., {Gronwall}, C., \& {Williams},
  A. 2012, \aj, 143, 145

\bibitem[{{Kenney} {et~al.}(1995){Kenney}, {Rubin}, {Planesas}, \&
  {Young}}]{1995/Kenney}
{Kenney}, J. D.~P., {Rubin}, V.~C., {Planesas}, P., \& {Young}, J.~S. 1995,
  \apj, 438, 135

\bibitem[{{Khostovan} {et~al.}(2020){Khostovan}, {Malhotra}, {Rhoads}, {Jiang},
  {Wang}, {Wold}, {Zheng}, {Barrientos}, {Coughlin}, {Harish}, {Hu}, {Infante},
  {Perez}, {Pharo}, {Valdes}, {Walker}, \& {Yang}}]{2020/Khostovan}
{Khostovan}, A.~A., {Malhotra}, S., {Rhoads}, J.~E., {et~al.} 2020, \mnras,
  493, 3966

\bibitem[{{Kleiner} {et~al.}(2021){Kleiner}, {Serra}, {Maccagni}, {Venhola},
  {Morokuma-Matsui}, {Peletier}, {Iodice}, {Raj}, {de Blok}, {Comrie},
  {J{\'o}zsa}, {Kamphuis}, {Loni}, {Loubser}, {Moln{\'a}r}, {Passmoor},
  {Ramatsoku}, {Sivitilli}, {Smirnov}, {Thorat}, \& {Vitello}}]{2021/Kleiner}
{Kleiner}, D., {Serra}, P., {Maccagni}, F.~M., {et~al.} 2021, \aap, 648, A32

\bibitem[{{Krabbe} {et~al.}(2024){Krabbe}, {Hernandez-Jimenez}, {Mendes de
  Oliveira}, {Jaffe}, {Oliveira}, {Cardoso}, {Smith Castelli}, {Dors},
  {Cortesi}, \& {Crossett}}]{2024/Krabbe}
{Krabbe}, A.~C., {Hernandez-Jimenez}, J.~A., {Mendes de Oliveira}, C., {et~al.}
  2024, \mnras, 528, 1125

\bibitem[{{Kreckel} {et~al.}(2016){Kreckel}, {Blanc}, {Schinnerer}, {Groves},
  {Adamo}, {Hughes}, \& {Meidt}}]{2016/Kreckel}
{Kreckel}, K., {Blanc}, G.~A., {Schinnerer}, E., {et~al.} 2016, \apj, 827, 103

\bibitem[{{Kron}(1980)}]{1980/Kron}
{Kron}, R.~G. 1980, \apjs, 43, 305

\bibitem[{{Lara-L{\'o}pez} {et~al.}(2022){Lara-L{\'o}pez}, {Gal{\'a}n-de Anta},
  {Sarzi}, {Iodice}, {Davis}, {Zabel}, {Corsini}, {de Zeeuw}, {Fahrion},
  {Falc{\'o}n-Barroso}, {Gadotti}, {McDermid}, {Pinna}, {Rodriguez-Gomez}, {van
  de Ven}, {Zhu}, {Coccato}, {Lyubenova}, \& {Mart{\'\i}n-Navarro}}]{2022/Lara}
{Lara-L{\'o}pez}, M.~A., {Gal{\'a}n-de Anta}, P.~M., {Sarzi}, M., {et~al.}
  2022, \aap, 660, A105

\bibitem[{{Loni} {et~al.}(2021){Loni}, {Serra}, {Kleiner}, {Cortese},
  {Catinella}, {Koribalski}, {Jarrett}, {Molnar}, {Davis}, {Iodice},
  {Lee-Waddell}, {Loi}, {Maccagni}, {Peletier}, {Popping}, {Ramatsoku},
  {Smith}, \& {Zabel}}]{2021/Loni}
{Loni}, A., {Serra}, P., {Kleiner}, D., {et~al.} 2021, \aap, 648, A31

\bibitem[{{Lopes} {et~al.}(2024){Lopes}, {Ribeiro}, \& {Brambila}}]{2024/Lopes}
{Lopes}, P. A.~A., {Ribeiro}, A. L.~B., \& {Brambila}, D. 2024, \mnras, 527,
  L19

\bibitem[{{Lotz} {et~al.}(2008){Lotz}, {Davis}, {Faber}, {Guhathakurta},
  {Gwyn}, {Huang}, {Koo}, {Le Floc'h}, {Lin}, {Newman}, {Noeske}, {Papovich},
  {Willmer}, {Coil}, {Conselice}, {Cooper}, {Hopkins}, {Metevier}, {Primack},
  {Rieke}, \& {Weiner}}]{2008/Lotz}
{Lotz}, J.~M., {Davis}, M., {Faber}, S.~M., {et~al.} 2008, \apj, 672, 177

\bibitem[{{Lotz} {et~al.}(2004){Lotz}, {Primack}, \& {Madau}}]{2004/Lotz}
{Lotz}, J.~M., {Primack}, J., \& {Madau}, P. 2004, \aj, 128, 163

\bibitem[{{Loubser} {et~al.}(2024){Loubser}, {Mosia}, {Serra}, {Kleiner},
  {Peletier}, {Kraan-Korteweg}, {Iodice}, {Loni}, {Kamphuis}, \&
  {Zabel}}]{2024/Lousber}
{Loubser}, S.~I., {Mosia}, K., {Serra}, P., {et~al.} 2024, \mnras, 527, 7158

\bibitem[{{Ly} {et~al.}(2011){Ly}, {Lee}, {Dale}, {Momcheva}, {Salim},
  {Staudaher}, {Moore}, \& {Finn}}]{2011/Ly}
{Ly}, C., {Lee}, J.~C., {Dale}, D.~A., {et~al.} 2011, \apj, 726, 109

\bibitem[{{Maccagni} {et~al.}(2020){Maccagni}, {Murgia}, {Serra}, {Govoni},
  {Morokuma-Matsui}, {Kleiner}, {Buchner}, {J{\'o}zsa}, {Kamphuis},
  {Makhathini}, {Moln{\'a}r}, {Prokhorov}, {Ramaila}, {Ramatsoku}, {Thorat}, \&
  {Smirnov}}]{2020/Maccagni}
{Maccagni}, F.~M., {Murgia}, M., {Serra}, P., {et~al.} 2020, \aap, 634, A9

\bibitem[{{Maddox} {et~al.}(2019){Maddox}, {Serra}, {Venhola}, {Peletier},
  {Loubser}, \& {Iodice}}]{2019/Maddox}
{Maddox}, N., {Serra}, P., {Venhola}, A., {et~al.} 2019, \mnras, 490, 1666

\bibitem[{{Mahajan} {et~al.}(2011){Mahajan}, {Mamon}, \&
  {Raychaudhury}}]{2011/Mahajan}
{Mahajan}, S., {Mamon}, G.~A., \& {Raychaudhury}, S. 2011, \mnras, 416, 2882

\bibitem[{{Makarov} {et~al.}(2014){Makarov}, {Prugniel}, {Terekhova},
  {Courtois}, \& {Vauglin}}]{2014/Makarov}
{Makarov}, D., {Prugniel}, P., {Terekhova}, N., {Courtois}, H., \& {Vauglin},
  I. 2014, \aap, 570, A13

\bibitem[{{Marinacci} {et~al.}(2018){Marinacci}, {Vogelsberger}, {Pakmor},
  {Torrey}, {Springel}, {Hernquist}, {Nelson}, {Weinberger}, {Pillepich},
  {Naiman}, \& {Genel}}]{Marinacci18}
{Marinacci}, F., {Vogelsberger}, M., {Pakmor}, R., {et~al.} 2018, \mnras, 480,
  5113

\bibitem[{{McGee} {et~al.}(2009){McGee}, {Balogh}, {Bower}, {Font}, \&
  {McCarthy}}]{2009/McGee}
{McGee}, S.~L., {Balogh}, M.~L., {Bower}, R.~G., {Font}, A.~S., \& {McCarthy},
  I.~G. 2009, \mnras, 400, 937

\bibitem[{{McNab} {et~al.}(2021){McNab}, {Balogh}, {van der Burg}, {Forestell},
  {Webb}, {Vulcani}, {Rudnick}, {Muzzin}, {Cooper}, {McGee}, {Biviano},
  {Cerulo}, {Chan}, {De Lucia}, {Demarco}, {Finoguenov}, {Forrest}, {Golledge},
  {Jablonka}, {Lidman}, {Nantais}, {Old}, {Pintos-Castro}, {Poggianti},
  {Reeves}, {Wilson}, {Yee}, \& {Zaritsky}}]{2021/McNab}
{McNab}, K., {Balogh}, M.~L., {van der Burg}, R. F.~J., {et~al.} 2021, \mnras,
  508, 157

\bibitem[{{Mendes de Oliveira} {et~al.}(2019){Mendes de Oliveira}, {Ribeiro},
  {Schoenell}, {Kanaan}, {Overzier}, {Molino}, {Sampedro}, {Coelho}, {Barbosa},
  {Cortesi}, {Costa-Duarte}, {Herpich}, {Hernandez-Jimenez}, {Placco},
  {Xavier}, {Abramo}, {Saito}, {Chies-Santos}, {Ederoclite}, {Lopes de
  Oliveira}, {Gon{\c{c}}alves}, {Akras}, {Almeida}, {Almeida-Fernandes},
  {Beers}, {Bonatto}, {Bonoli}, {Cypriano}, {Vinicius-Lima}, {de Souza},
  {Fabiano de Souza}, {Ferrari}, {Gon{\c{c}}alves}, {Gonzalez},
  {Guti{\'e}rrez-Soto}, {Hartmann}, {Jaffe}, {Kerber}, {Lima-Dias}, {Lopes},
  {Menendez-Delmestre}, {Nakazono}, {Novais}, {Ortega-Minakata}, {Pereira},
  {Perottoni}, {Queiroz}, {Reis}, {Santos}, {Santos-Silva}, {Santucci},
  {Barbosa}, {Siffert}, {Sodr{\'e}}, {Torres-Flores}, {Westera}, {Whitten},
  {Alcaniz}, {Alonso-Garc{\'\i}a}, {Alencar}, {Alvarez-Candal}, {Amram},
  {Azanha}, {Barb{\'a}}, {Bernardinelli}, {Borges Fernandes}, {Branco},
  {Brito-Silva}, {Buzzo}, {Caffer}, {Campillay}, {Cano}, {Carvano}, {Castejon},
  {Cid Fernandes}, {Dantas}, {Daflon}, {Damke}, {de la Reza}, {de Melo de
  Azevedo}, {De Paula}, {Diem}, {Donnerstein}, {Dors}, {Dupke}, {Eikenberry},
  {Escudero}, {Faifer}, {Far{\'\i}as}, {Fernandes}, {Fernandes}, {Fontes},
  {Galarza}, {Hirata}, {Katena}, {Gregorio-Hetem},
  {Hern{\'a}ndez-Fern{\'a}ndez}, {Izzo}, {Jaque Arancibia}, {Jatenco-Pereira},
  {Jim{\'e}nez-Teja}, {Kann}, {Krabbe}, {Labayru}, {Lazzaro}, {Lima Neto},
  {Lopes}, {Magalh{\~a}es}, {Makler}, {de Menezes}, {Miralda-Escud{\'e}},
  {Monteiro-Oliveira}, {Montero-Dorta}, {Mu{\~n}oz-Elgueta}, {Nemmen}, {Nilo
  Castell{\'o}n}, {Oliveira}, {Ort{\'\i}z}, {Pattaro}, {Pereira}, {Quint},
  {Riguccini}, {Rocha Pinto}, {Rodrigues}, {Roig}, {Rossi}, {Saha}, {Santos},
  {Schnorr M{\"u}ller}, {Sesto}, {Silva}, {Smith Castelli}, {Teixeira},
  {Telles}, {Thom de Souza}, {Th{\"o}ne}, {Trevisan}, {de Ugarte Postigo},
  {Urrutia-Viscarra}, {Veiga}, {Vika}, {Vitorelli}, {Werle}, {Werner}, \&
  {Zaritsky}}]{2019/Mendes}
{Mendes de Oliveira}, C., {Ribeiro}, T., {Schoenell}, W., {et~al.} 2019,
  \mnras, 489, 241

\bibitem[{{Menezes} {et~al.}(2014){Menezes}, {Steiner}, \&
  {Ricci}}]{2014/Menezes}
{Menezes}, R.~B., {Steiner}, J.~E., \& {Ricci}, T.~V. 2014, \mnras, 438, 2597

\bibitem[{{Moles} {et~al.}(2008){Moles}, {Ben{\'\i}tez}, {Aguerri}, {Alfaro},
  {Broadhurst}, {Cabrera-Ca{\~n}o}, {Castander}, {Cepa}, {Cervi{\~n}o},
  {Crist{\'o}bal-Hornillos}, {Fern{\'a}ndez-Soto}, {Gonz{\'a}lez Delgado},
  {Infante}, {M{\'a}rquez}, {Mart{\'\i}nez}, {Masegosa}, {del Olmo}, {Perea},
  {Prada}, {Quintana}, \& {S{\'a}nchez}}]{2008/Moles}
{Moles}, M., {Ben{\'\i}tez}, N., {Aguerri}, J.~A.~L., {et~al.} 2008, \aj, 136,
  1325

\bibitem[{{Morokuma-Matsui} {et~al.}(2022){Morokuma-Matsui}, {Bekki}, {Wang},
  {Serra}, {Koyama}, {Morokuma}, {Egusa}, {For}, {Nakanishi}, {Koribalski},
  {Okamoto}, {Kodama}, {Lee}, {Maccagni}, {Miura}, {Espada}, {Takeuchi},
  {Yang}, {Lee}, {Ueda}, \& {Matsushita}}]{2022/Morokuma-Matsui}
{Morokuma-Matsui}, K., {Bekki}, K., {Wang}, J., {et~al.} 2022, \apjs, 263, 40

\bibitem[{{Morokuma-Matsui} {et~al.}(2019){Morokuma-Matsui}, {Serra},
  {Maccagni}, {For}, {Wang}, {Bekki}, {Morokuma}, {Egusa}, {Espada}, {Miura},
  {Nakanishi}, {Koribalski}, \& {Takeuchi}}]{2019/Morokuma-Matsui}
{Morokuma-Matsui}, K., {Serra}, P., {Maccagni}, F.~M., {et~al.} 2019, \pasj,
  71, 85

\bibitem[{{Muzzin} {et~al.}(2014){Muzzin}, {van der Burg}, {McGee}, {Balogh},
  {Franx}, {Hoekstra}, {Hudson}, {Noble}, {Taranu}, {Webb}, {Wilson}, \&
  {Yee}}]{2014/Muzzin}
{Muzzin}, A., {van der Burg}, R.~F.~J., {McGee}, S.~L., {et~al.} 2014, \apj,
  796, 65

\bibitem[{{Nasonova} {et~al.}(2011){Nasonova}, {de Freitas Pacheco}, \&
  {Karachentsev}}]{2011/Nasonova}
{Nasonova}, O.~G., {de Freitas Pacheco}, J.~A., \& {Karachentsev}, I.~D. 2011,
  \aap, 532, A104

\bibitem[{{Nelson} {et~al.}(2019){Nelson}, {Pillepich}, {Springel}, {Pakmor},
  {Weinberger}, {Genel}, {Torrey}, {Vogelsberger}, {Marinacci}, \&
  {Hernquist}}]{Nelson19}
{Nelson}, D., {Pillepich}, A., {Springel}, V., {et~al.} 2019, \mnras, 490, 3234

\bibitem[{{Nelson} {et~al.}(2018){Nelson}, {Pillepich}, {Springel},
  {Weinberger}, {Hernquist}, {Pakmor}, {Genel}, {Torrey}, {Vogelsberger},
  {Kauffmann}, {Marinacci}, \& {Naiman}}]{Nelson18}
{Nelson}, D., {Pillepich}, A., {Springel}, V., {et~al.} 2018, \mnras, 475, 624

\bibitem[{{Oman} \& {Hudson}(2016)}]{2016/Oman}
{Oman}, K.~A. \& {Hudson}, M.~J. 2016, \mnras, 463, 3083

\bibitem[{{Pallero} {et~al.}(2022){Pallero}, {G{\'o}mez}, {Padilla},
  {Bah{\'e}}, {Vega-Mart{\'\i}nez}, \& {Torres-Flores}}]{Pallero22}
{Pallero}, D., {G{\'o}mez}, F.~A., {Padilla}, N.~D., {et~al.} 2022, \mnras,
  511, 3210

\bibitem[{{Pallero} {et~al.}(2019){Pallero}, {G{\'o}mez}, {Padilla},
  {Torres-Flores}, {Demarco}, {Cerulo}, \& {Olave-Rojas}}]{Pallero19}
{Pallero}, D., {G{\'o}mez}, F.~A., {Padilla}, N.~D., {et~al.} 2019, \mnras,
  488, 847

\bibitem[{{Pascual} {et~al.}(2007){Pascual}, {Gallego}, \&
  {Zamorano}}]{2007/Pascual}
{Pascual}, S., {Gallego}, J., \& {Zamorano}, J. 2007, \pasp, 119, 30

\bibitem[{{Pasquali} {et~al.}(2019){Pasquali}, {Smith}, {Gallazzi}, {De Lucia},
  {Zibetti}, {Hirschmann}, \& {Yi}}]{2019/Pasquali}
{Pasquali}, A., {Smith}, R., {Gallazzi}, A., {et~al.} 2019, \mnras, 484, 1702

\bibitem[{{Pedrini} {et~al.}(2022){Pedrini}, {Fossati}, {Gavazzi}, {Fumagalli},
  {Boselli}, {Consolandi}, {Sun}, {Yagi}, \& {Yoshida}}]{2022/Pedrini}
{Pedrini}, A., {Fossati}, M., {Gavazzi}, G., {et~al.} 2022, \mnras, 511, 5180

\bibitem[{{P{\'e}rez-Gonz{\'a}lez} {et~al.}(2013){P{\'e}rez-Gonz{\'a}lez},
  {Cava}, {Barro}, {Villar}, {Cardiel}, {Ferreras}, {Rodr{\'\i}guez-Espinosa},
  {Alonso-Herrero}, {Balcells}, {Cenarro}, {Cepa}, {Charlot}, {Cimatti},
  {Conselice}, {Daddi}, {Donley}, {Elbaz}, {Espino}, {Gallego}, {Gobat},
  {Gonz{\'a}lez-Mart{\'\i}n}, {Guzm{\'a}n}, {Hern{\'a}n-Caballero},
  {Mu{\~n}oz-Tu{\~n}{\'o}n}, {Renzini}, {Rodr{\'\i}guez-Zaur{\'\i}n}, {Tresse},
  {Trujillo}, \& {Zamorano}}]{2013/PerezGonzalez}
{P{\'e}rez-Gonz{\'a}lez}, P.~G., {Cava}, A., {Barro}, G., {et~al.} 2013, \apj,
  762, 46

\bibitem[{{Pillepich} {et~al.}(2018){Pillepich}, {Nelson}, {Hernquist},
  {Springel}, {Pakmor}, {Torrey}, {Weinberger}, {Genel}, {Naiman}, {Marinacci},
  \& {Vogelsberger}}]{Pillepich18}
{Pillepich}, A., {Nelson}, D., {Hernquist}, L., {et~al.} 2018, \mnras, 475, 648

\bibitem[{{Pillepich} {et~al.}(2019){Pillepich}, {Nelson}, {Springel},
  {Pakmor}, {Torrey}, {Weinberger}, {Vogelsberger}, {Marinacci}, {Genel}, {van
  der Wel}, \& {Hernquist}}]{Pillepich19}
{Pillepich}, A., {Nelson}, D., {Springel}, V., {et~al.} 2019, \mnras, 490, 3196

\bibitem[{{Pinna} {et~al.}(2019{\natexlab{a}}){Pinna}, {Falc{\'o}n-Barroso},
  {Martig}, {Coccato}, {Corsini}, {de Zeeuw}, {Gadotti}, {Iodice}, {Leaman},
  {Lyubenova}, {Mart{\'\i}n-Navarro}, {Morelli}, {Sarzi}, {van de Ven},
  {Viaene}, \& {McDermid}}]{2019B/Pinna}
{Pinna}, F., {Falc{\'o}n-Barroso}, J., {Martig}, M., {et~al.}
  2019{\natexlab{a}}, \aap, 625, A95

\bibitem[{{Pinna} {et~al.}(2019{\natexlab{b}}){Pinna}, {Falc{\'o}n-Barroso},
  {Martig}, {Sarzi}, {Coccato}, {Iodice}, {Corsini}, {de Zeeuw}, {Gadotti},
  {Leaman}, {Lyubenova}, {McDermid}, {Minchev}, {Morelli}, {van de Ven}, \&
  {Viaene}}]{2019/Pinna}
{Pinna}, F., {Falc{\'o}n-Barroso}, J., {Martig}, M., {et~al.}
  2019{\natexlab{b}}, \aap, 623, A19

\bibitem[{{Poci} {et~al.}(2021){Poci}, {McDermid}, {Lyubenova}, {Zhu}, {van de
  Ven}, {Iodice}, {Coccato}, {Pinna}, {Corsini}, {Falc{\'o}n-Barroso},
  {Gadotti}, {Grand}, {Fahrion}, {Mart{\'\i}n-Navarro}, {Sarzi}, {Viaene}, \&
  {de Zeeuw}}]{2021/Poci}
{Poci}, A., {McDermid}, R.~M., {Lyubenova}, M., {et~al.} 2021, \aap, 647, A145

\bibitem[{{Raj} {et~al.}(2020){Raj}, {Iodice}, {Napolitano}, {Hilker},
  {Spavone}, {Peletier}, {Su}, {Falc{\'o}n-Barroso}, {van de Ven}, {Cantiello},
  {Kleiner}, {Venhola}, {Mieske}, {Paolillo}, {Capaccioli}, \&
  {Schipani}}]{2020/Raj}
{Raj}, M.~A., {Iodice}, E., {Napolitano}, N.~R., {et~al.} 2020, \aap, 640, A137

\bibitem[{{Raj} {et~al.}(2019){Raj}, {Iodice}, {Napolitano}, {Spavone}, {Su},
  {Peletier}, {Davis}, {Zabel}, {Hilker}, {Mieske}, {Falcon Barroso},
  {Cantiello}, {van de Ven}, {Watkins}, {Salo}, {Schipani}, {Capaccioli}, \&
  {Venhola}}]{2019/Raj}
{Raj}, M.~A., {Iodice}, E., {Napolitano}, N.~R., {et~al.} 2019, \aap, 628, A4

\bibitem[{{Rhee} {et~al.}(2017){Rhee}, {Smith}, {Choi}, {Yi}, {Jaff{\'e}},
  {Candlish}, \& {S{\'a}nchez-J{\'a}nssen}}]{2017/Rhee}
{Rhee}, J., {Smith}, R., {Choi}, H., {et~al.} 2017, \apj, 843, 128

\bibitem[{{Rodriguez-Gomez} {et~al.}(2019){Rodriguez-Gomez}, {Snyder}, {Lotz},
  {Nelson}, {Pillepich}, {Springel}, {Genel}, {Weinberger}, {Tacchella},
  {Pakmor}, {Torrey}, {Marinacci}, {Vogelsberger}, {Hernquist}, \&
  {Thilker}}]{2019/RodriguezGomez}
{Rodriguez-Gomez}, V., {Snyder}, G.~F., {Lotz}, J.~M., {et~al.} 2019, \mnras,
  483, 4140

\bibitem[{{Romero-G{\'o}mez} {et~al.}(2024){Romero-G{\'o}mez}, {Peletier},
  {Aguerri}, \& {Smith}}]{2024/Romero}
{Romero-G{\'o}mez}, J., {Peletier}, R.~F., {Aguerri}, J.~A.~L., \& {Smith}, R.
  2024, \aap, 689, A40

\bibitem[{{Rosa} {et~al.}(2018){Rosa}, {Milone}, {Krabbe}, \&
  {Rodrigues}}]{2018Ap&SS.363..131R}
{Rosa}, D.~A., {Milone}, A.~C., {Krabbe}, A.~C., \& {Rodrigues}, I. 2018,
  \apss, 363, 131

\bibitem[{{Salzer} {et~al.}(2023){Salzer}, {Carr}, {Sieben}, {Brunker}, \&
  {Hirschauer}}]{2023/Salzer}
{Salzer}, J.~J., {Carr}, D.~J., {Sieben}, J., {Brunker}, S.~W., \&
  {Hirschauer}, A.~S. 2023, \aj, 166, 81

\bibitem[{{Sampaio} {et~al.}(2024){Sampaio}, {de Carvalho},
  {Arag{\'o}n-Salamanca}, {Merrifield}, {Ferreras}, \&
  {Cornwell}}]{2024/Sampaio}
{Sampaio}, V.~M., {de Carvalho}, R.~R., {Arag{\'o}n-Salamanca}, A., {et~al.}
  2024, \mnras, 532, 982

\bibitem[{{Sarzi} {et~al.}(2018){Sarzi}, {Iodice}, {Coccato}, {Corsini}, {de
  Zeeuw}, {Falc{\'o}n-Barroso}, {Gadotti}, {Lyubenova}, {McDermid}, {van de
  Ven}, {Fahrion}, {Pizzella}, \& {Zhu}}]{2018/Sarzi}
{Sarzi}, M., {Iodice}, E., {Coccato}, L., {et~al.} 2018, \aap, 616, A121

\bibitem[{{Scharf} {et~al.}(2005){Scharf}, {Zurek}, \& {Bureau}}]{2005/Scharf}
{Scharf}, C.~A., {Zurek}, D.~R., \& {Bureau}, M. 2005, \apj, 633, 154

\bibitem[{{Serra} {et~al.}(2019){Serra}, {Maccagni}, {Kleiner}, {de Blok}, {van
  Gorkom}, {Hugo}, {Iodice}, {J{\'o}zsa}, {Kamphuis}, {Kraan-Korteweg}, {Loni},
  {Makhathini}, {Moln{\'a}r}, {Oosterloo}, {Peletier}, {Ramaila}, {Ramatsoku},
  {Smirnov}, {Smith}, {Spavone}, {Thorat}, {Trager}, \& {Venhola}}]{2019/Serra}
{Serra}, P., {Maccagni}, F.~M., {Kleiner}, D., {et~al.} 2019, \aap, 628, A122

\bibitem[{{Smith Castelli} {et~al.}(2024){Smith Castelli}, {Cortesi}, {Haack},
  {Lopes}, {Thain{\'a}-Batista}, {Cid Fernandes}, {Lomel{\'\i}-N{\'u}{\~n}ez},
  {Ribeiro}, {de Bom}, {Cernic}, {Sodr{\'e}}, {Zenocratti}, {De Rossi},
  {Calder{\'o}n}, {Herpich}, {Telles}, {Saha}, {Lopes}, {Lopes-Silva},
  {Gon{\c{c}}alves}, {Bambrila}, {Cardoso}, {Buzzo}, {Astudillo Sotomayor},
  {Demarco}, {Leigh}, {Sarzi}, {Men{\'e}ndez-Delmestre}, {Faifer},
  {Jim{\'e}nez-Teja}, {Grossi}, {Hern{\'a}ndez-Jim{\'e}nez}, {Krabbe},
  {Guti{\'e}rrez Soto}, {Brand{\~a}o}, {Espinosa}, {Olave-Rojas}, {Oliveira
  Schwarz}, {Almeida-Fernandes}, {Schoenell}, {Ribeiro}, {Kanaan}, \& {Mendes
  de Oliveira}}]{PaperI}
{Smith Castelli}, A.~V., {Cortesi}, A., {Haack}, R.~F., {et~al.} 2024, \mnras,
  530, 3787

\bibitem[{{Sobral} {et~al.}(2009){Sobral}, {Best}, {Geach}, {Smail}, {Kurk},
  {Cirasuolo}, {Casali}, {Ivison}, {Coppin}, \& {Dalton}}]{2009/Sobral}
{Sobral}, D., {Best}, P.~N., {Geach}, J.~E., {et~al.} 2009, \mnras, 398, 75

\bibitem[{{Spavone} {et~al.}(2022){Spavone}, {Iodice}, {D'Ago}, {van de Ven},
  {Morelli}, {Corsini}, {Sarzi}, {Coccato}, {Fahrion}, {Falc{\'o}n-Barroso},
  {Gadotti}, {Lyubenova}, {Mart{\'\i}n-Navarro}, {McDermid}, {Pinna},
  {Pizzella}, {Poci}, {de Zeeuw}, \& {Zhu}}]{2022/Spavone}
{Spavone}, M., {Iodice}, E., {D'Ago}, G., {et~al.} 2022, \aap, 663, A135

\bibitem[{{Spavone} {et~al.}(2020){Spavone}, {Iodice}, {van de Ven},
  {Falc{\'o}n-Barroso}, {Raj}, {Hilker}, {Peletier}, {Capaccioli}, {Mieske},
  {Venhola}, {Napolitano}, {Cantiello}, {Paolillo}, \&
  {Schipani}}]{2020/Spavone}
{Spavone}, M., {Iodice}, E., {van de Ven}, G., {et~al.} 2020, \aap, 639, A14

\bibitem[{{Spilker} {et~al.}(2022){Spilker}, {Suess}, {Setton}, {Bezanson},
  {Feldmann}, {Greene}, {Kriek}, {Lower}, {Narayanan}, \&
  {Verrico}}]{2022/Spilker}
{Spilker}, J.~S., {Suess}, K.~A., {Setton}, D.~J., {et~al.} 2022, \apjl, 936,
  L11

\bibitem[{{Spriggs} {et~al.}(2021){Spriggs}, {Sarzi}, {Gal{\'a}n-de Anta},
  {Napiwotzki}, {Viaene}, {Nedelchev}, {Coccato}, {Corsini}, {Fahrion},
  {Falc{\'o}n-Barroso}, {Gadotti}, {Iodice}, {Lyubenova},
  {Mart{\'\i}n-Navarro}, {McDermid}, {Morelli}, {Pinna}, {van de Ven}, {de
  Zeeuw}, \& {Zhu}}]{2021/Spriggs}
{Spriggs}, T.~W., {Sarzi}, M., {Gal{\'a}n-de Anta}, P.~M., {et~al.} 2021, \aap,
  653, A167

\bibitem[{{Spriggs} {et~al.}(2020){Spriggs}, {Sarzi}, {Napiwotzki},
  {Gal{\'a}n-de Anta}, {Viaene}, {Nedelchev}, {Coccato}, {Corsini}, {de Zeeuw},
  {Falc{\'o}n-Barroso}, {Gadotti}, {Iodice}, {Lyubenova},
  {Mart{\'\i}n-Navarro}, {McDermid}, {Pinna}, {van de Ven}, \&
  {Zhu}}]{2020/Spriggs}
{Spriggs}, T.~W., {Sarzi}, M., {Napiwotzki}, R., {et~al.} 2020, \aap, 637, A62

\bibitem[{{Springel} {et~al.}(2018){Springel}, {Pakmor}, {Pillepich},
  {Weinberger}, {Nelson}, {Hernquist}, {Vogelsberger}, {Genel}, {Torrey},
  {Marinacci}, \& {Naiman}}]{Springel18}
{Springel}, V., {Pakmor}, R., {Pillepich}, A., {et~al.} 2018, \mnras, 475, 676

\bibitem[{{Springel} {et~al.}(2001){Springel}, {White}, {Tormen}, \&
  {Kauffmann}}]{Springel01}
{Springel}, V., {White}, S. D.~M., {Tormen}, G., \& {Kauffmann}, G. 2001,
  \mnras, 328, 726

\bibitem[{{Su} {et~al.}(2021){Su}, {Salo}, {Janz}, {Laurikainen}, {Venhola},
  {Peletier}, {Iodice}, {Hilker}, {Cantiello}, {Napolitano}, {Spavone}, {Raj},
  {van de Ven}, {Mieske}, {Paolillo}, {Capaccioli}, {Valentijn}, \&
  {Watkins}}]{2021/Su}
{Su}, A.~H., {Salo}, H., {Janz}, J., {et~al.} 2021, \aap, 647, A100

\bibitem[{{Taniguchi} {et~al.}(2015){Taniguchi}, {Kajisawa}, {Kobayashi},
  {Shioya}, {Nagao}, {Capak}, {Aussel}, {Ichikawa}, {Murayama}, {Scoville},
  {Ilbert}, {Salvato}, {Sanders}, {Mobasher}, {Miyazaki}, {Komiyama}, {Le
  F{\`e}vre}, {Tasca}, {Lilly}, {Carollo}, {Renzini}, {Rich}, {Schinnerer},
  {Kaifu}, {Karoji}, {Arimoto}, {Okamura}, {Ohta}, {Shimasaku}, \&
  {Hayashino}}]{2015/Taniguchi}
{Taniguchi}, Y., {Kajisawa}, M., {Kobayashi}, M. A.~R., {et~al.} 2015, \pasj,
  67, 104

\bibitem[{{Taylor} {et~al.}(2011){Taylor}, {Hopkins}, {Baldry}, {Brown},
  {Driver}, {Kelvin}, {Hill}, {Robotham}, {Bland-Hawthorn}, {Jones}, {Sharp},
  {Thomas}, {Liske}, {Loveday}, {Norberg}, {Peacock}, {Bamford}, {Brough},
  {Colless}, {Cameron}, {Conselice}, {Croom}, {Frenk}, {Gunawardhana},
  {Kuijken}, {Nichol}, {Parkinson}, {Phillipps}, {Pimbblet}, {Popescu},
  {Prescott}, {Sutherland}, {Tuffs}, {van Kampen}, \&
  {Wijesinghe}}]{2011/Taylor}
{Taylor}, E.~N., {Hopkins}, A.~M., {Baldry}, I.~K., {et~al.} 2011, \mnras, 418,
  1587

\bibitem[{{Thain{\'a}-Batista} {et~al.}(2023){Thain{\'a}-Batista}, {Cid
  Fernandes}, {Herpich}, {Mendes de Oliveira}, {Werle}, {Espinosa}, {Lopes},
  {Smith Castelli}, {Sodr{\'e}}, {Telles}, {Kanaan}, {Ribeiro}, \&
  {Schoenell}}]{2023/ThainaBatista}
{Thain{\'a}-Batista}, J., {Cid Fernandes}, R., {Herpich}, F.~R., {et~al.} 2023,
  \mnras, 526, 1874

\bibitem[{{van der Burg} {et~al.}(2020){van der Burg}, {Rudnick}, {Balogh},
  {Muzzin}, {Lidman}, {Old}, {Shipley}, {Gilbank}, {McGee}, {Biviano},
  {Cerulo}, {Chan}, {Cooper}, {De Lucia}, {Demarco}, {Forrest}, {Gwyn},
  {Jablonka}, {Kukstas}, {Marchesini}, {Nantais}, {Noble}, {Pintos-Castro},
  {Poggianti}, {Reeves}, {Stefanon}, {Vulcani}, {Webb}, {Wilson}, {Yee}, \&
  {Zaritsky}}]{2020/vanderBurg}
{van der Burg}, R. F.~J., {Rudnick}, G., {Balogh}, M.~L., {et~al.} 2020, \aap,
  638, A112

\bibitem[{{Venhola} {et~al.}(2018){Venhola}, {Peletier}, {Laurikainen}, {Salo},
  {Iodice}, {Mieske}, {Hilker}, {Wittmann}, {Lisker}, {Paolillo}, {Cantiello},
  {Janz}, {Spavone}, {D'Abrusco}, {van de Ven}, {Napolitano}, {Verdoes Kleijn},
  {Maddox}, {Capaccioli}, {Grado}, {Valentijn}, {Falc{\'o}n-Barroso}, \&
  {Limatola}}]{2018/Venhola}
{Venhola}, A., {Peletier}, R., {Laurikainen}, E., {et~al.} 2018, \aap, 620,
  A165

\bibitem[{{Venhola} {et~al.}(2019){Venhola}, {Peletier}, {Laurikainen}, {Salo},
  {Iodice}, {Mieske}, {Hilker}, {Wittmann}, {Paolillo}, {Cantiello}, {Janz},
  {Spavone}, {D'Abrusco}, {van de Ven}, {Napolitano}, {Verdoes Kleijn},
  {Capaccioli}, {Grado}, {Valentijn}, {Falc{\'o}n-Barroso}, \&
  {Limatola}}]{2019/Venhola}
{Venhola}, A., {Peletier}, R., {Laurikainen}, E., {et~al.} 2019, \aap, 625,
  A143

\bibitem[{{Venhola} {et~al.}(2017){Venhola}, {Peletier}, {Laurikainen}, {Salo},
  {Lisker}, {Iodice}, {Capaccioli}, {Verdois Kleijn}, {Valentijn}, {Mieske},
  {Hilker}, {Wittmann}, {van de Ven}, {Grado}, {Spavone}, {Cantiello},
  {Napolitano}, {Paolillo}, \& {Falc{\'o}n-Barroso}}]{2017/Venhola}
{Venhola}, A., {Peletier}, R., {Laurikainen}, E., {et~al.} 2017, \aap, 608,
  A142

\bibitem[{{Venhola} {et~al.}(2022){Venhola}, {Peletier}, {Salo}, {Laurikainen},
  {Janz}, {Haigh}, {Wilkinson}, {Iodice}, {Hilker}, {Mieske}, {Cantiello}, \&
  {Spavone}}]{2022/Venhola}
{Venhola}, A., {Peletier}, R.~F., {Salo}, H., {et~al.} 2022, \aap, 662, A43

\bibitem[{{Vilella-Rojo} {et~al.}(2015){Vilella-Rojo}, {Viironen},
  {L{\'o}pez-Sanjuan}, {Cenarro}, {Varela}, {D{\'\i}az-Garc{\'\i}a},
  {Crist{\'o}bal-Hornillos}, {Ederoclite}, {Mar{\'\i}n-Franch}, \&
  {Moles}}]{2015/VilellaRojo}
{Vilella-Rojo}, G., {Viironen}, K., {L{\'o}pez-Sanjuan}, C., {et~al.} 2015,
  \aap, 580, A47

\bibitem[{{Vollmer} {et~al.}(2009){Vollmer}, {Soida}, {Chung}, {Chemin},
  {Braine}, {Boselli}, \& {Beck}}]{2009/Vollmer}
{Vollmer}, B., {Soida}, M., {Chung}, A., {et~al.} 2009, \aap, 496, 669

\bibitem[{{Waugh} {et~al.}(2002){Waugh}, {Drinkwater}, {Webster},
  {Staveley-Smith}, {Kilborn}, {Barnes}, {Bhathal}, {de Blok}, {Boyce},
  {Disney}, {Ekers}, {Freeman}, {Gibson}, {Henning}, {Jerjen}, {Knezek},
  {Koribalski}, {Marquarding}, {Minchin}, {Price}, {Putman}, {Ryder}, {Sadler},
  {Stootman}, \& {Zwaan}}]{2002/Waugh}
{Waugh}, M., {Drinkwater}, M.~J., {Webster}, R.~L., {et~al.} 2002, \mnras, 337,
  641

\bibitem[{{Wetzel}(2011)}]{2011/Wetzel}
{Wetzel}, A.~R. 2011, \mnras, 412, 49

\bibitem[{{Wetzel} {et~al.}(2013){Wetzel}, {Tinker}, {Conroy}, \& {van den
  Bosch}}]{2013/Wetzel}
{Wetzel}, A.~R., {Tinker}, J.~L., {Conroy}, C., \& {van den Bosch}, F.~C. 2013,
  \mnras, 432, 336

\bibitem[{{Wolf} {et~al.}(2003){Wolf}, {Meisenheimer}, {Rix}, {Borch}, {Dye},
  \& {Kleinheinrich}}]{2003/Wolf}
{Wolf}, C., {Meisenheimer}, K., {Rix}, H.~W., {et~al.} 2003, \aap, 401, 73

\bibitem[{{Wright} {et~al.}(2018){Wright}, {Lagos}, {Davies}, {Power},
  {Trayford}, \& {Wong}}]{Wright18}
{Wright}, R.~J., {Lagos}, C.~d.~P., {Davies}, L.~J.~M., {et~al.} 2018, ArXiv
  e-prints [\eprint[arXiv]{1810.07335}]

\bibitem[{{Zabel} {et~al.}(2019){Zabel}, {Davis}, {Smith}, {Maddox}, {Bendo},
  {Peletier}, {Iodice}, {Venhola}, {Baes}, {Davies}, {de Looze}, {Gomez},
  {Grossi}, {Kenney}, {Serra}, {van de Voort}, {Vlahakis}, \&
  {Young}}]{2019/Zabel}
{Zabel}, N., {Davis}, T.~A., {Smith}, M. W.~L., {et~al.} 2019, \mnras, 483,
  2251

\bibitem[{{Zabludoff} {et~al.}(1996){Zabludoff}, {Zaritsky}, {Lin}, {Tucker},
  {Hashimoto}, {Shectman}, {Oemler}, \& {Kirshner}}]{1996/Zabludoff}
{Zabludoff}, A.~I., {Zaritsky}, D., {Lin}, H., {et~al.} 1996, \apj, 466, 104

\end{thebibliography}

\begin{appendix}
\onecolumn
\section{Emission line maps and galaxy features}
\label{app:maps}
In this appendix, we present the 77 \HalphaNII{} maps obtained by the \texttt{PELE} code, as described in Section \ref{sec:PELE}. The galaxies located within 1 \Rvir{} of Fornax\,A include NGC\,1341, NGC\,1326B, NGC\,1326A, FCC\,35, NGC\,1326, NGC\,1316C, NGC\,1317, FCC\,28, NGC\,1316, NGC\,1310 and FCC\,B112, and shown in Fig.\,\ref{fig:1.2Rvir_1.6Rvir} and Fig.\,\ref{fig:1.6Rvir_1.99Rvir}. In Table\,\ref{tab:data_morph1}, we present the main properties displayed by the 77 \HalphaNII{} emitters. In the case of the morphology of the emission, we consider as {\it central} emission that corresponding to a central compact region, {\it knots} indicates emission in several discrete compact regions, and {\it extended} emission corresponds to contiguous regions displaying \HalphaNII{} emission. If two types can be associated with the morphology of the emission, we indicate the one we consider to be dominant.

\begin{figure*}[h!]
    \centering
    \includegraphics[width=0.49\textwidth]{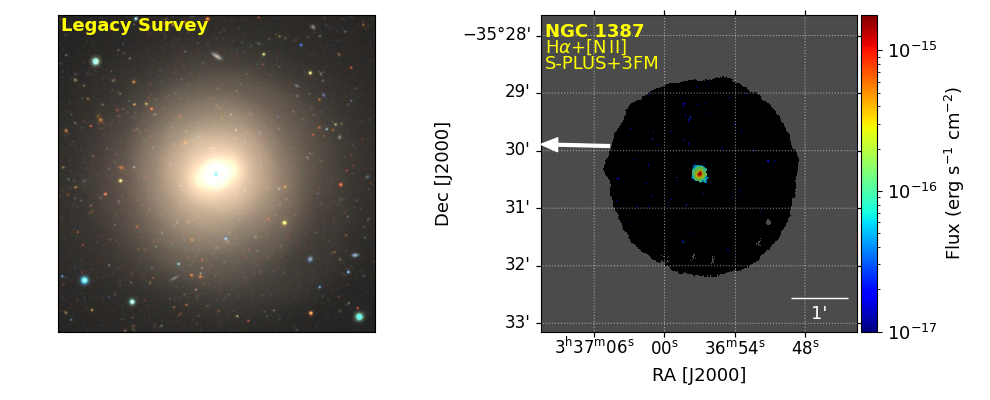}
    \includegraphics[width=0.49\textwidth]{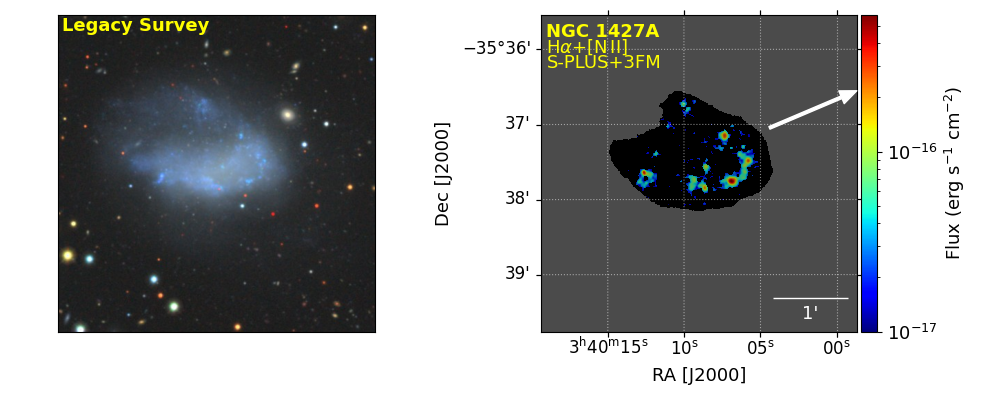}
    \includegraphics[width=0.49\textwidth]{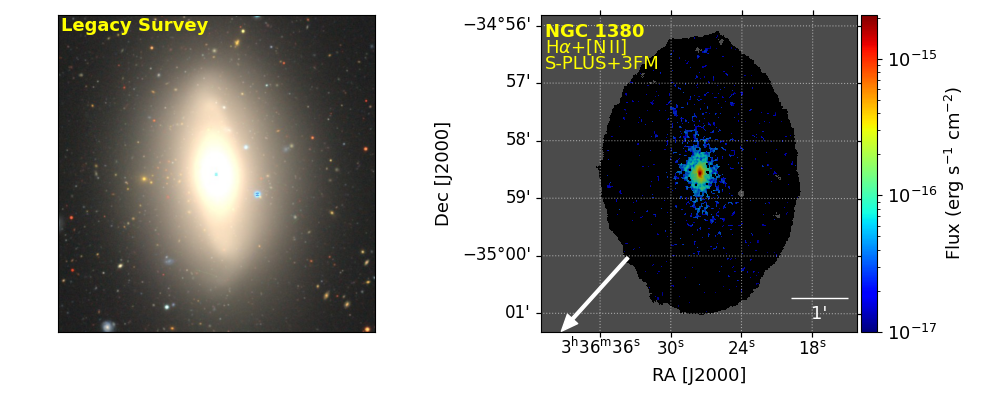}
    \includegraphics[width=0.49\textwidth]{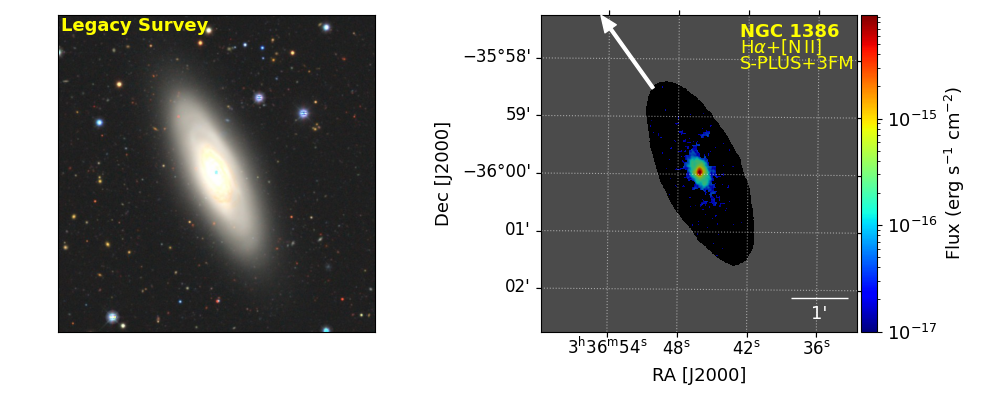}
    \includegraphics[width=0.49\textwidth]{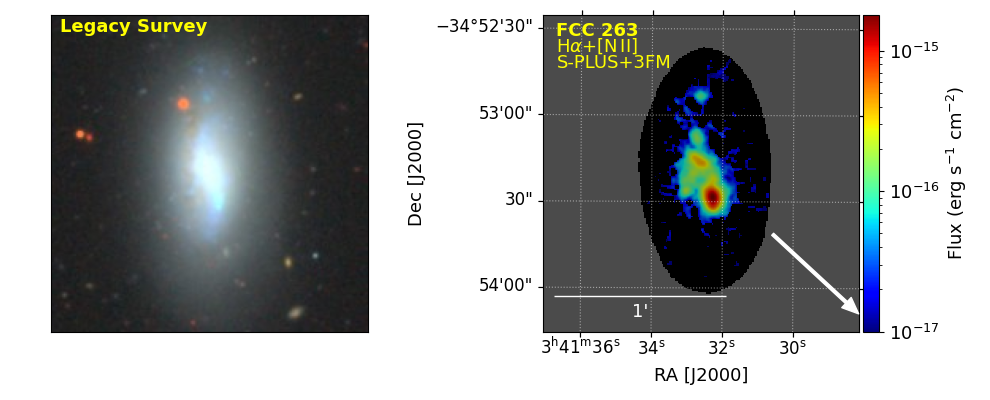}
    \includegraphics[width=0.49\textwidth]{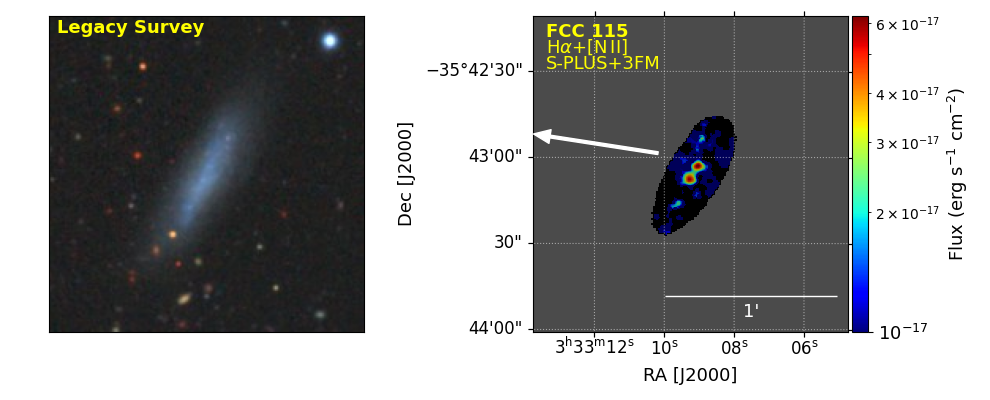}
    \includegraphics[width=0.49\textwidth]{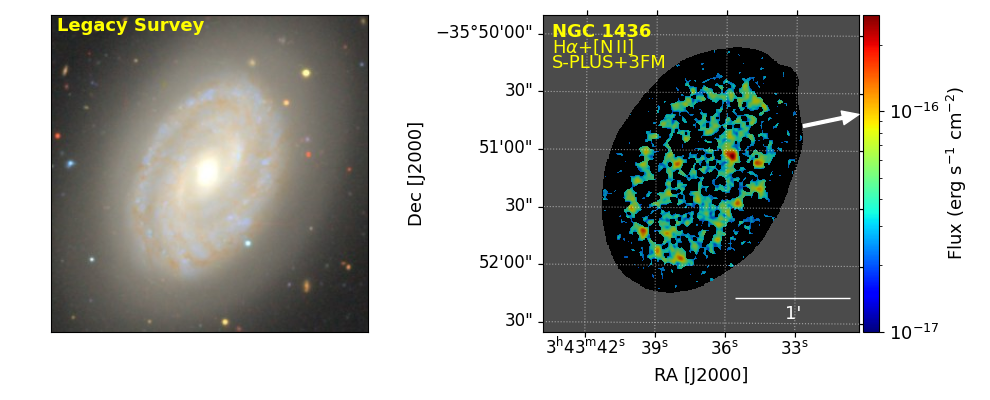}
    \includegraphics[width=0.49\textwidth]{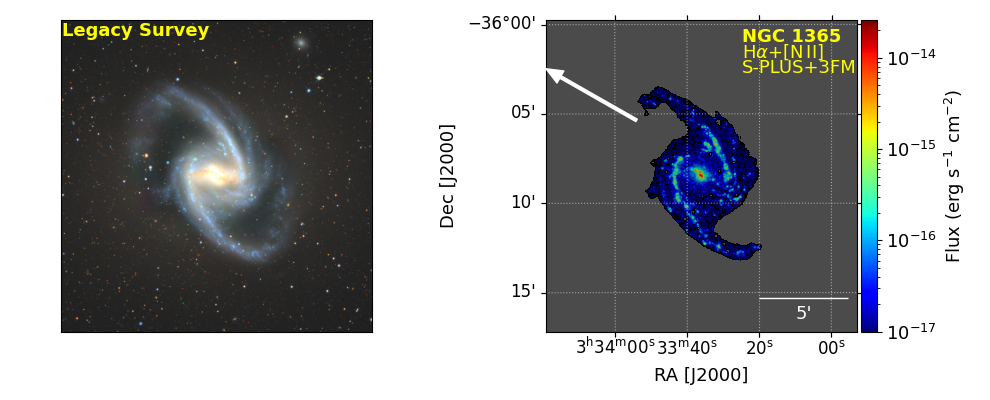}
    \includegraphics[width=0.49\textwidth]{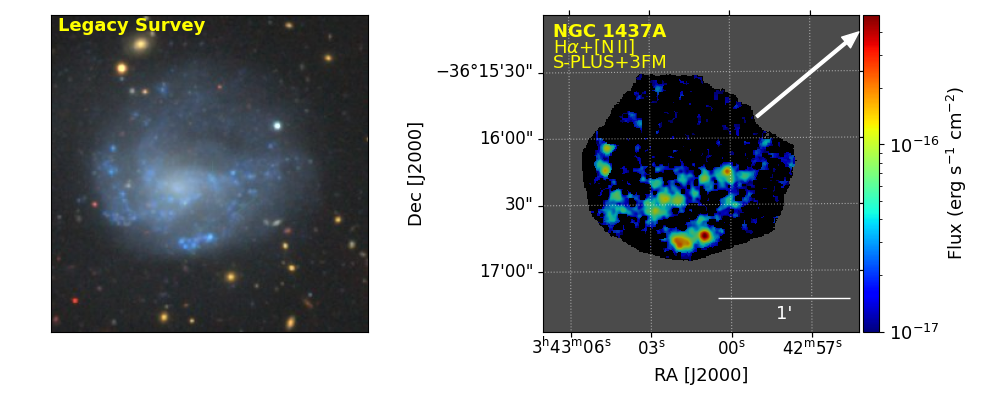}
    \includegraphics[width=0.49\textwidth]{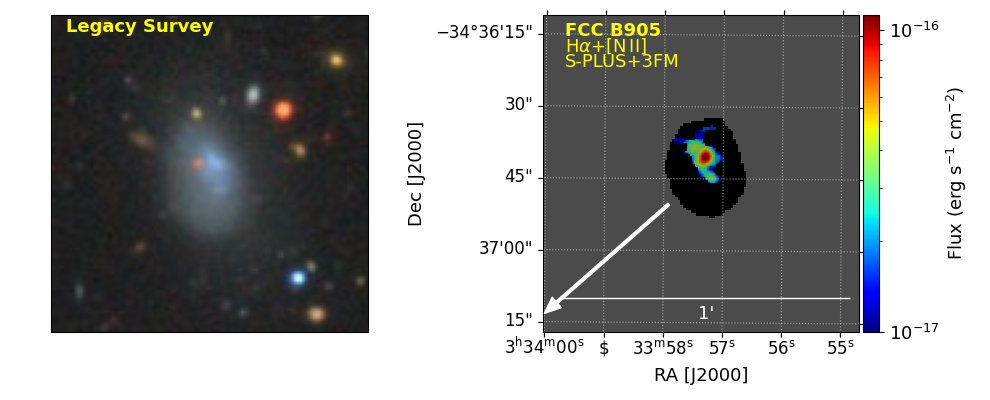}
    \caption{Images from DESI Legacy Imaging Surveys (left panels) and the corresponding \HalphaNII{} maps (right panels) for all emitters located at a projected radius R < 0.62 \Rvir{} from NGC\,1399. From left to right and from top to bottom, we show the galaxies ordered from the closest to NGC\,1399 to the farthest. The white arrow points to the center of the Fornax cluster.}
    \label{fig:0Rvir_0.7Rvir}
\end{figure*}

\begin{figure*}
    \centering
    \includegraphics[width=0.49\textwidth]{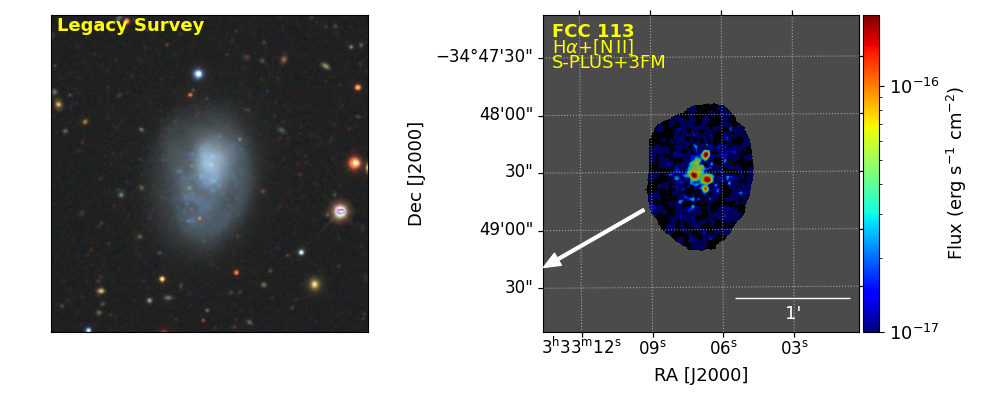}
    \includegraphics[width=0.49\textwidth]{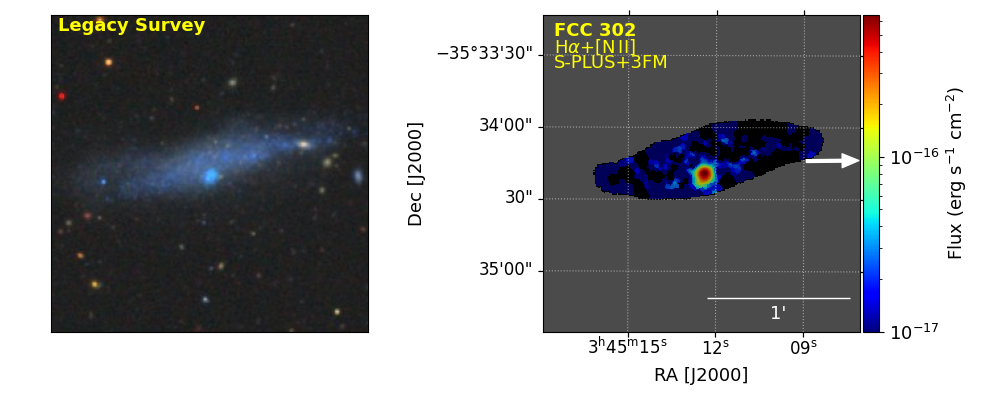}
    \includegraphics[width=0.49\textwidth]{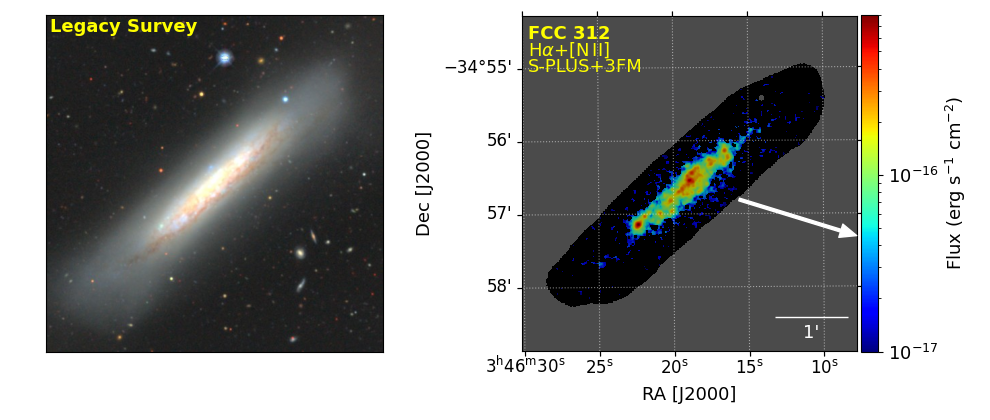}
    \includegraphics[width=0.49\textwidth]{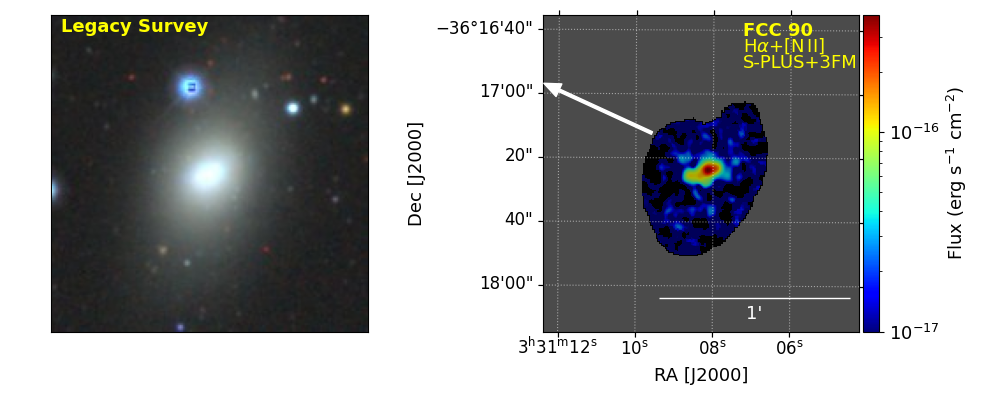}
    \includegraphics[width=0.49\textwidth]{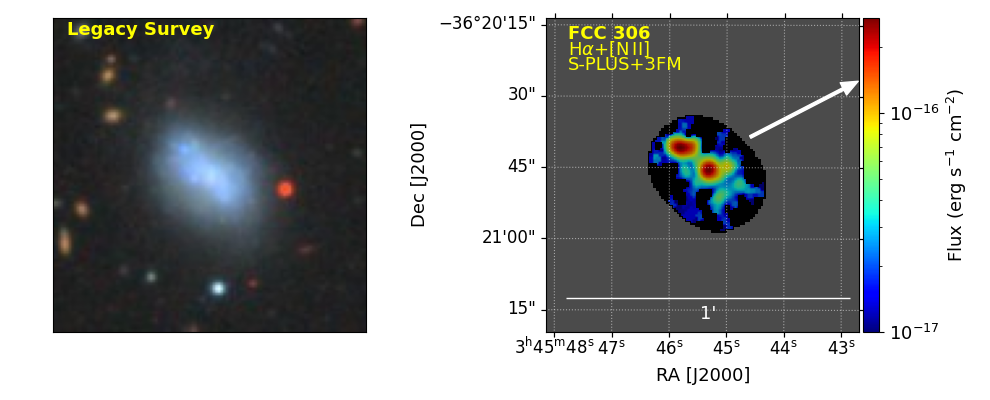}
    \includegraphics[width=0.49\textwidth]{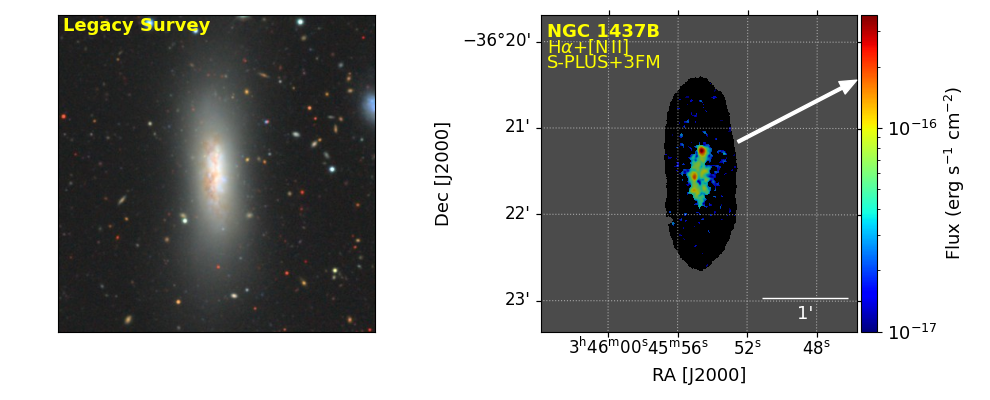}
    \includegraphics[width=0.49\textwidth]{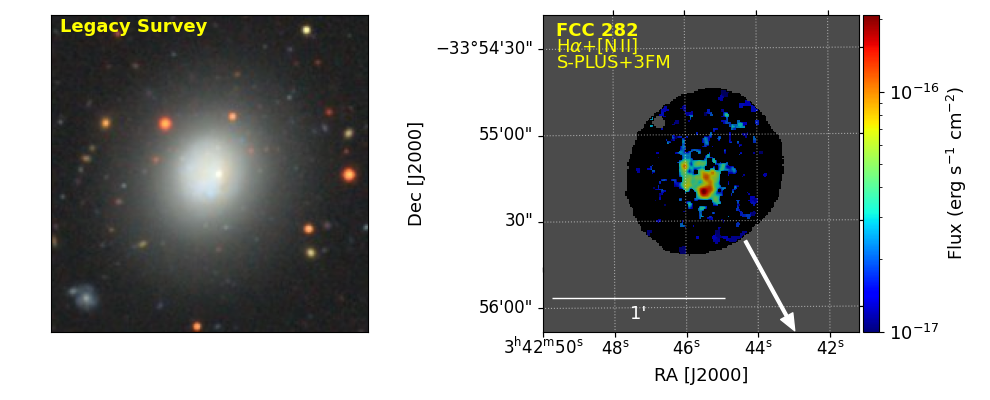}
    \includegraphics[width=0.49\textwidth]{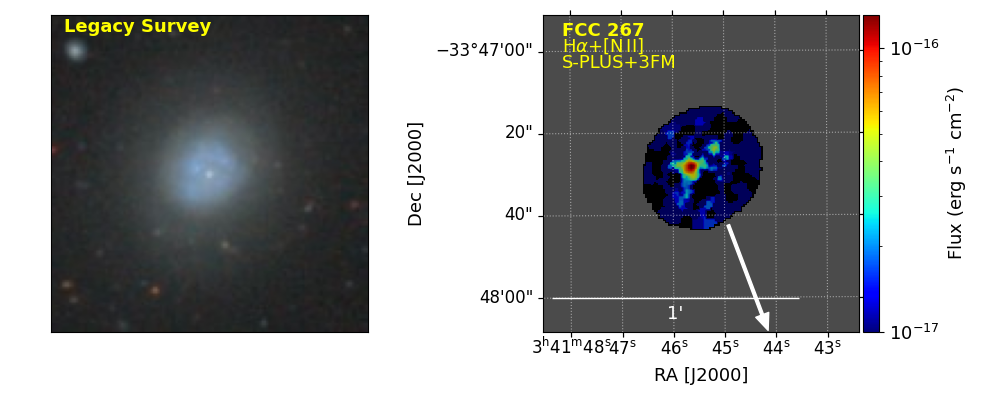}
    \includegraphics[width=0.49\textwidth]{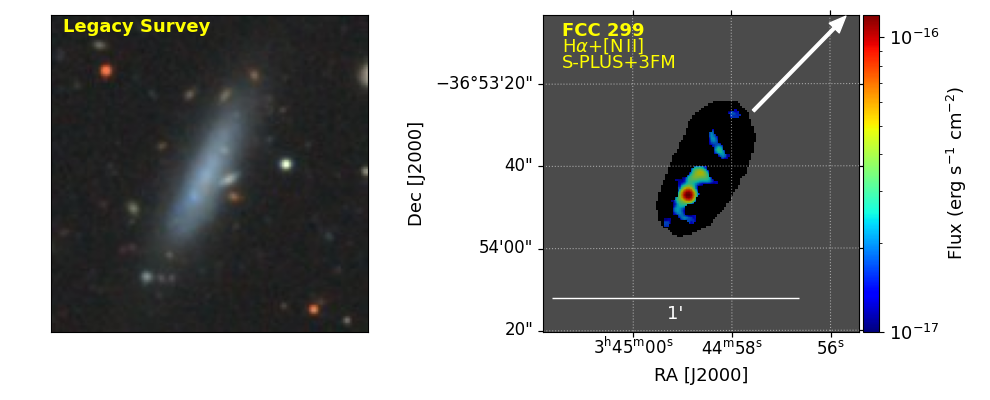}
    \includegraphics[width=0.49\textwidth]{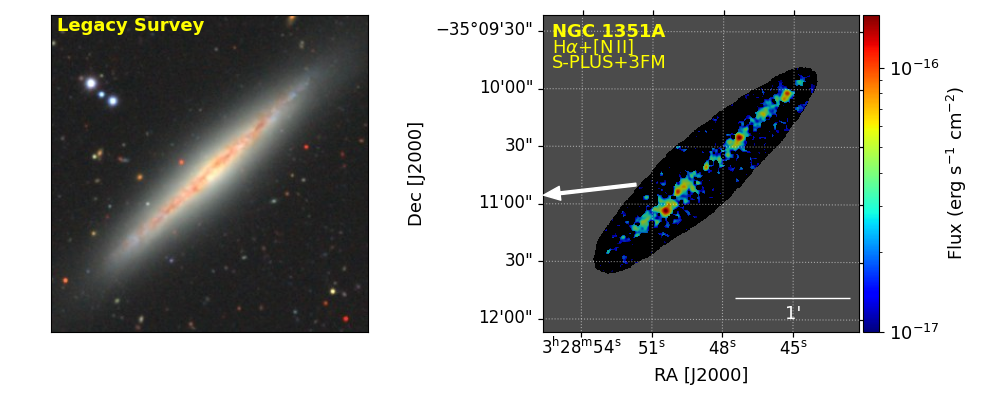}
    \includegraphics[width=0.49\textwidth]{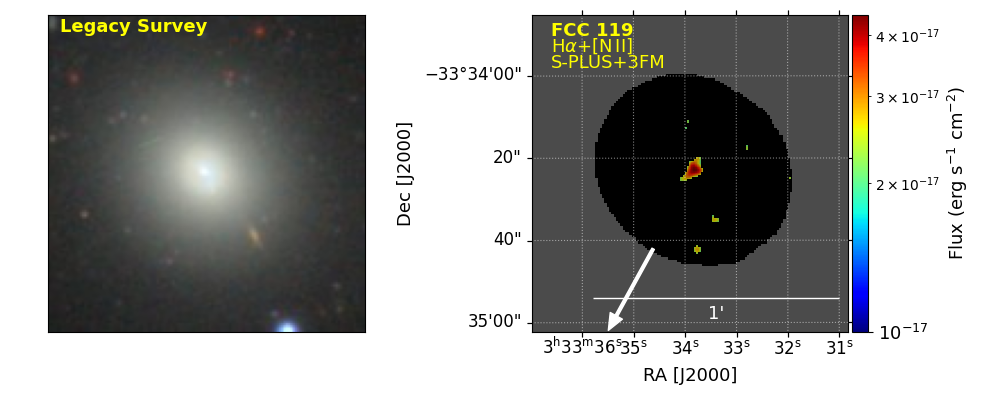}
    \includegraphics[width=0.49\textwidth]{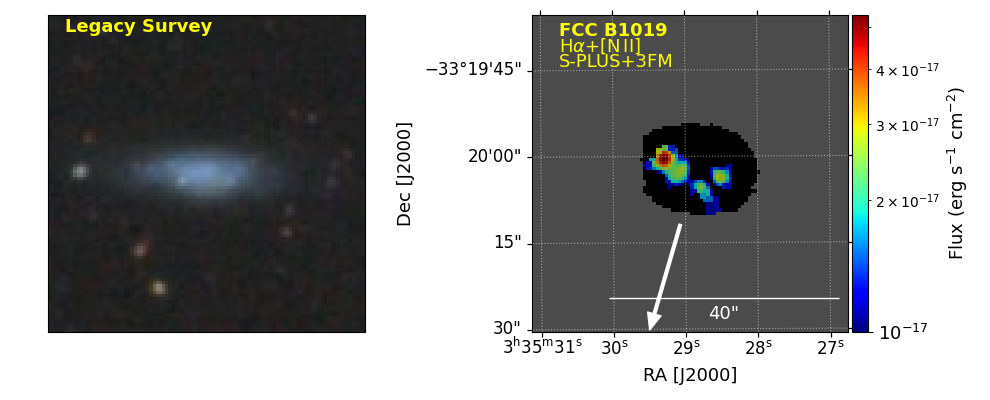}
 \caption{Same as Figure\,\ref{fig:0Rvir_0.7Rvir} but for emitters located at 0.62 \Rvir{} $\leq$ R $\leq$ 1.1 \Rvir{}.}
     \label{fig:0.7Rvir_1.2Rvir}
\end{figure*}

\begin{figure*}
    \centering
    \includegraphics[width=0.49\textwidth]{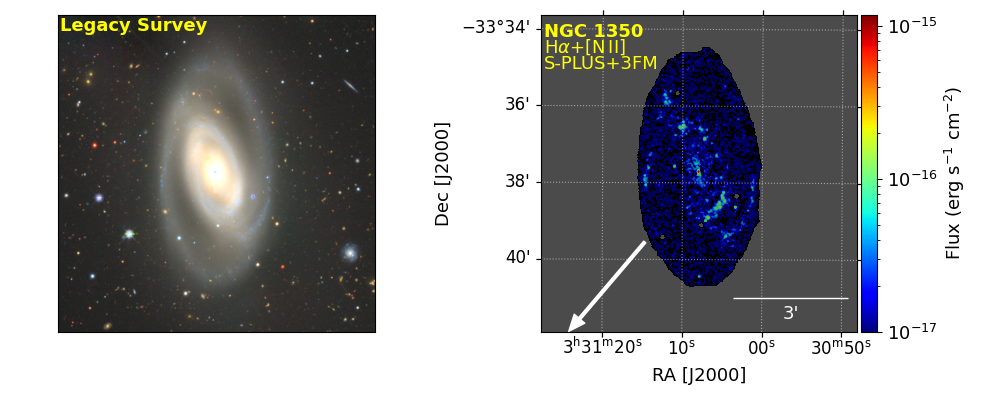}
    \includegraphics[width=0.49\textwidth]{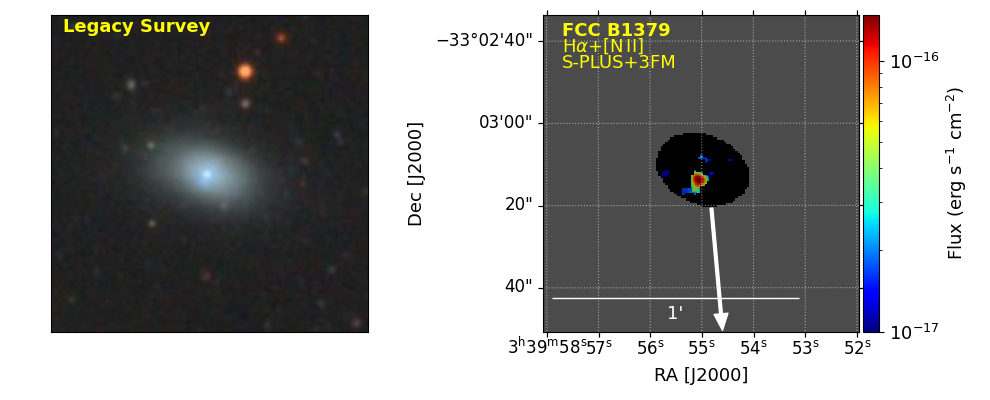}
    \includegraphics[width=0.49\textwidth]{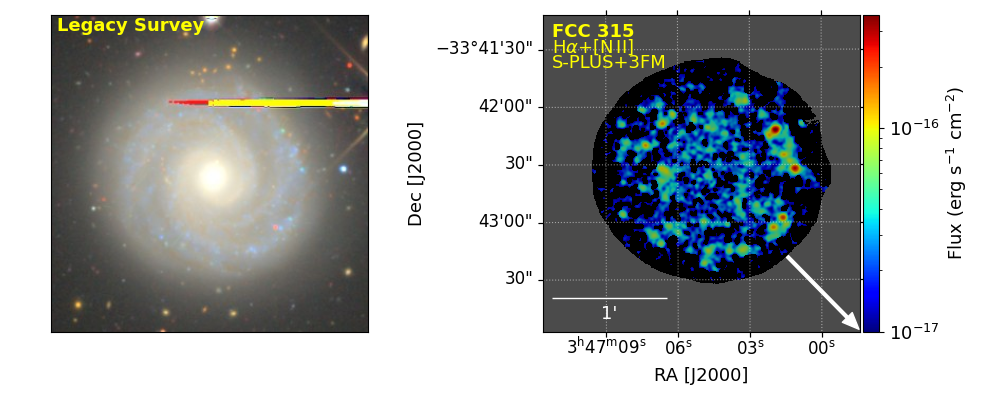}
    \includegraphics[width=0.49\textwidth]{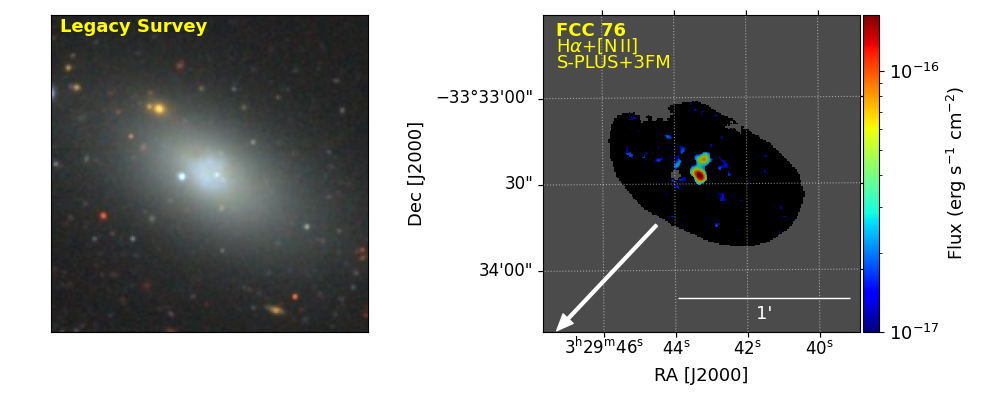}
    \includegraphics[width=0.49\textwidth]{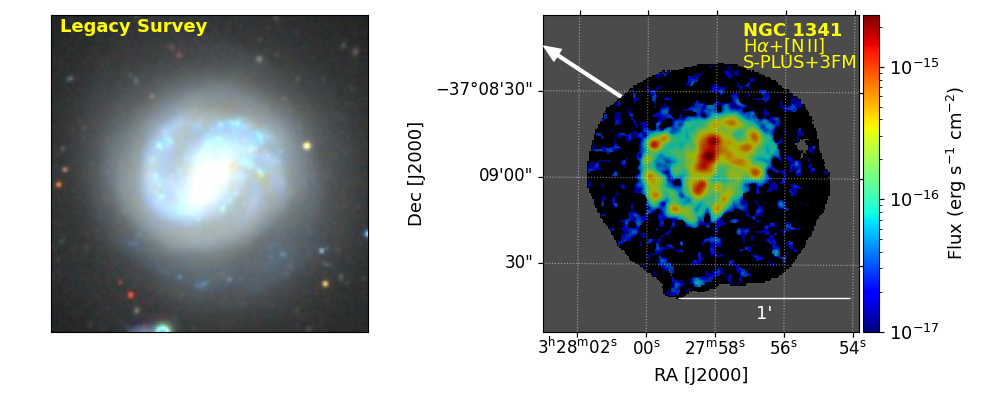}
    \includegraphics[width=0.49\textwidth]{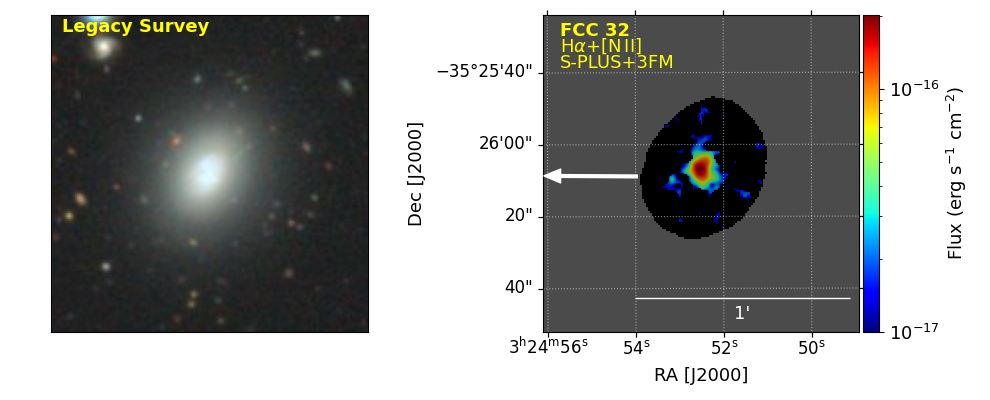}
    \includegraphics[width=0.49\textwidth]{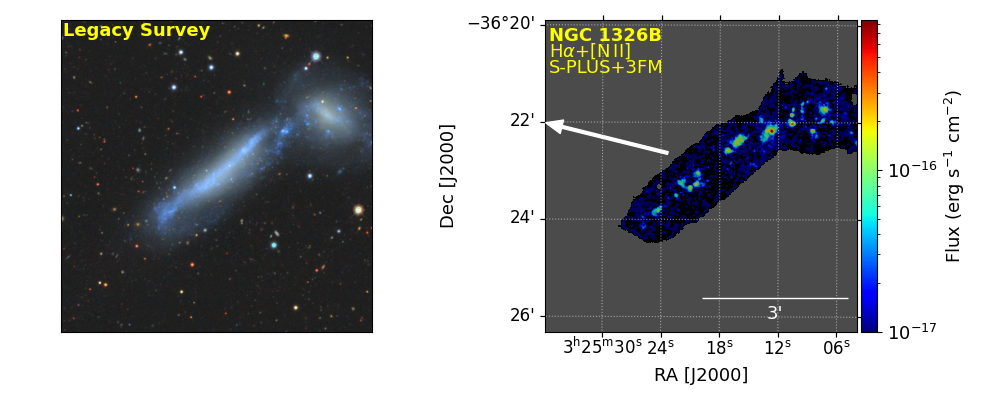}
    \includegraphics[width=0.49\textwidth]{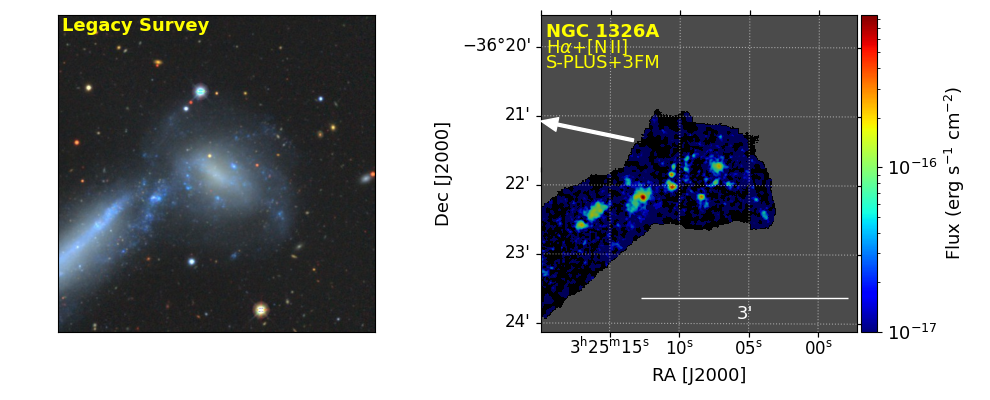}
    \includegraphics[width=0.49\textwidth]{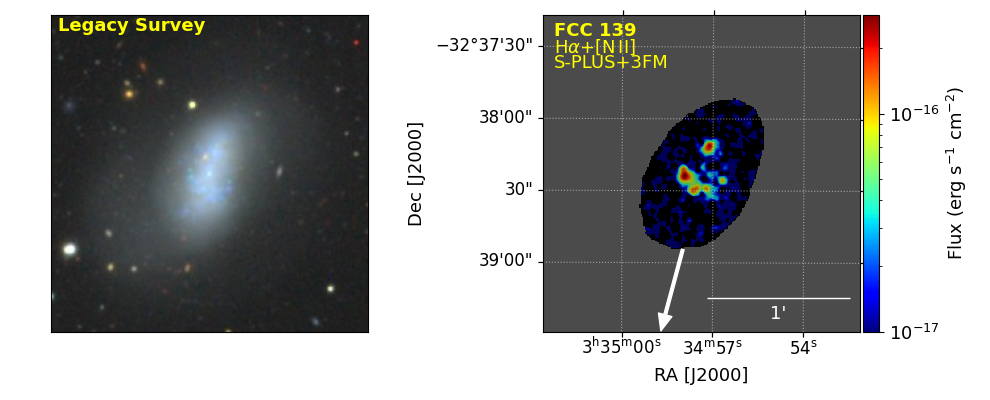}
    \includegraphics[width=0.49\textwidth]{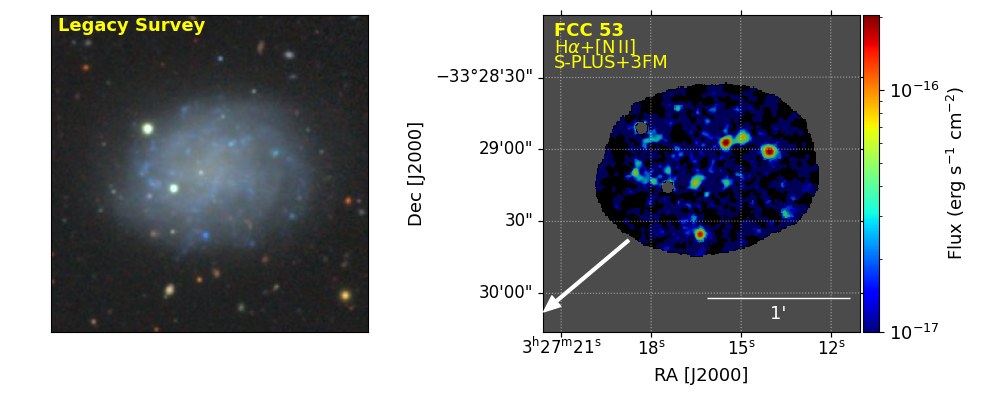}
    \includegraphics[width=0.49\textwidth]{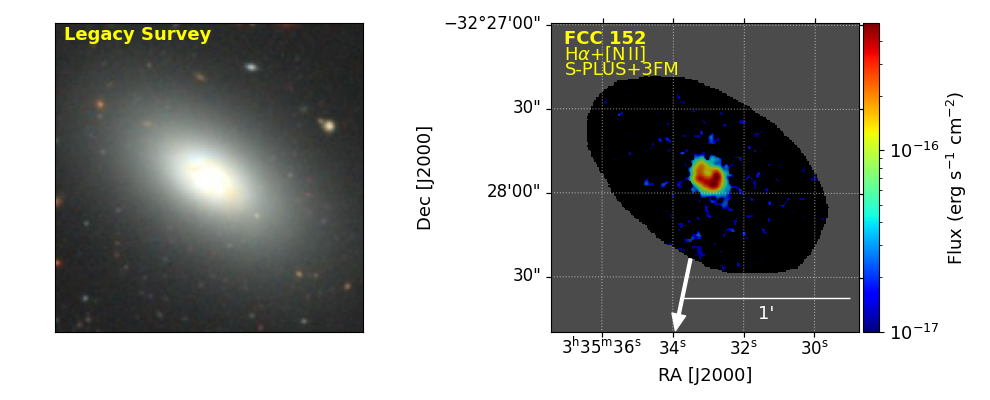}
    \includegraphics[width=0.49\textwidth]{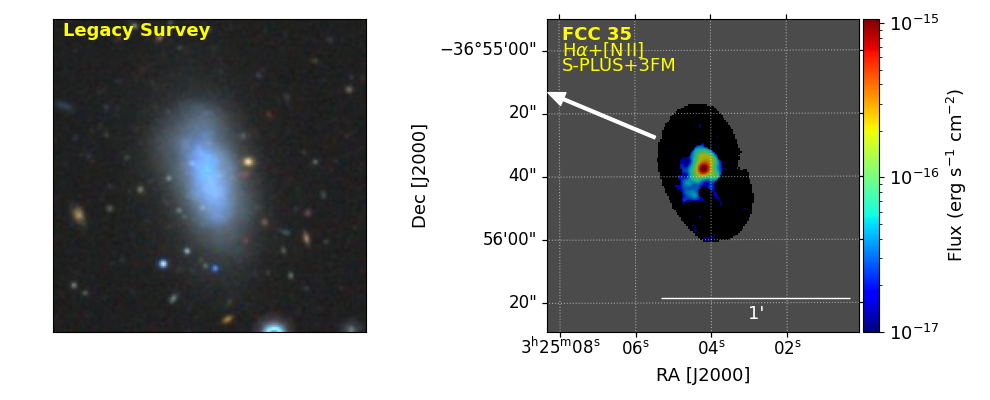}
 \caption{Same as Figure\,\ref{fig:0Rvir_0.7Rvir} but for emitters located at 1.1 \Rvir{} < R < 1.54 \Rvir{}.}
 \label{fig:1.2Rvir_1.6Rvir}
\end{figure*}

\begin{figure*}
    \centering
    \includegraphics[width=0.49\textwidth]{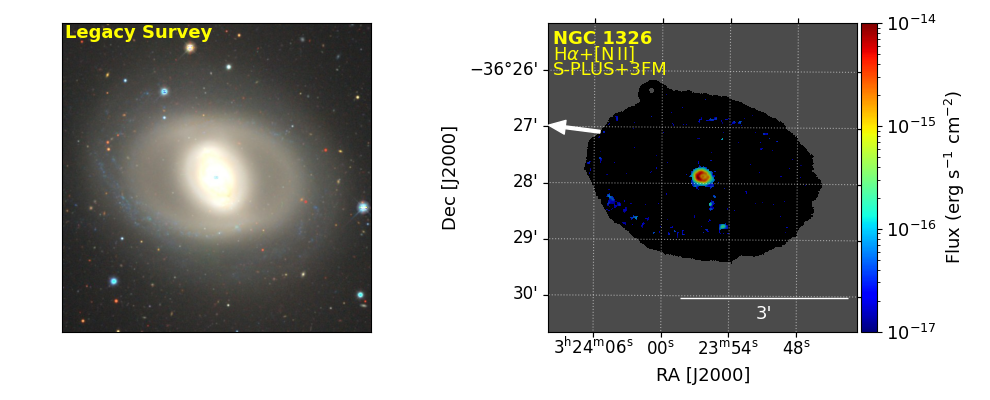}
    \includegraphics[width=0.49\textwidth]{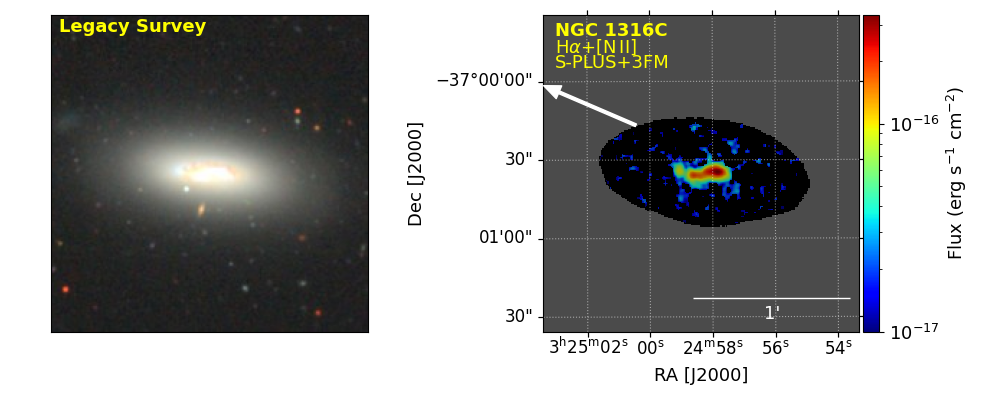}
    \includegraphics[width=0.49\textwidth]{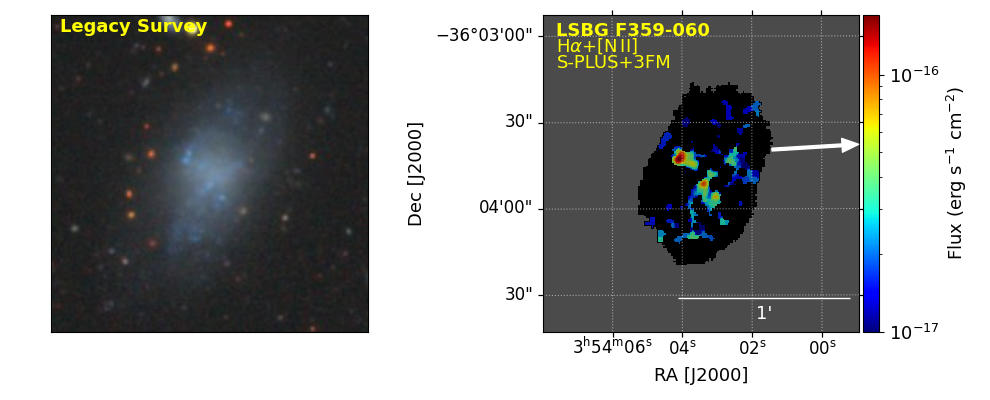}
    \includegraphics[width=0.49\textwidth]{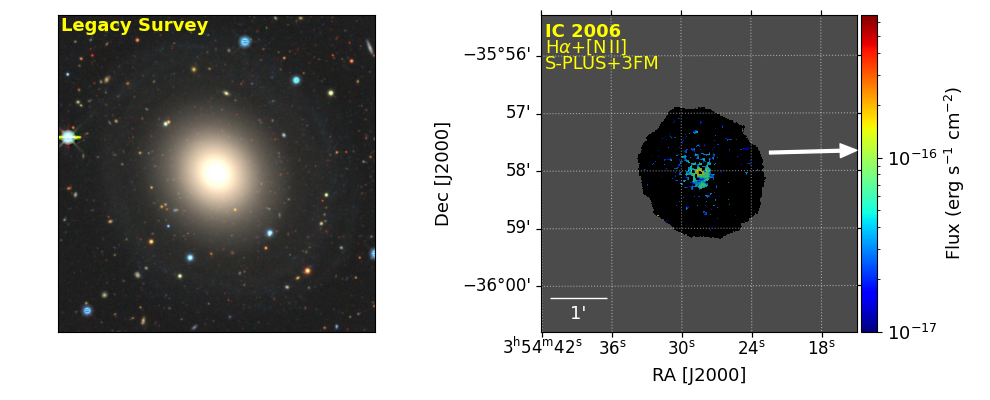}
    \includegraphics[width=0.49\textwidth]{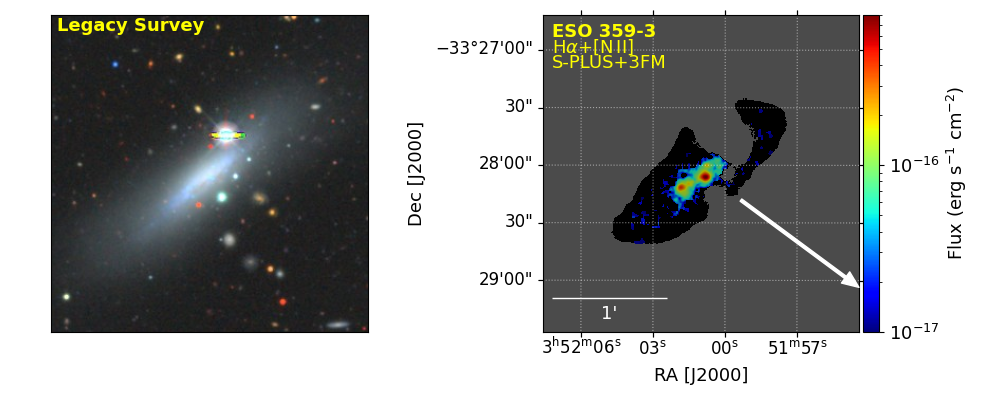}
    \includegraphics[width=0.49\textwidth]{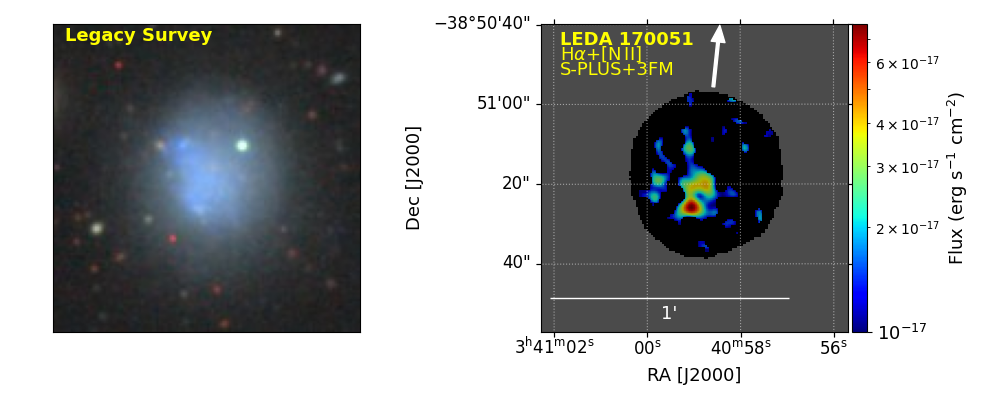}
    \includegraphics[width=0.49\textwidth]{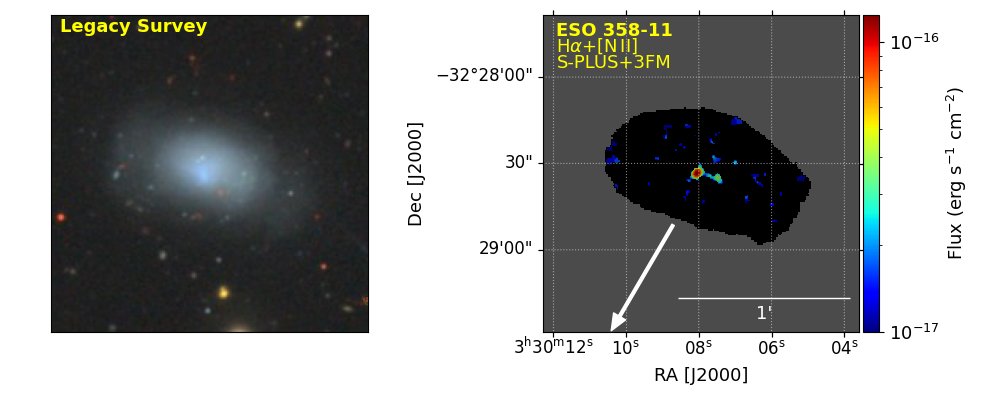}
    \includegraphics[width=0.49\textwidth]{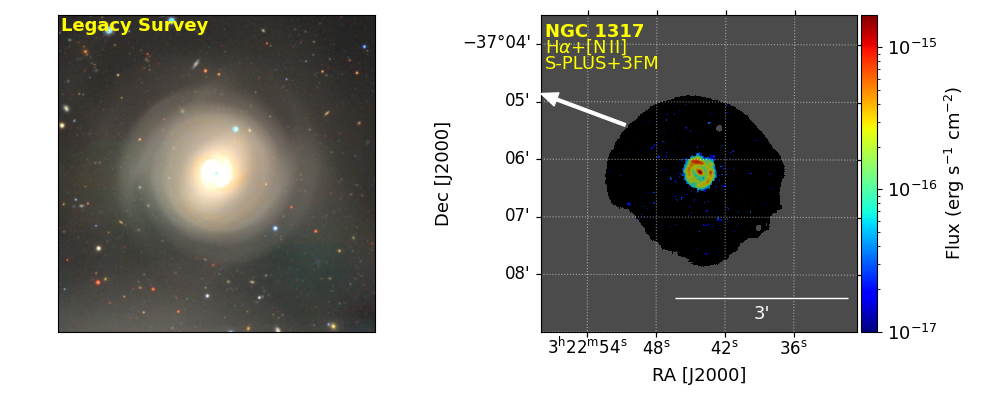}
    \includegraphics[width=0.49\textwidth]{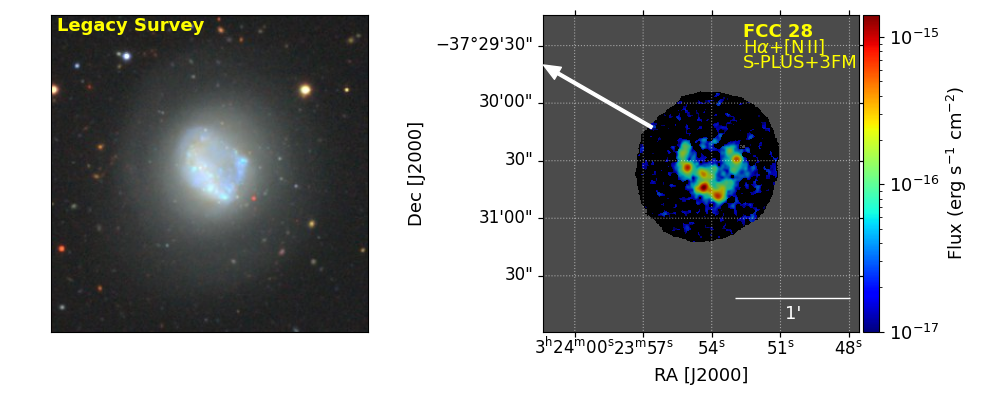}
    \includegraphics[width=0.49\textwidth]{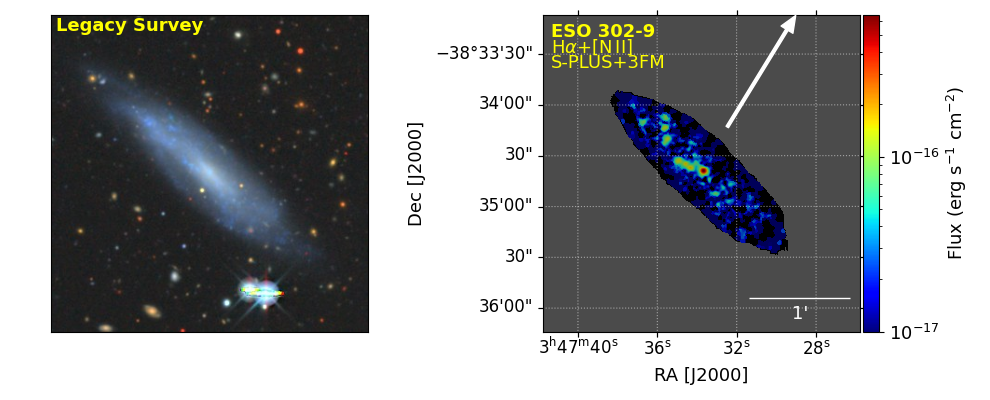}
    \includegraphics[width=0.49\textwidth]{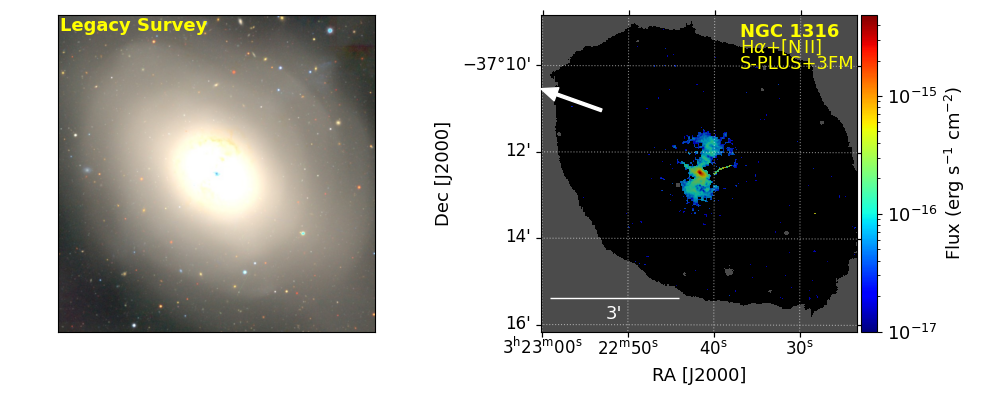}
    \includegraphics[width=0.49\textwidth]{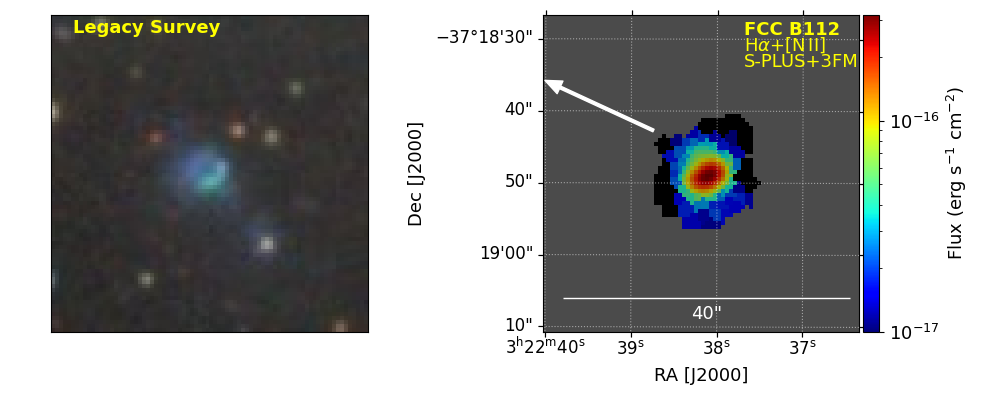}
 \caption{Same as Figure\,\ref{fig:0Rvir_0.7Rvir} but for emitters located at 1.54 \Rvir{} $\leq$ R < 1.85 \Rvir{}.}
 \label{fig:1.6Rvir_1.99Rvir}
\end{figure*}

\begin{figure*}
    \centering
    \includegraphics[width=0.49\textwidth]{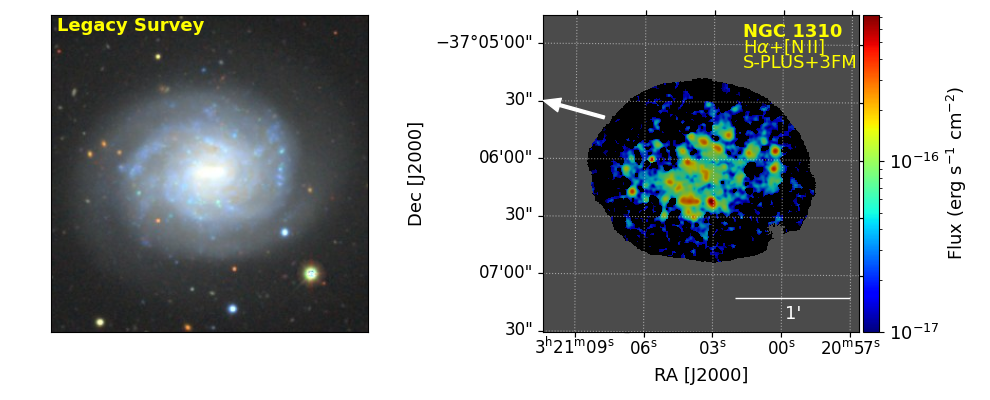}
    \includegraphics[width=0.49\textwidth]{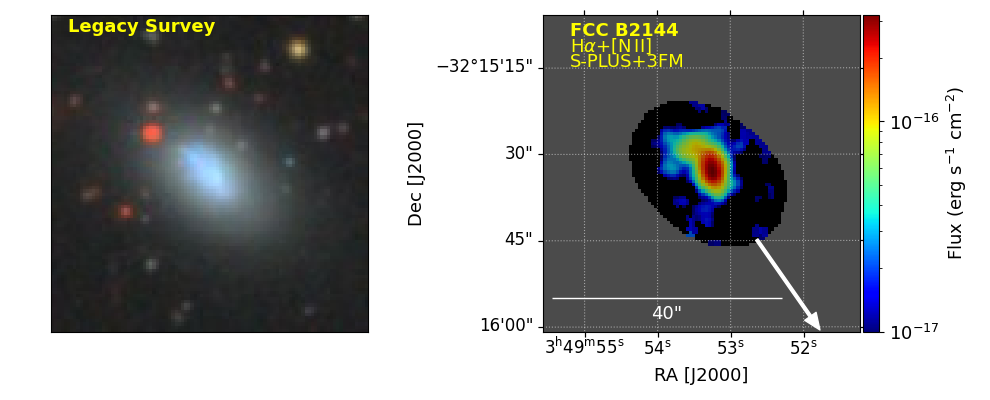}
    \includegraphics[width=0.49\textwidth]{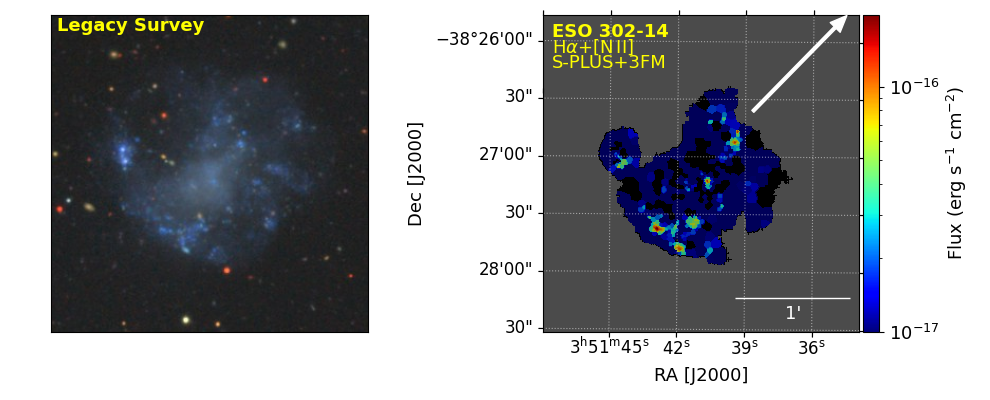}
    \includegraphics[width=0.49\textwidth]{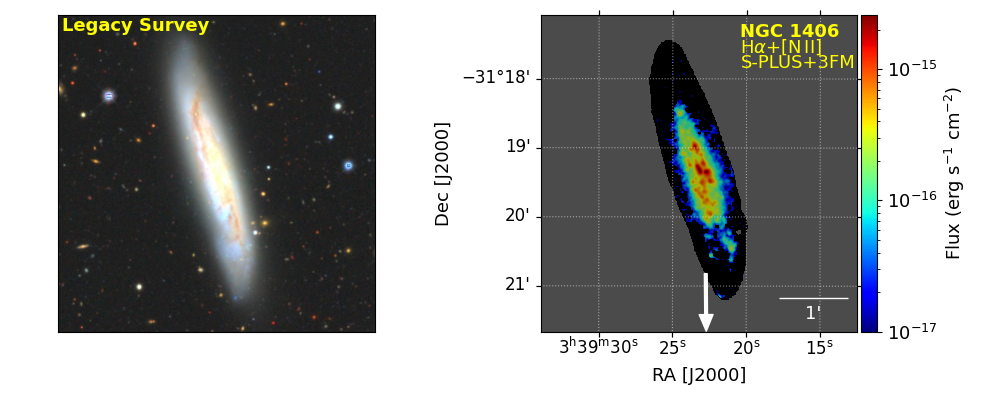}
    \includegraphics[width=0.49\textwidth]{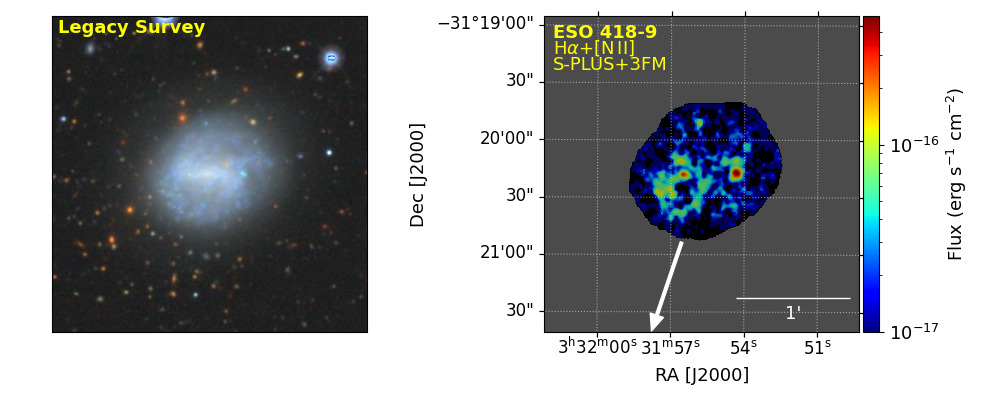}
    \includegraphics[width=0.49\textwidth]{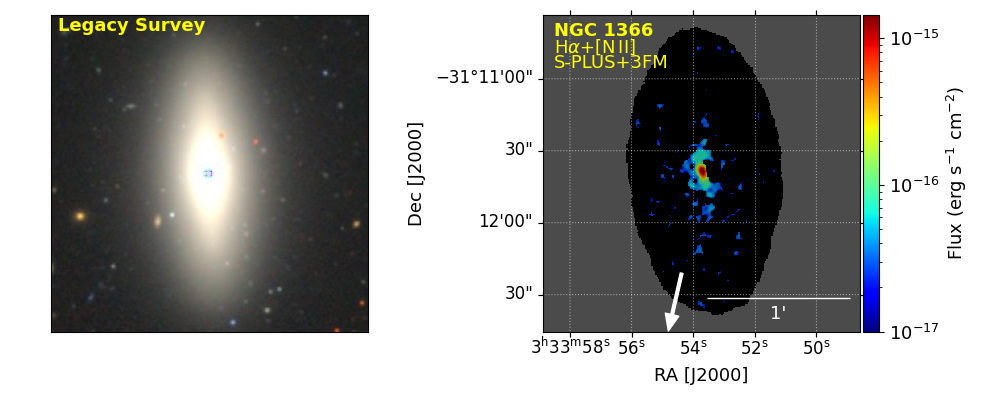}
    \includegraphics[width=0.49\textwidth]{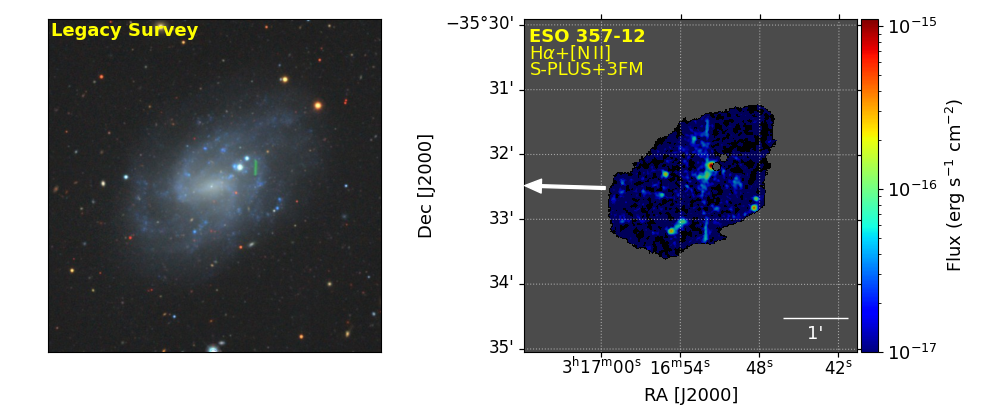}
    \includegraphics[width=0.49\textwidth]{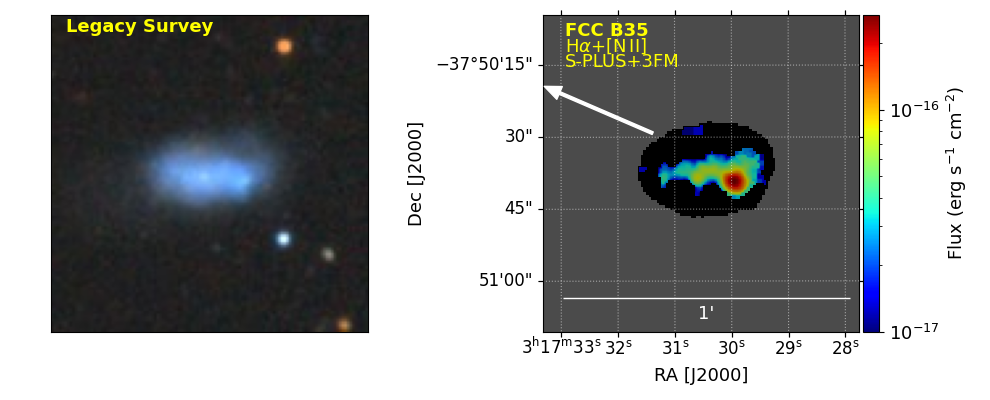}
    \includegraphics[width=0.49\textwidth]{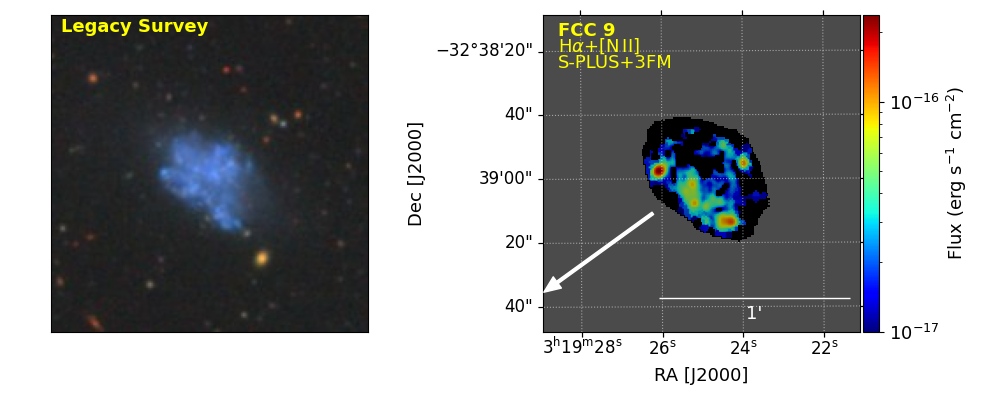}
    \includegraphics[width=0.49\textwidth]{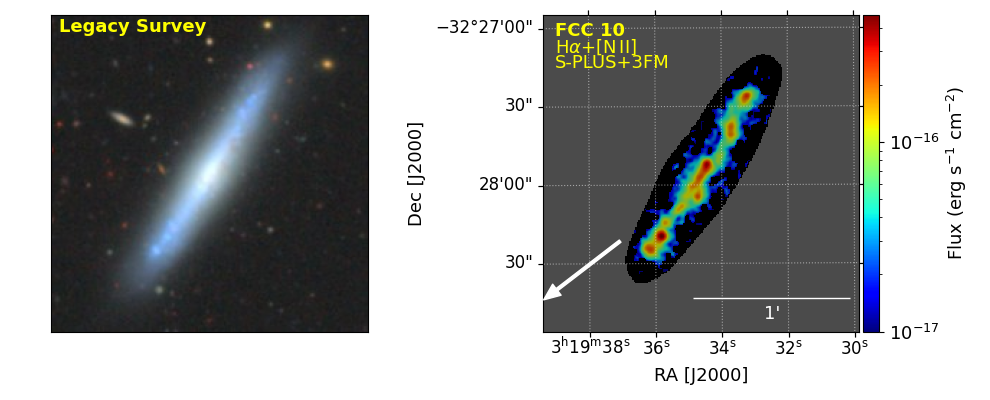}
    \includegraphics[width=0.49\textwidth]{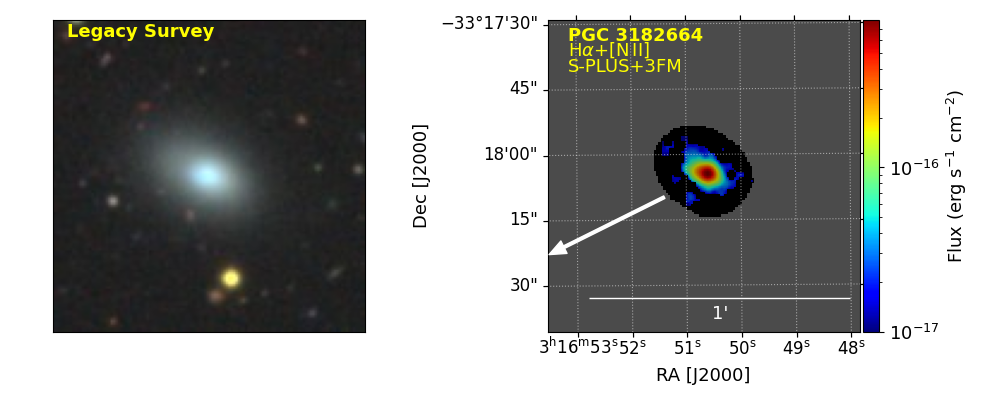}
    \includegraphics[width=0.49\textwidth]{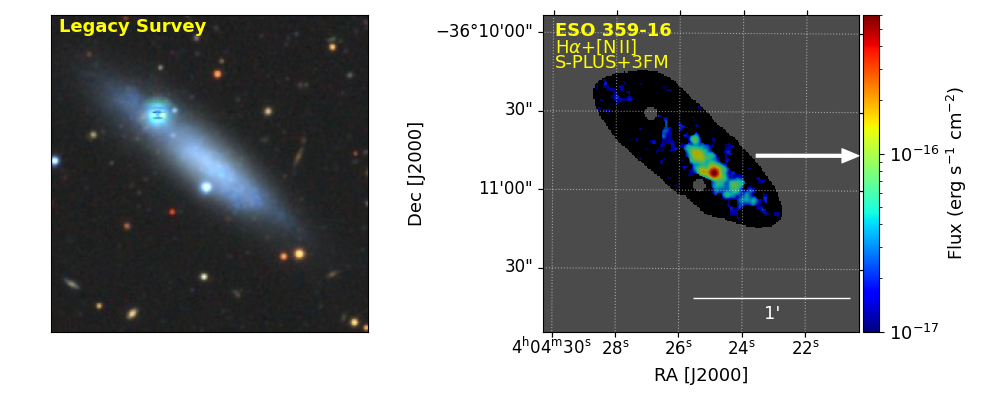}
 \caption{Same as Figure\,\ref{fig:0Rvir_0.7Rvir} but for emitters located at 1.85 \Rvir{} $\leq$ R < 2.65 \Rvir{}.}
\label{fig:1.99Rvir_2.7Rvir}
\end{figure*}

\begin{figure*}
    \centering
    \includegraphics[width=0.49\textwidth]{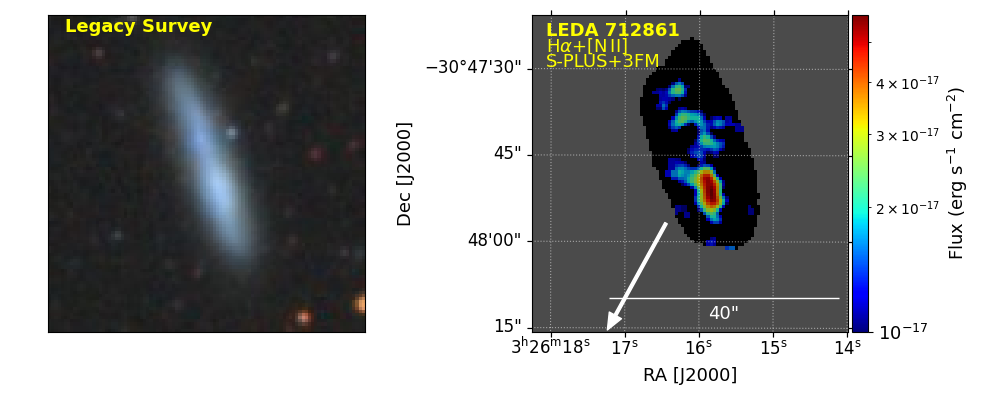}
    \includegraphics[width=0.49\textwidth]{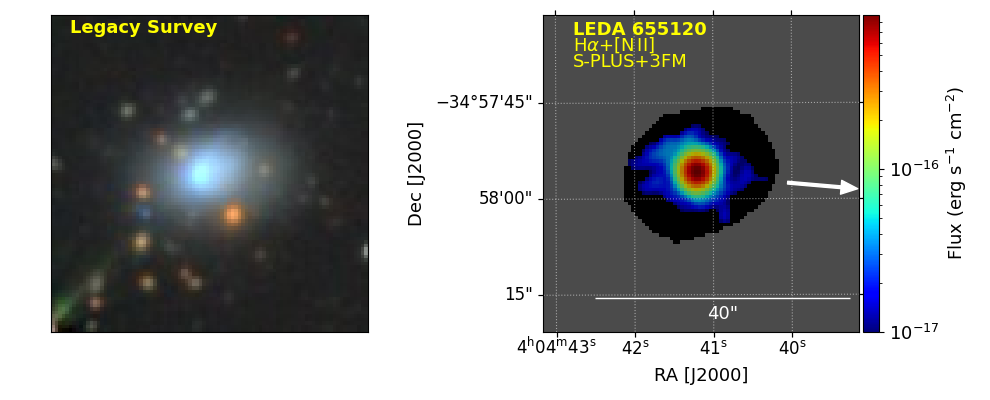}
    \includegraphics[width=0.49\textwidth]{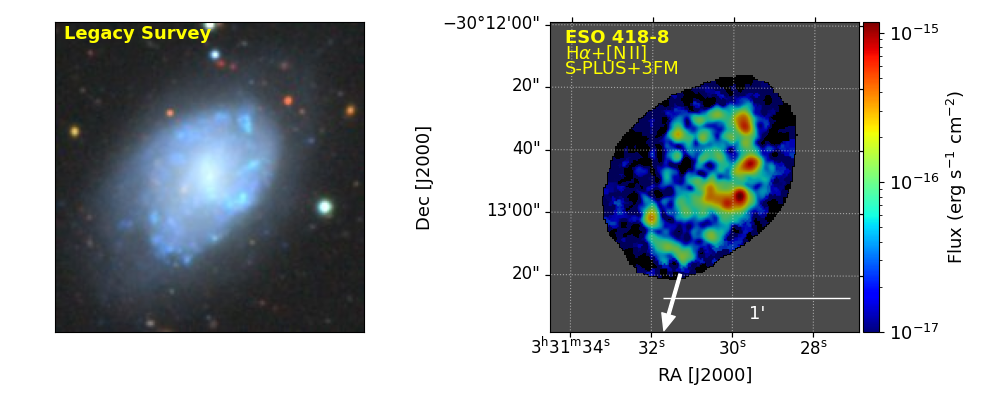}
    \includegraphics[width=0.49\textwidth]{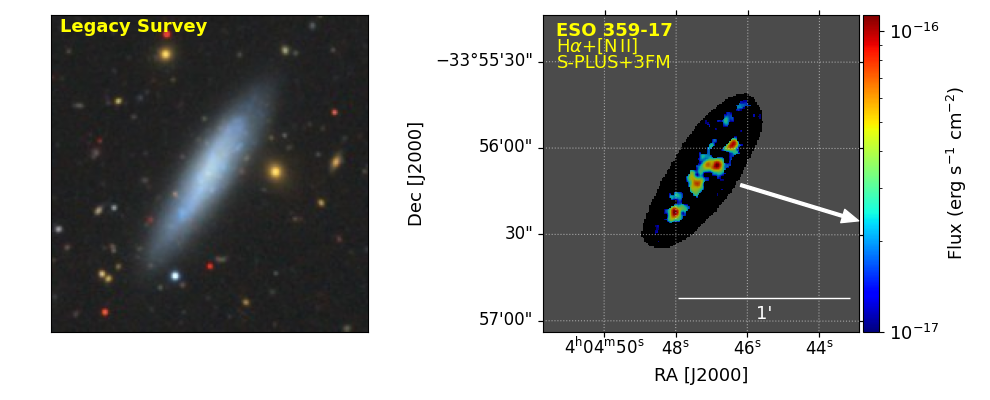}
    \includegraphics[width=0.49\textwidth]{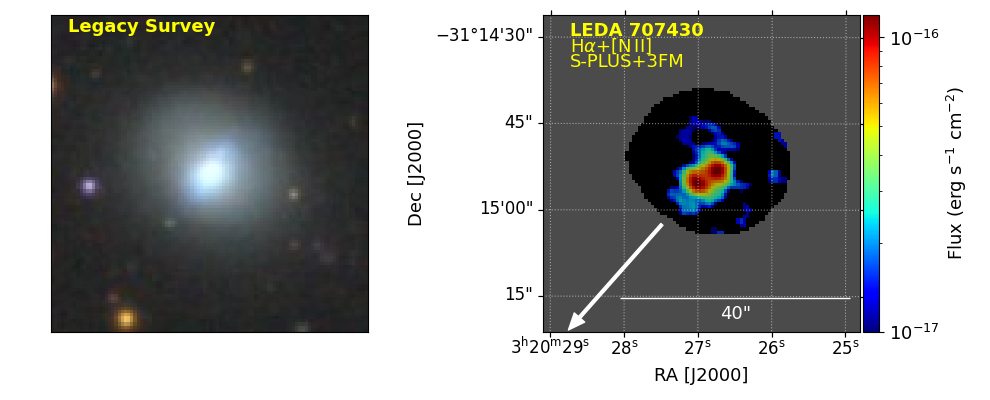}
    \includegraphics[width=0.49\textwidth]{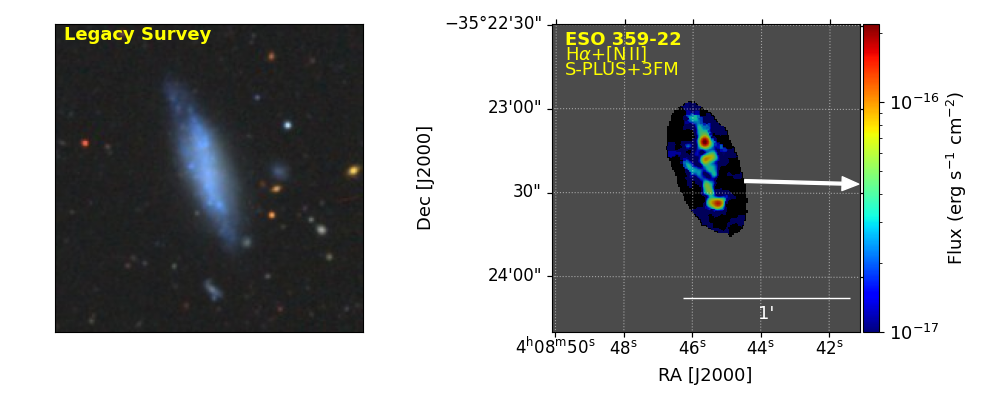}
    \includegraphics[width=0.49\textwidth]{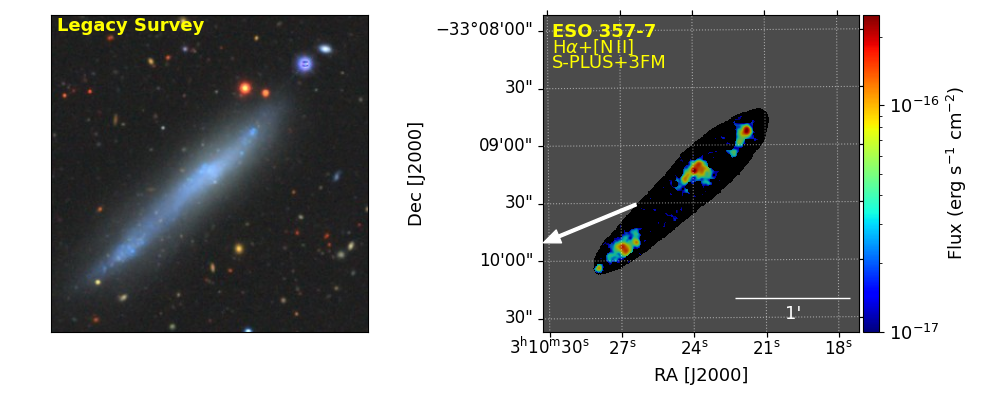}    
    \includegraphics[width=0.49\textwidth]{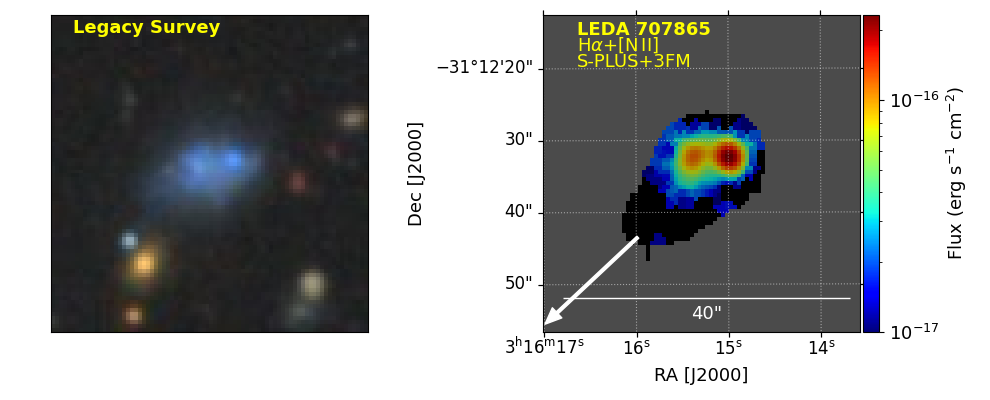}
    \includegraphics[width=0.49\textwidth]{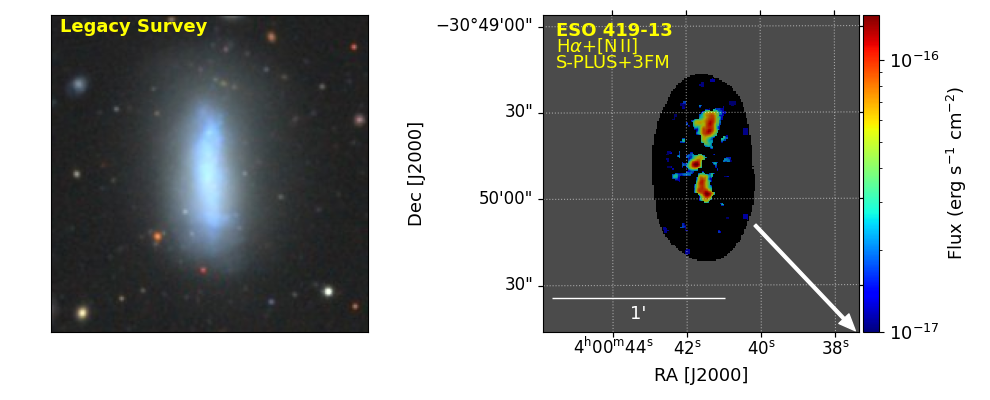}
    \includegraphics[width=0.49\textwidth]{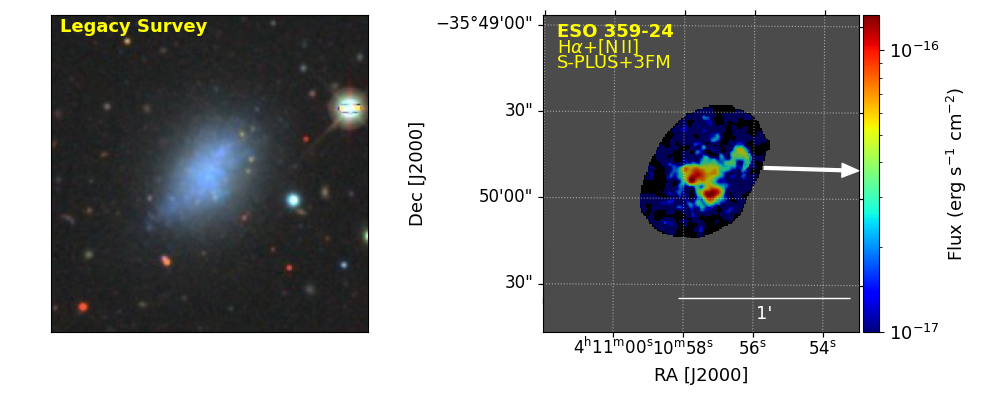}
    \includegraphics[width=0.49\textwidth]{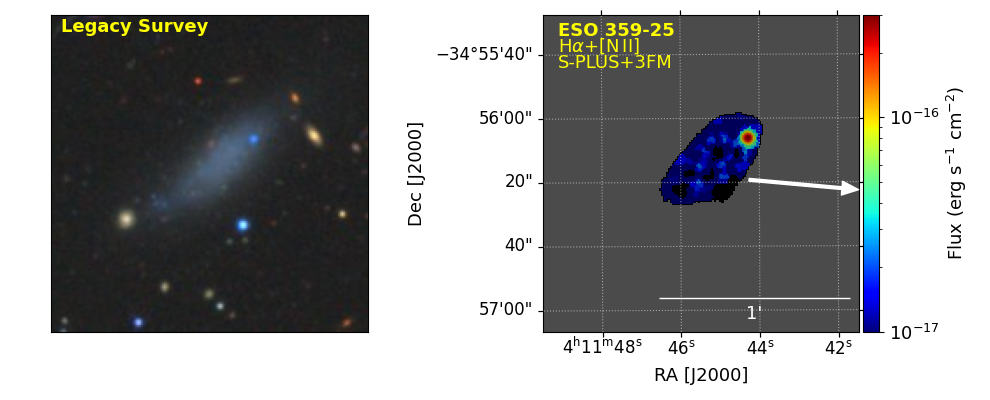}
    \includegraphics[width=0.49\textwidth]{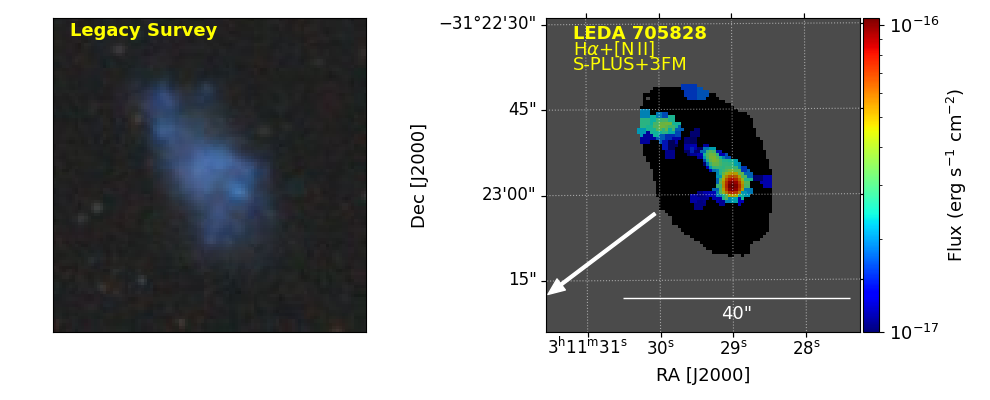}
  \caption{Same as Figure\,\ref{fig:0Rvir_0.7Rvir} but for emitters located at 2.65 \Rvir{} $\leq$ R < 3.47 \Rvir{}.}
\label{fig:2.7Rvir_3.7Rvir}
\end{figure*}

\begin{figure*}
    \centering
    \includegraphics[width=0.49\textwidth]{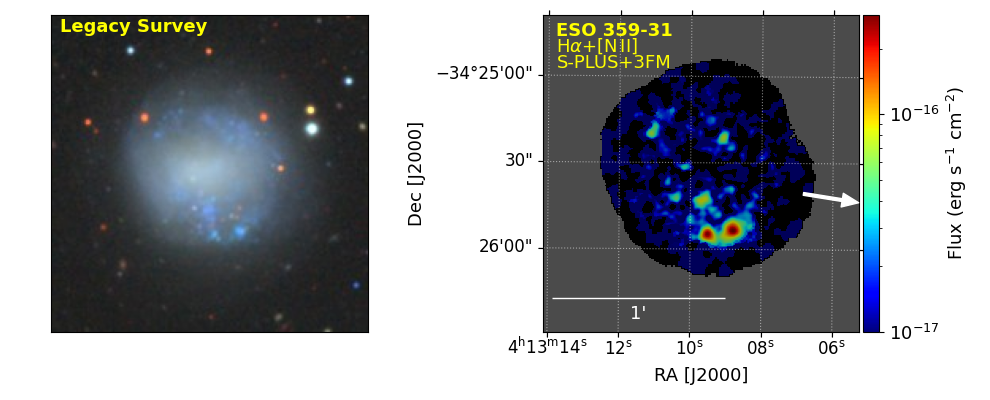}    
    \includegraphics[width=0.49\textwidth]{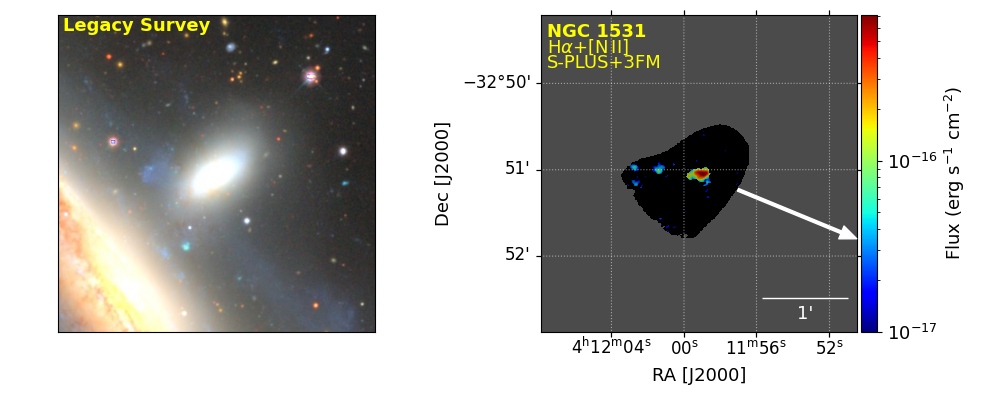}
    \includegraphics[width=0.49\textwidth]{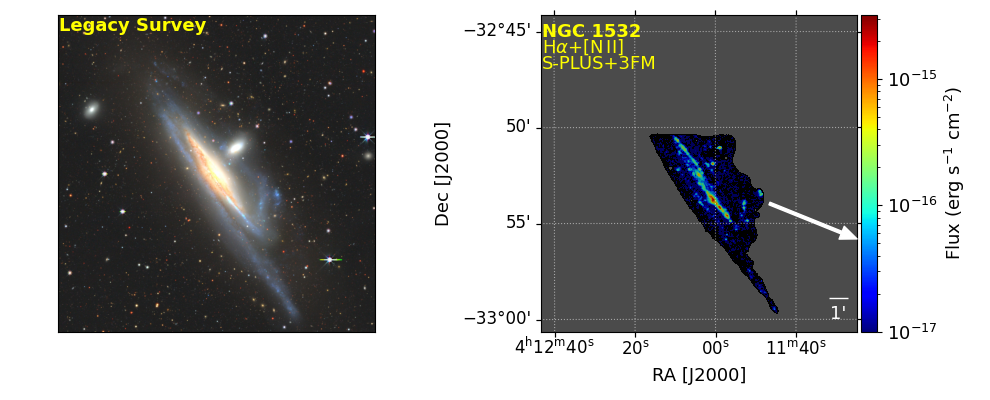}
    \includegraphics[width=0.49\textwidth]{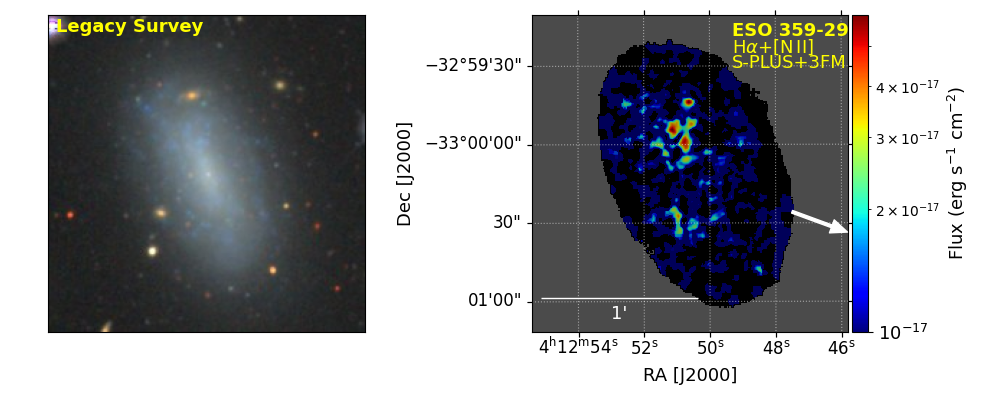}
    \includegraphics[width=0.49\textwidth]{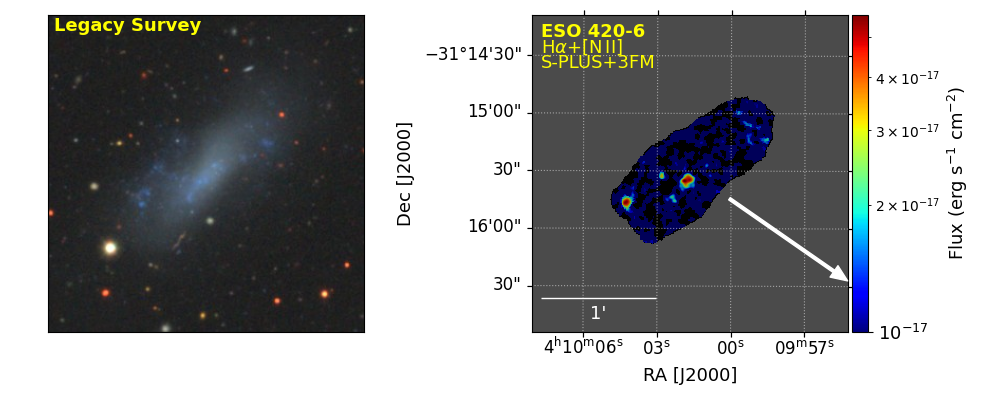}
    \includegraphics[width=0.49\textwidth]{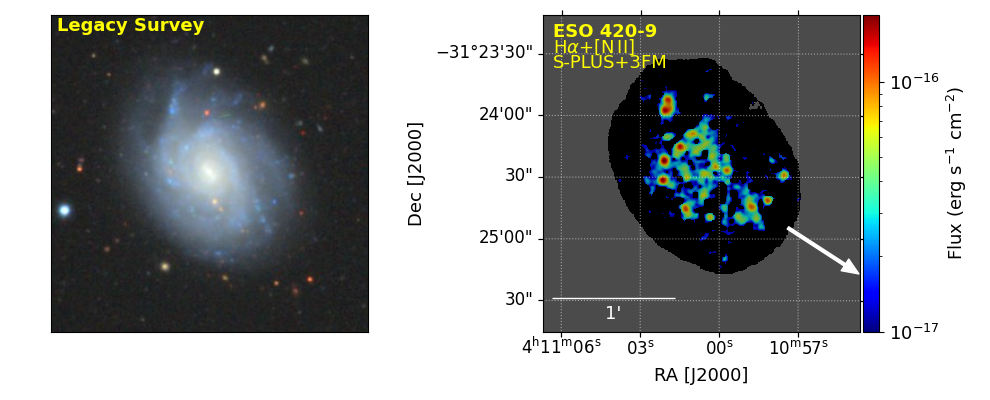}
    \includegraphics[width=0.49\textwidth]{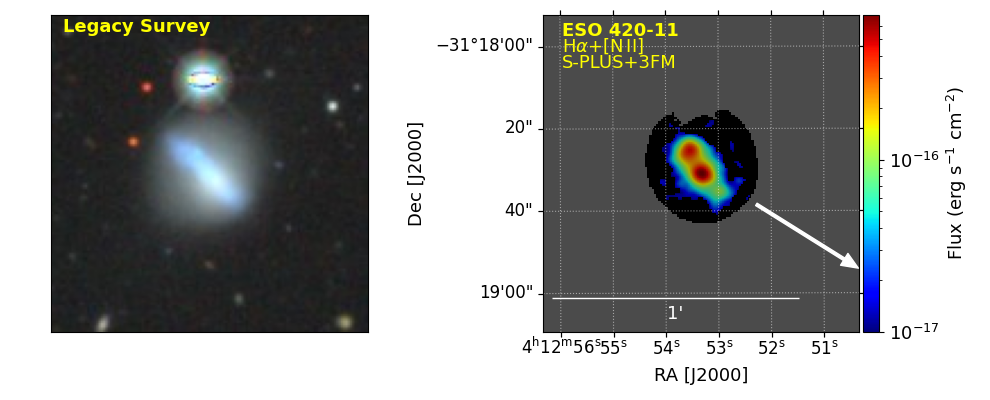}
 \caption{Same as Figure\,\ref{fig:0Rvir_0.7Rvir} but for emitters located at  R $\geq$ 3.47 \Rvir{}.}
 \label{fig:3.7Rvir_out}
\end{figure*}

\begin{landscape}
\begin{longtable}{lccccccc}

\caption{List of galaxies with derived \HalphaNII{} maps using 3FM with S-PLUS images.}\\
\label{tab:data_morph1}
{Name} & RA  & DEC & Vr & Optical Morphology  & Distance & Morphological Features & Emission Morphology \\
& (h m s) & ($^{\circ}$ $\arcmin$ $\arcsec$) & (km s$^{-1}$) & (Early/Late)  & ($R_\mathrm{vir}$)  & (unperturbed/perturbed/merger) & (central/knots/extended) \\
\hline
\endfirsthead
\caption{continued.}\\
\hline
{Name} & RA  & DEC & Vr & Optical Morphology  & Distance & Morphological Features & Emission Morphology \\
& (h m s) & ($^{\circ}$ $\arcmin$ $\arcsec$) & (km s$^{-1}$) & (Early/Late)  & ($R_\mathrm{vir}$)  & (unperturbed/perturbed/merger) & (central/knots/extended) \\
\hline
\endhead
\hline
\endfoot

NGC\,1387 & 03 36 57.0 & $-35$ 30 23 & 1302 & Early & 0.16 & unperturbed & central \\
NGC\,1427A & 03 40 09.0 & $-35$ 37 39 & 2028 & Late & 0.19 & perturbed/merger & knots \\
NGC\,1380 & 03 36 27.6 & $-34$ 58 33 & 1877 & Early & 0.31 & unperturbed & extended \\
NGC\,1386 & 03 36 46.2 & $-35$ 59 58 & 868 & Late & 0.32 & unperturbed & extended \\
FCC\,263 & 03 41 32.5 & $-34$ 53 20 & 1724 & Late & 0.42 & unperturbed & extended \\
FCC\,115 & 03 33 09.2 & $-35$ 43 05 & 1700 & Late & 0.56 & unperturbed & knots \\
NGC\,1436 & 03 43 37.0 & $-35$ 51 11 & 1387 & Late & 0.56 & unperturbed & extended \\
NGC\,1365 & 03 33 36.0 & $-36$ 08 28 & 1636 & Late & 0.60 & perturbed & extended \\
NGC\,1437A & 03 43 02.2 & $-36$ 16 23 & 886 & Late & 0.62 & perturbed & knots \\
FCCB\,905 & 03 33 57.3 & $-34$ 36 43 & 1242 & Late & 0.62 & unperturbed & extended \\
FCC\,113 & 03 33 06.8 & $-34$ 48 30 & 1388 & Late & 0.64 & perturbed & knots \\
FCC\,302 & 03 45 12.5 & $-35$ 34 19 & 803 & Late & 0.69 & perturbed/merger & knots \\
FCC\,312 & 03 46 18.9 & $-34$ 56 34 & 1929 & Late & 0.84 & perturbed/merger & extended \\
FCC\,90 & 03 31 08.3 & $-36$ 17 24 & 1813 & Early & 0.85 & unperturbed & extended \\
FCC\,306 & 03 45 45.4 & $-36$ 20 46 & 886 & Late & 0.86 & unperturbed & knots \\
NGC\,1437B & 03 45 54.7 & $-36$ 21 31 & 1497 & Late & 0.88 & perturbed/merger & extended \\
FCC\,282 & 03 42 45.5 & $-33$ 55 13 & 1225 & Early & 0.88 & unperturbed & extended \\
FCC\,267 & 03 41 45.4 & $-33$ 47 29 & 834 & Late & 0.90 & unperturbed & knots \\
FCC\,299 & 03 44 58.6 & $-36$ 53 41 & 2151 & Late & 0.97 & merger & knots \\
NGC\,1351A & 03 28 48.9 & $-35$ 10 43 & 1353 & Late & 0.99 & unperturbed & extended \\
FCC\,119 & 03 33 34.0 & $-33$ 34 23 & 1374 & Early & 1.07 & unperturbed & central \\
FCC\,B1019 & 03 35 28.9 & $-33$ 20 02 & 2013 & Late & 1.10 & unperturbed & knots\\
NGC\,1350 & 03 31 07.9 & $-33$ 37 44 & 1905 & Late & 1.18 & unperturbed & extended \\
FCC\,B1379 & 03 39 55.0 & $-33$ 03 11 & 709 & Early & 1.21 & unperturbed & knots \\
FCC\,315 & 03 47 05.0 & $-33$ 42 34 & 1080 & Late & 1.24 & unperturbed & extended \\
FCC\,76 & 03 29 43.2 & $-33$ 33 26 & 1808 & Early & 1.30 & unperturbed & knots \\
NGC\,1341 & 03 27 58.4 & $-37$ 08 58 & 1876 & Late & 1.36 & perturbed & extended \\
FCC\,32 & 03 24 52.4 & $-35$ 26 07 & 1318 & Early & 1.39 & unperturbed & central \\
NGC\,1326B & 03 25 19.8 & $-36$ 23 05 & 999 & Late & 1.41 & perturbed/merger & knots \\
NGC\,1326A & 03 25 08.6 & $-36$ 21 49 & 1831 & Late & 1.42 & unperturbed & knots \\
FCC\,139 & 03 34 57.4 & $-32$ 38 22 & 1769 & Late & 1.45 & merger & knots \\
FCC\,53 & 03 27 16.3 & $-33$ 29 09 & 1628 & Late & 1.52 & unperturbed & knots \\
FCC\,152 & 03 35 33.0 & $-32$ 27 54 & 1389 & Early & 1.53 & unperturbed & extended \\
FCC\,35 & 03 25 04.1 & $-36$ 55 39 & 1800 & Late & 1.54 & unperturbed & extended \\
NGC\,1326 & 03 23 56.3 & $-36$ 27 52 & 1360 & Late & 1.56 & unperturbed & central \\
NGC\,1316C & 03 24 58.3 & $-37$ 00 35 & 1800 & Early & 1.57 & unperturbed & extended \\
LSBG\,F359-060 & 03 54 03.4 & $-36$ 03 47 & 1389 & Late & 1.61 & unperturbed & knots \\
IC\,2006 & 03 54 28.4 & $-35$ 58 02 & 1382 & Early & 1.64 & unperturbed & extended \\
ESO\,359-3 & 03 52 00.9 & $-33$ 28 03 & 1574 & Late & 1.71 & merger & knots \\
LEDA\,170051 & 03 40 58.9 & $-38$ 51 18 & 829 & Late & 1.72 & unperturbed & knots \\
ESO\,358-11 & 03 30 07.9 & $-32$ 28 33 & $-$ & Early & 1.72 & unperturbed & central \\
NGC\,1317 & 03 22 44.3 & $-37$ 06 13 & 1941 & Late & 1.79 & unperturbed & extended \\
FCC\,28 & 03 23 54.4 & $-37$ 30 36 & 1405 & Late & 1.79 & perturbed/merger & knots \\
ESO\,302-9 & 03 47 33.7 & $-38$ 34 40 & 987 & Late & 1.81 & perturbed/merger & knots \\
NGC\,1316 & 03 22 41.7 & $-37$ 12 28 & 1803 & Early & 1.82 &  unperturbed & extended \\
FCC\,B112 & 03 22 38.1 & $-37$ 18 48 & 1686 & Late & 1.85 & merger & extended \\
NGC\,1310 & 03 21 03.5 & $-37$ 06 07 & 1805 & Late &1.94 & unperturbed & extended \\
FCC\,B2144 & 03 49 53.3 & $-32$ 15 33 & 1155 & Early & 1.99 & merger & extended \\
ESO\,302-14 & 03 51 40.9 & $-38$ 27 08 & 872 & Late & 2.00 & perturbed/merger & knots \\
NGC\,1406 & 03 39 23.2 & $-31$ 19 22 & 1075 & Late & 2.07 & perturbed/merger & extended \\
ESO\,418-9 & 03 31 55.7 & $-31$ 20 17 & 972 & Late & 2.17 & unperturbed & extended \\
NGC\,1366 & 03 33 53.7 & $-31$ 11 39 & 1231 & Early & 2.18 & unperturbed & central \\
ESO\,357-12 & 03 16 53.2 & $-35$ 32 28 & 1567 & Late & 2.20 & perturbed/merger & knots \\
FCC\,B35 & 03 17 30.5 & $-37$ 50 37 & 996 & Late & 2.42 & unperturbed & extended \\
FCC\,9 & 03 19 25.0 & $-32$ 38 58 & 1751 & Late & 2.42 & perturbed & extended \\
FCC\,10 & 03 19 34.6 & $-32$ 27 55 & 1443 & Late & 2.46 & merger & extended \\
PGC\,3182664 & 03 16 50.6 & $-33$ 18 04 & 1679 & Early & 2.48 & unperturbed & extended \\
ESO\,359-16 & 04 04 25.3 & $-36$ 10 53 & 1406 & Late & 2.65 & merger & extended \\
LEDA\,712861 & 03 26 16.1 & $-30$ 47 47 & 1529 & Late & 2.65 & unperturbed & extended \\
LEDA\,655120 & 04 04 41.1 & $-34$ 57 55 & 2341 & Early & 2.69 & perturbed/merger & extended \\
ESO\,418-8 & 03 31 30.6 & $-30$ 12 48 & 1195 & Late & 2.72 & perturbed & extended \\
ESO\,359-17 & 04 04 47.3 & $-33$ 56 08 & 1218 & Late & 2.81 & unperturbed & knots \\
LEDA\,707430 & 03 20 26.9 & $-31$ 14 53 & 1829 & Early & 2.82 & perturbed & extended \\
ESO\,359-22 & 04 08 45.6 & $-35$ 23 24 & 1430 & Late & 3.08 & merger & knots \\
ESO\,357-7 & 03 10 23.6 & $-33$ 09 14 & 1121 & Late & 3.12 & perturbed/merger & knots \\
LEDA\,707865 & 03 16 15.2 & $-31$ 12 34 & 1113 & Late & 3.14 & perturbed/merger & extended \\
ESO\,419-13 & 04 00 41.6 & $-30$ 49 50 & 1490 & Late & 3.28 & unperturbed & knots \\
ESO\,359-24 & 04 10 57.5 & $-35$ 49 51 & 849 & Late & 3.30 & unperturbed & extended \\
ESO\,359-25 & 04 11 45.4 & $-34$ 56 17 & $-$ & Late & 3.41 & unperturbed & knots \\
LEDA\,705828 & 03 11 29.4 & $-31$ 22 56 & 1559 & Late & 3.47 & unperturbed & extended \\
ESO\,359-31 & 04 13 09.7 & $-34$ 25 33 & 1495 & Late & 3.59 & unperturbed & knots \\
NGC\,1531 & 04 11 59.1 & $-32$ 51 02 & 1190 & Early & 3.70 &  unperturbed & knots \\
NGC\,1532 & 04 12 04.0 & $-32$ 52 22 & 1041 & Late & 3.70 & perturbed/merger & extended \\
ESO\,359-29 & 04 12 50.6 & $-33$ 00 10 & 879 & Late & 3.75 & unperturbed & knots \\
ESO\,420-6 & 04 10 01.6 & $-31$ 15 30 & 1388 & Late & 3.90 & unperturbed & knots \\
ESO\,420-9 & 04 11 00.6 & $-31$ 24 27 & 1367 & Late & 3.95 & unperturbed & extended \\
ESO\,420-11 & 04 12 53.3 & $-31$ 18 30 & 1387 & Late & 4.14 & unperturbed & extended \\
\end{longtable}
\end{landscape}
\FloatBarrier
\end{appendix}

\end{document}